\documentclass{aa}
\usepackage{txfonts}
\usepackage{float,epsfig,psfig}
\usepackage{pstricks}

\begin{document}
% \thesaurus{11.03.1, 12.03.3, 12.12.1
% }

\title{XMM-Newton study of the two-dimensional structure of the REFLEX-DXL galaxy clusters.}
\author{Alexis Finoguenov, Hans B\"ohringer, Yu-Ying Zhang}

%\offprints{Alexis Finoguenov, alexis@mpe.mpg.de}

\institute{Max-Planck-Institut f\"ur extraterrestrische Physik,
             Giessenbachstra\ss e, 85748 Garching, Germany }

\date{Received April 25, 2005; accepted June 24 2005}
\authorrunning{Finoguenov et al.}
\titlerunning{XMM-Newton maps of DXL clusters}

\abstract{We carry out a two-dimensional study of temperature, entropy and
pressure distributions in a nearly volume-limited subsample of REFLEX
clusters at redshift 0.3, the REFLEX-DXL. We use the observations gained by
XMM-Newton, which cover the central $1-2\times r_{500}$. We define the
substructure in both entropy and pressure as a deviation from the mean
profile of the sample. The non-parametric locally weighted regression
suggests a broken power law approximation to the entropy profile with inner
and outer slopes of 0.78 and 0.52, respectively, and a break at
$0.5r_{500}$. The characterization of the pressure profile is more complex,
requiring three power laws, with slopes $-0.64$ at $r<0.3r_{500}$, $-2.47$
at $r>0.5r_{500}$ and a slope of $-1.50$ in between. An analysis of the
substructure in the pressure and entropy maps reveals somewhat larger
fluctuations around the mean pressure profile compared to the
entropy. Typically, pressure fluctuations are found on the 30\% level, while
the entropy fluctuations are at the 20\% level (r.m.s.). We compare the
cumulative distribution of the substructure level in the REFLEX-DXL sample
with the results of numerical simulation and by means of KS test show that
they are in agreement. A discussion of the origin of the substructure is
provided on individual cluster basis.   }

\maketitle
%\oneandhalfspace

\section{Introduction}

Use of clusters of galaxies for cosmological tests in undergoing and future
surveys rely on the understanding of the cluster physics. A number of
observational and theoretical work has been devoted to studying the effects
of non-gravitational heating in explaining the observed scaling of clusters
with mass (see Voit 2004 for a review). Another long recognized issue
consists in the effect of merging on the appearance of clusters
(e.g. Randall et al. 2002; Rowley et al. 2004). It is widely accepted that
the later effect dominates the appearance of the most massive clusters,
where it also could be more pronounced, due to dependence of the accretion
history of clusters on their mass. Thus, a study of the dynamical state of a
representative sample of massive clusters appears to be well grounded. In
this paper we explore the structure of individual clusters from a
volume-limited sample of most massive clusters at a redshift of 0.3 to
provide an in-depth coverage of the properties that define both the X-ray
luminosity as well as the thermal Sunyaev-Zel'dovich effect (SZE) of
clusters.

\subsection{The REFLEX-DXL sample}

The REFLEX-DXL galaxy cluster sample, comprising {\bf d}istant {\bf X}-ray
{\bf l}uminous objects within REFLEX, was constructed from the REFLEX galaxy
cluster survey covering the ROSAT detected galaxy clusters above a flux
limit of $3 \cdot 10^{-12}$ erg s$^{-1}$ cm$^{-2}$ in the 0.1 to 2.4 keV
band in an 4.24 ster region of the southern sky (see B\"ohringer et al. 2001
for details). The REFLEX-DXL clusters form a nearly volume limited subset of
REFLEX in the redshift range 0.27 to 0.31 including 13 members plus a
fourteenth cluster which was previously assigned to the sample but had to be
excluded after a revision of the optically determined redshift. We have
included this cluster here since it nicely fits into the homogeneous
observational data properties of the REFLEX-DXL sample. All clusters exceed
a luminosity of $10^{45}~ h_{50}^{-1}$ erg s$^{-1}$ in the ROSAT band (0.1
to 2.4 keV). For five of the clusters the XMM observations were affected by
strong proton flares. Their re-observation, which will be reported
elsewhere, have provided us with good results except for one cluster,
RXCJ2011--5725. One of the clusters included in the present study,
RXCJ0658--5557, was a verification phase target with a significantly longer
nominal observing time. Further properties of the REFLEX-DXL clusters are
described in Zhang et al. (2004) and B\"ohringer et al. (2005, in prep.).

In Table \ref{t:ol} we summarize the properties of the clusters used in this
study. Col. (1) gives the RXCJ designation of the source, (2--4) optical
redshift, corresponding luminosity distance, and the conversion factor from
apparent angular to physical scale (calculated for
$\Omega_M=1-\Omega_\Lambda=0.3$, and a Hubble constant of 70 km sec$^{-1}$
Mpc$^{-1}$). Col. (5) gives the measured position of the cluster center,
used in the analysis of cluster structure, (6) is the mean temperature $T_w$
from Table 4 of Zhang et al. (2004), derived using the
$0.5^\prime<r<4^\prime$ region and the energy band $1-10$ keV. For the
bullet cluster (RXCJ0658--5557), we use the 0.4--10 keV band and remove the
soft excess by an additional spectral component, for the reasons explained
in detail below. (7) is $r_{500}$ used for the scaling plots (an estimated
radius encompassing the density equal to $500\rho_{\rm crit}$), (8) lists
the other names of the clusters.

The calculation of $r_{500}$ is following $r_{500}=0.45 {\rm Mpc} \times
\sqrt{kT_w/{\rm keV}} h_{70}^{-1} h(z)^{-1}$, where the scaling in
Finoguenov et al. (2001) for $h_{50}$ ($r_{500}\approx 0.63 {\rm
Mpc}\sqrt{kT/{\rm keV}}$) is translated into our assumption for
$h_{70}=1$. We use \mbox{$h(z)=(\Omega_M (1+z)^3 + \Omega_\Lambda)^{1/2}$},
suitable for our choice of the cosmological model. In Finoguenov et
al. (2001) it has been demonstrated that the cosmological corrections are
negligible in deriving the scaling for $r_{500}$ in their sample of local
clusters. These corrections are, however important for REFLEX-DXL.

The suggested modified entropy scaling reads $S \sim T_w^{0.65\sim 2/3}
h(z)^{-4/3}$ (Ponman et al. 2003). In the analysis of clusters, we will also
present scaled pressure plots. As entropy, $T n^{-2/3}$, scales as
$T_w^{2/3}$, the density scales as $T_w^{1/2}$ and the pressure $T n \sim
T_w^{3/2}$. Finally the correction for the evolution of the critical density
is $h(z)^2$. Throughout the paper we will use the corrections by $T_{10}=
{T_w \over 10 {\rm keV}}$. For the temperature profiles, however, we simply
correct for the $T_w$ to match other studies.

\begin{table*}[ht]
\begin{center}
\renewcommand{\arraystretch}{1.1}\renewcommand{\tabcolsep}{0.12cm}
\caption{\footnotesize
\centerline{Scaling assumptions for the analyzed clusters.}}
\label{t:ol}%

\begin{tabular}{ccccccccc}
\hline
\hline
RXCJ  &   & $D_l$& plate             &cluster center&$kT_{ew}$& $r_{500}$ & other\\
name  &     z  & Mpc  & scale        &R.A., Decl.   &keV& Mpc & names\\
      &        &      & kpc/$^\prime$&FK5 (Eq.J2000)&&  & \\
\hline
0014.3--3022&  0.3066 &933& 271& 00:14:19.288 --30:23:07.41  & $8.3\pm0.4$  & 1.11  &     A2744, AC118\\
0043.4--2037&  0.2924 &903& 263& 00:43:24.446 --20:37:30.72  & $6.8\pm0.4$  & 1.01  &     A2813\\
0232.2--4420&  0.2836 &884& 257& 02:32:18.561 --44:20:48.40  & $7.6\pm0.4$  & 1.07  &     \\
0307.0--2840&  0.2578 &825& 240& 03:07:02.084 --28:40:00.21  & $6.6\pm0.3$  & 1.01  &     A3088\\
0528.9--3927&  0.2839 &885& 257& 05:28:52.731 --39:28:24.82  & $7.7\pm0.6$  & 1.08  &     \\
0532.9--3701&  0.2747 &864& 251& 05:32:56.043 --37:01:33.52  & $7.8\pm0.6$  & 1.09  &     \\
0658--5557  &  0.2965 &912& 265& 06:58:31.453 --55:56:16.74  & $12.3\pm0.3$ & 1.37  &     1ES0657-558\\
1131.9--1955&  0.3075 &935& 272& 11:31:55.742 --19:55:42.82  & $7.4\pm0.5$  & 1.05  &        A1300\\
2337.6+0016&  0.2779 &871& 253& 23:37:38.323  +00:16:05.07  & $7.5\pm0.4$  & 1.07  &     A2631\\
\hline
\end{tabular}
\end{center}
\end{table*}

\section{Analysis}

The main goal of the primary data reduction is to produce soft and hard
images based on merging the data obtained by all EPIC detectors. Initial
steps of the data reduction include the XMMSAS event processing, and light
curve screening, which is similar to the approach adopted by other groups
(e.g. De Luca \& Molendi 2003). The background subtraction has been
described in Zhang et al. (2004).

With these screened photon event files we produced MOS and pn images of the
individual observations in the energy bands from 0.5 to 2.0 and 2.0 to 7.5
keV with more details available in Briel et al. (2004).

The broad-band images can be used for making visible intensity structures,
and variations of the temperature of the X-ray emitting plasma by producing
hardness-ratio maps, which can be uniquely translated into a plasma
temperature distributions. In addition, one can produce pressure and entropy
maps of the plasma by combining the surface brightness map and the hardness
ratio map. Useful hardness ratio maps can only be produced from smoothed
surface brightness maps. There exists a variety of different smoothing
procedures like top-hat smoothing, Gaussian smoothing or adaptive smoothing
(e.g.  Churazov et al. 1999). We applied the wavelet decomposition method,
which is described in detail by Vikhlinin et al. (1998). The advantage of
using wavelets consists in background removal by spatial filtering and a
control over the statistical significance of the detected
structures. Complications arise due to splitting the image into discrete
scales, which we overcome by additional smoothing applied before producing
the hardness ratio map. The use of wavelets provides us with a decent method
to identify the regions susceptible to temperature variations.  Another
important feature is the high spatial resolution achieved in detecting the
structure, as wavelets do not smear the data.

For every cluster, we show the results of the broad-band image
investigation, an image in the 0.5--2 keV energy band, the hardness of the
emission, deduced from the ratio of the wavelet-reconstructed images in the
0.5--2 and 2--7.5 keV bands, as well as the projected pressure and entropy
maps, constructed using the wavelet-smoothed surface-brightness map in the
0.5--2 keV energy range as an indicator of the electron density squared and
the hardness ratio map as temperature distribution, and using the
definitions of the pressure $P\sim T\times \sqrt{I}$ and of the entropy
$S\sim T / \sqrt[3]{I}$. In addition to visualizing the temperature
distribution through hardness ratios we also determined the local
temperatures spectroscopically.

The spectral analysis was carried out using only the pn data. Background
subtraction here is more demanding and based on the XMM observations of the
Chandra Deep Field South, as described in Zhang et al. (2004). We have
selected the 0.4--7.9 keV band for the analysis of all the clusters and
added a soft ($kT\sim0.2 keV$) spectral component to remove any soft
excess. For the most affected system, the bullet cluster (RXCJ0658--5557),
we find a good agreement between the mean temperatures using this method
($12.3\pm0.3$) and the 1--10 keV band ($11.8\pm0.3$).

In the spectral analysis we produced two masks per cluster defining the
spectral extraction region, one to confirm the temperature structure,
combined with the image, the other -- to confirm pressure and entropy
structure. We combine the regions so that counting statistics are not the
limiting factor in our derivation of the cluster properties. We use the
wavelet-based maps to identify regions with similar X-ray colors and
intensity levels. To generate the mask file for use in the subsequent
spectral analysis, we sample the changes in the intensity and hardness ratio
at the precision allowed by the statistics of the data. We then examine each
of the isolated regions with approximately equal color and intensity,
imposing an additional criterion that the regions should be larger than the
PSF width ($15^{\prime\prime}$) and contain more than 500 counts in the
background corrected pn image.  When sampling the pressure and entropy
structure, the region selection should be fine to achieve a nearly equal
temperature within the region, as the X-ray data is only sensitive to
pressure and entropy by measuring the temperature and normalization of the
spectra. So, we used finer region selection with a reduced threshold on a
number of counts (300), resulting in a higher number of regions at expense
of larger uncertainty in the parameter determination.  So, in the paper we
tabulate mostly the confirmation of the temperature structure, with
exception of RXCJ0232.2--4420 and the bullet cluster, where the other mask
was more useful.

For every cluster we provide a table, containing the measured values with
their $\pm1\sigma$ errors for one parameter of interest. As all these tables
are similar, we give a single description for all here. Col. (1) labels the
region according to region selection shown in the accompanied figure,
column (2) lists the temperature in keV. Derived quantities, that use an
estimate of the projected length, as described below are reported in
cols. (3--6). These are electron density, entropy, pressure, and the (local)
gas mass. Cols. 7--8 report the minimal ($r_{min}$) and maximal ($r_{max}$)
radii of the extraction area, col. (9) provides remarks on the region.

For this detailed analysis we also perform an estimate of the projection
length of each analyzed region to obtain actual gas properties at these
locations, as described at length in Henry et al. (2004) and Mahdavi et
al. (2005). To avoid the importance of the projection effects, we discard
the regions having a ratio of the minimal to the maximal radii of values
exceeding 0.8.

\section{Results}

\begin{figure}
\includegraphics[width=8cm]{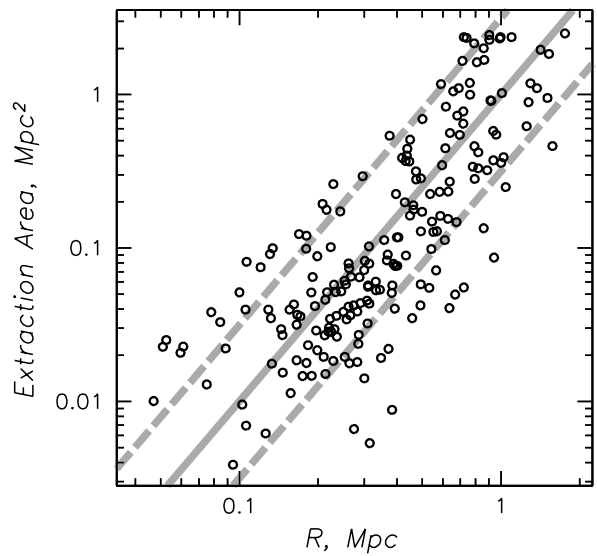}

\caption{Characterization of the ability of the current masking
technique to reveal spatial variations in the spectral properties of gas as
a function of clustercentric distance (window function). The points
indicate an extraction area in Mpc$^2$ as a function of radius of the zone,
calculated as $(r_{\rm min}+r_{\rm max})/2$. The grey lines indicate a
quadratic dependence of the area on the radius. The data is in remarkably
good agreement with these curves, indicating that the sampling corresponds
roughly to a grid in cylindrical coordinates. The upper dashed line
correspond to the case of no sampling on azimuth, solid line -- having 3
regions and low dashed line -- 10 to sample the azimuthal scale.}
\label{f:tf}%
\end{figure}

The statistics achieved in the observation of the REFLEX-DXL clusters allows
us only to recognize the strongest fluctuations in either temperature,
entropy, or pressure.  In the presentation of the results we indicate the
features seen in the hardness ratio based maps and discuss how much it is
possible to confirm them through a direct spectroscopic analysis. In
selecting the regions according to their properties or according to their
statistics, we implicitly perform an adaptive filtering of the signal. It is
therefore important to characterize the spatial frequencies sampled in the
analysis, which is also a way to characterize the analysis carried out
and the cluster spatial scales sampled. In Fig.\ref{f:tf} we present such
an analysis, where it becomes clear that the choice of the regions
corresponds to a grid in cylindrical coordinates, sampling the azimuthal
angle with typically 3 sectors on radial scales from 0.1 to 1 Mpc.

Since there is a discussion on the reliability of the temperature
determination, we have provided a plot for each cluster, where a comparison
to the average cluster temperature profile of Vikhlinin et al. (2005) is
presented. One can see in each case that there is a good agreement in the
results. We also point out that a few clusters where we probed the region
outside $r_{500}$, reveal strong asymmetries indicating accretion from a
filament. It may be that the presence or removal of such zones could be the
underlying reason for some of the reported disagreement in temperature
profiles.

We define a cool core of a cluster or its debris as the gas with entropy
significantly below 200--300 keV cm$^2$, which according to Voit \& Bryan
(2001) could cool in a Hubble time. As local examples of cool cores have
entropies lower than 100 keV cm$^2$, we have used the later criterion for
the cooling core identification.

\begin{figure*}
\includegraphics[width=8cm]{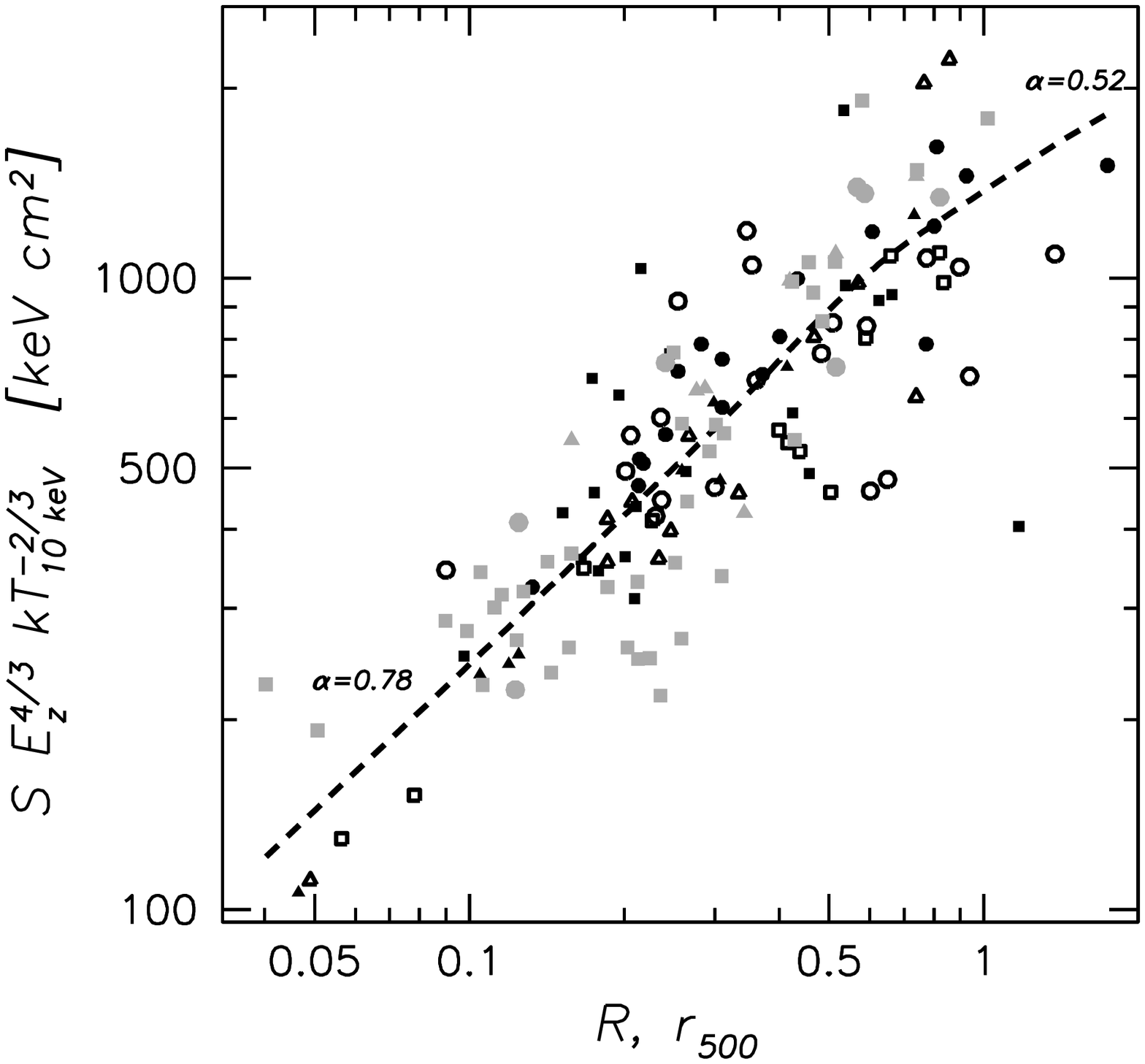}\hfill\includegraphics[width=8cm]{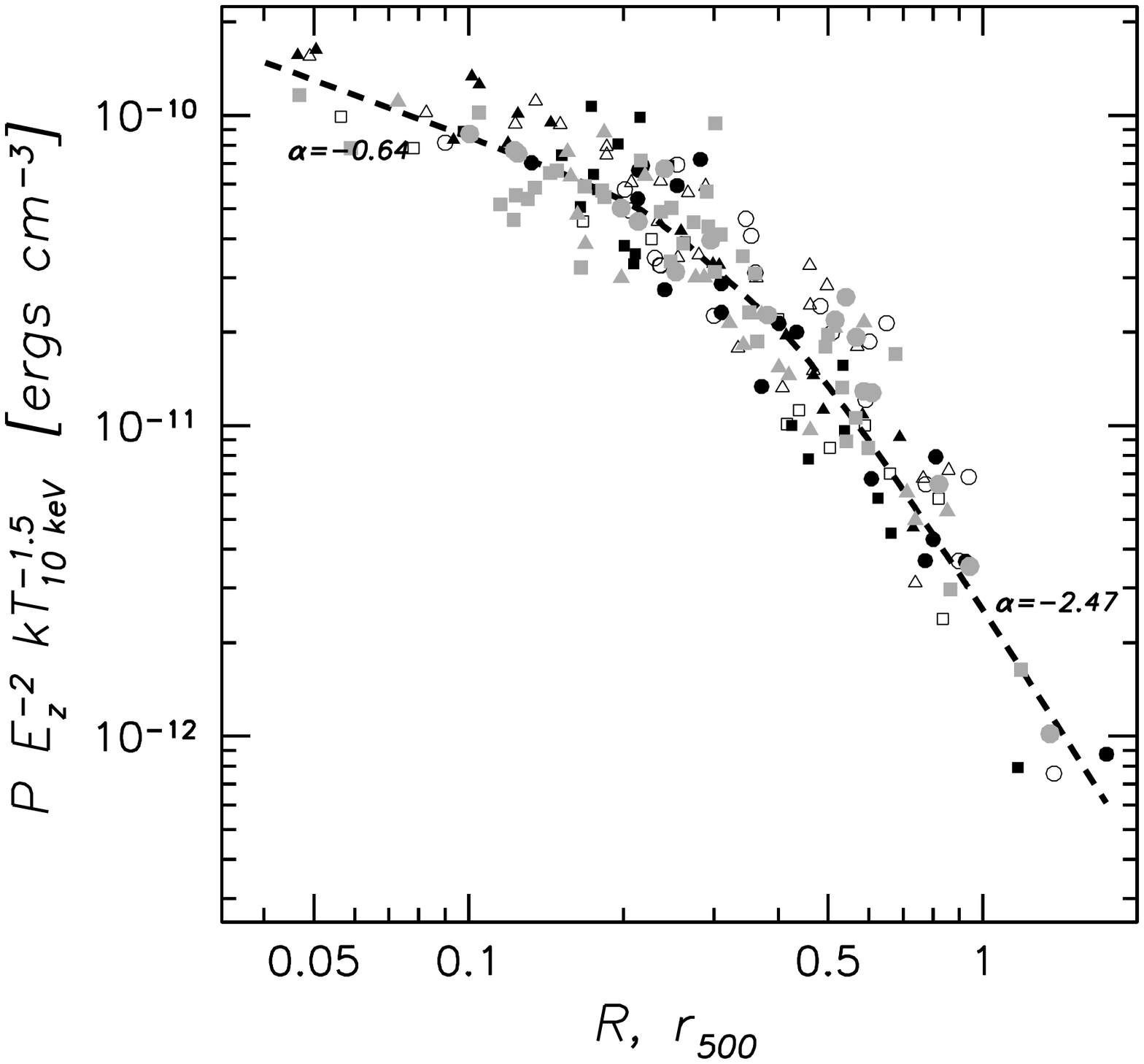}

\caption{Comparison between the entropy and pressure of the sample and
the analytical approximation, used to study the dispersion. The entropy and
pressure points corresponding to the same cluster are shown using the same
symbol. The dashed line show the results of the fit using the non-parametric
locally weighted regression method.}
\label{f:comp}%
\end{figure*}

Before proceeding with the description of individual systems, we summarize
the results by presenting the fits to the entropy and pressure profiles. To
define the shape of the entropy and pressure profiles, we applied the
non-parametric locally weighted regression, following Sanderson et al. (2005
and references therein). This analysis results in the non-parametric curve,
which we approximate below with power laws. Our analysis illustrated in
Fig.\ref{f:comp} suggests a broken power law approximation to the entropy
profile with an inner and an outer slopes of 0.78 and 0.52, respectively,
and a break at $0.5r_{500}$. The amplitude of fluctuations around the best
exceeds the effect of the statistics and is a measure of the important of
substructure, as discussed below. The average level of fluctuations, which
is 15\% (20\%) in the case of entropy (pressure), could be taken as the
accuracy to which the approximation to the entropy distribution could be
determined.  The entropy profiles with exclusion of the substructure has
been analyzed in Zhang et al. (2005), yielding a steeper index of 0.95. The
characterization of the pressure profile is more complex, requiring three
power laws, with slopes $-0.64$ at $r<0.3r_{500}$, $-2.47$ at $r>0.5r_{500}$
and a slope of $-1.50$ in between. For completeness, we present in
Tab.\ref{t:cl-sp-all} a standard approach of using the orthogonal regression
and assuming a power law shape to approximate the shape of the entropy and
pressure profiles. We present the results obtained using different masks and
also combine the clusters with and without the rescaling, described above.
As could be easily seen from Tab.\ref{t:sc0}, this approach results in a
much larger amplitude of the residuals, even taking into account that the
power law shape was fitted separately for each cluster, while in the
non-parametric approach, one shape is used to approximate all clusters.

\begin{table}[ht]
\begin{center}
\renewcommand{\arraystretch}{0.9}\renewcommand{\tabcolsep}{0.09cm}
\caption{\footnotesize Results of the orthogonal regression analysis of
  entropy and pressure profiles.} 
\label{t:cl-sp-all}%
\begin{tabular}{rccccc}
\hline
\hline
 \multicolumn{1}{c}{Name} & S($0.2r_{500}$) &   S slope  &  P($0.2r_{500}$) 
& P slope\\
RXCJ & keV cm$^2$ & & \multicolumn{2}{l}{$10^{-11}$ ergs cm$^{-3}$} \\
\hline
0014.3--3022 & $373\pm21$& $0.30\pm0.12$& $7.8\pm1.1$& $-1.70\pm0.34$ \\
            & $352\pm29$& $0.40\pm0.13$&$13.9\pm1.6$& $-2.28\pm0.23$ \\
0043.4--2037 & $416\pm32$& $0.05\pm0.30$& $5.2\pm0.7$& $-1.69\pm0.21$ \\
            & $287\pm17$& $0.61\pm0.16$& $5.1\pm0.3$& $-1.92\pm0.12$ \\
0232.2--4420 & $280\pm18$& $0.60\pm0.19$& $4.3\pm0.4$& $-1.35\pm0.16$ \\
            & $236\pm46$& $0.50\pm0.62$& $4.2\pm0.6$& $-0.91\pm0.40$ \\
0307.0--2840 & $222\pm 7$& $0.91\pm0.12$& $2.8\pm0.2$& $-1.41\pm0.12$ \\
            & $243\pm 3$& $0.93\pm0.06$& $3.3\pm0.2$& $-1.35\pm0.20$ \\
0528.9--3927 & $173\pm17$& $0.86\pm0.21$& $4.4\pm0.4$& $-1.55\pm0.20$ \\
            & $124\pm24$& $1.10\pm0.36$& $4.9\pm0.5$& $-1.64\pm0.43$ \\
1131.9--1955 & $279\pm22$& $0.73\pm0.24$& $6.2\pm0.7$& $-1.72\pm0.24$ \\
            & $238\pm10$& $0.79\pm0.18$& $5.9\pm0.5$& $-1.83\pm0.21$ \\
2337.6+0016 & $324\pm20$& $0.80\pm0.15$& $3.9\pm0.2$& $-1.20\pm0.15$ \\
            & $354\pm35$& $0.63\pm0.30$& $4.3\pm0.7$& $-1.39\pm0.48$ \\
0532.9--3701 & $435\pm33$& $0.11\pm0.10$& $2.8\pm0.7$& $-0.64\pm0.52$ \\
            & $296\pm13$& $0.21\pm0.23$& $6.8\pm0.3$& $-2.18\pm0.13$ \\
0658--5557   & $382\pm15$& $0.86\pm0.11$& $8.3\pm0.5$& $-1.34\pm0.21$ \\
            & $379\pm16$& $0.94\pm0.09$& $7.5\pm0.4$& $-1.62\pm0.17$ \\
all         & $290\pm 8$& $0.62\pm0.09$& $5.4\pm0.2$& $-1.54\pm0.10$ \\
$r>0.25r_{500}$&$381\pm53$& $0.38\pm0.27$& $11.4\pm1.0$& $-2.15\pm0.17$\\
$r<0.25r_{500}$&$388\pm31$& $1.33\pm0.37$& $6.0\pm0.5$& $-0.77\pm0.44$\\
T\&z-corrected &$452\pm11$& $0.64\pm0.08$& $6.8\pm0.2$& $-1.43\pm0.10$\\
$r>0.25r_{500}$&$570\pm72$& $0.44\pm0.24$& $11.5\pm1.1$& $-1.88\pm0.18$\\
$r<0.25r_{500}$&$509\pm31$& $1.10\pm0.27$& $5.8\pm0.5$& $-1.09\pm0.45$\\
\hline
\end{tabular}
\end{center}
\end{table}

\begin{table*}[ht]
\begin{center}
\renewcommand{\arraystretch}{0.9}\renewcommand{\tabcolsep}{0.05cm}
\caption{\footnotesize Entropy and pressure fluctuations
around the power law approximation, reported in Tab.\ref{t:cl-sp-all}.}
\label{t:sc0}%

\begin{tabular}{rccccccc}
\hline
\hline
 Name & $\sigma$S &   $\sigma$P &  $\sigma$S &   $\sigma$P& $\sigma$S &   $\sigma$P \\
RXCJ &\multicolumn{2}{c}{$r<0.3r_{500}$}&\multicolumn{2}{c}{$r>0.2r_{500}$}
& \multicolumn{2}{c}{from individual fits}\\
\hline
0014.3--3022 &$0.58\pm0.07$&$0.62\pm0.07$&$0.26\pm0.07$&$0.45\pm0.05$&$0.26\pm0.07$&$0.75\pm0.08$\\
            &$0.47\pm0.07$&$0.74\pm0.06$&$0.10\pm0.09$&$0.47\pm0.07$&$0.12\pm0.05$&$1.05\pm0.08$\\
0043.4--2037 &$0.49\pm0.06$&$0.64\pm0.11$&$0.44\pm0.12$&$0.44\pm0.20$&$0.47\pm0.14$&$1.04\pm0.14$\\
            &$0.44\pm0.11$&$0.64\pm0.09$&$0.08\pm0.08$&$0.11\pm0.08$&$3.61\pm0.25$&$0.91\pm0.42$\\
0232.2--4420 &$0.34\pm0.16$&$0.52\pm0.06$&$0.46\pm0.21$&$0.47\pm0.16$&$0.40\pm0.20$&$0.58\pm0.05$\\
            &$0.21\pm0.04$&$0.62\pm0.15$&$0.25\pm0.05$&$0.41\pm0.16$&$0.25\pm0.05$&$1.00\pm0.11$\\
0307.0--2840 &$0.29\pm0.06$&$0.28\pm0.05$&$0.26\pm0.07$&$0.24\pm0.07$&$0.28\pm0.06$&$0.25\pm0.06$\\
            &$0.10\pm0.03$&$0.45\pm0.05$&$0.16\pm0.04$&$0.17\pm0.04$&$0.04\pm0.02$&$0.48\pm0.05$\\
0528.9--3927 &$0.28\pm0.12$&$0.46\pm0.08$&$0.22\pm0.11$&$0.36\pm0.13$&$0.33\pm0.11$&$0.39\pm0.10$\\
            &$0.16\pm0.06$&$0.67\pm0.08$&$0.14\pm0.10$&$0.40\pm0.09$&$0.17\pm0.06$&$0.53\pm0.09$\\
1131.9--1955 &$0.42\pm0.05$&$0.45\pm0.07$&$0.29\pm0.05$&$0.32\pm0.06$&$0.37\pm0.04$&$0.50\pm0.07$\\
            &$0.57\pm0.05$&$0.74\pm0.08$&$0.29\pm0.07$&$0.44\pm0.08$&$0.78\pm0.05$&$0.81\pm0.07$\\
2337.6+0016 &$0.31\pm0.08$&$0.44\pm0.18$&$0.26\pm0.08$&$0.38\pm0.20$&$0.23\pm0.09$&$0.49\pm0.16$\\
            &$0.22\pm0.05$&$0.41\pm0.11$&$0.15\pm0.05$&$0.23\pm0.16$&$0.15\pm0.05$&$0.54\pm0.09$\\
0532.9--3701 &$0.58\pm0.34$&$0.75\pm0.52$&$0.43\pm0.33$&$0.44\pm0.30$&$0.41\pm0.31$&$1.41\pm0.72$\\
            &$0.52\pm0.05$&$0.73\pm0.09$&$0.35\pm0.06$&$0.28\pm0.07$&$0.36\pm0.05$&$0.62\pm0.08$\\
0658--5557   &$0.27\pm0.05$&$0.54\pm0.07$&$0.25\pm0.07$&$0.40\pm0.11$&$0.19\pm0.07$&$0.57\pm0.07$\\
            &$0.18\pm0.04$&$0.62\pm0.06$&$0.21\pm0.06$&$0.25\pm0.04$&$0.13\pm0.04$&$0.55\pm0.05$\\
\hline
\end{tabular}
\end{center}
\end{table*}

One of the most important results is an observation of a flattening in the
entropy profile at outer radii in DXL clusters, changing from 0.78 within
the $0.5 r_{500}$ to 0.54 outside. As the sample consists of the most massive
clusters in the Universe, we believe that the explanation of the observed
trend should be searched in the details of the accretion. As summarized in
Voit (2004), the index of the entropy profile is driven by the effects of
mass growth as well as evolution of the virial density. Under the assumption
of a smooth accretion, the entropy grows with radius as $S \sim
M_{gas}^{1-4/3}$ (Voit 2004). With a canonical cluster characteristic of the
surface brightness profile, $\beta=2/3$, $M_{\rm gas} \sim r$, where $M_{\rm
gas}$ is enclosed gas mass. However, as gas mass fraction tends to level off
at high radii, the cumulative gas mass starts to follow the mass of the dark
matter and so $M_{\rm gas} \sim r^{0.5}$. A similar flattening in the
entropy distribution is then expected and is observed in our data at $r\ge
0.4 r_{500}$. If this indeed is the explanation to our data, one would not
necessarily expect the same trend to be observed in low-mass clusters, where
the baryon fraction is growing with the radius even at $r_{500}$.

In addition to the provided explanation, we would like to mention that with
faster mass growth of the cluster (as during a major merger), the entropy
profile should become flatter (Voit 2004). Yet another effect, associated
with the survival of the low entropy gas during the merger (Motl et
al. 2004), reduces the increase in the entropy associated with the mass
growth and should effectively lead to flatter entropy profiles, however, the
full details should be obtained from simulations.

Once the general behavior of the entropy and pressure is found, we studied
the amplitude of the deviations from the average profile, allowing for the
change in the normalization and compared that with the statistical errors in
Table \ref{t:scatter}.

\begin{table*}[ht]
\begin{center}
\renewcommand{\arraystretch}{0.9}\renewcommand{\tabcolsep}{0.05cm}
\caption{\footnotesize Entropy and pressure fluctuations
around the mean sample trend. The results of the spectral analysis
using two different masks are shown. The I-T is the mask based on
cross-sections of the isothermal and isodensity regions. The S-P is the mask
based on the cross-sections of the isentropic and isobaric regions.}
\label{t:scatter}%

\begin{tabular}{rccccccc}
\hline
\hline
 Name & $\sigma$S &   $\sigma$P& $\sigma$S &   $\sigma$P \\
RXCJ &\multicolumn{2}{c}{I-T-mask} & \multicolumn{2}{c}{S-P-mask}\\
\hline
0014.3--3022 & $0.22\pm0.06$& $0.25\pm0.05$& $0.15\pm0.06$& $0.31\pm0.09$\\
0043.4--2037 & $0.18\pm0.13$& $0.33\pm0.25$& $0.04\pm0.04$& $0.26\pm0.25$\\
0232.2--4420 & $0.42\pm0.20$& $0.42\pm0.20$& $0.15\pm0.05$& $0.30\pm0.13$\\
0307.0--2840 & $0.20\pm0.06$& $0.17\pm0.06$& $0.09\pm0.05$& $0.10\pm0.05$\\
0528.9--3927 & $0.16\pm0.05$& $0.20\pm0.13$& $0.12\pm0.09$& $0.29\pm0.08$\\
1131.9--1955 & $0.12\pm0.04$& $0.13\pm0.05$& $0.28\pm0.07$& $0.26\pm0.08$\\
2337.6+0016  & $0.22\pm0.07$& $0.33\pm0.17$& $0.10\pm0.03$& $0.18\pm0.13$\\
0532.9--3701 & $0.41\pm0.34$& $0.41\pm0.34$& $0.35\pm0.05$& $0.25\pm0.07$\\
0658--5557   & $0.24\pm0.10$& $0.35\pm0.13$& $0.15\pm0.05$& $0.25\pm0.11$\\
\hline                                                             
\end{tabular}                                                      
\end{center}                                                       
\end{table*}

\begin{figure}
\includegraphics[width=6cm]{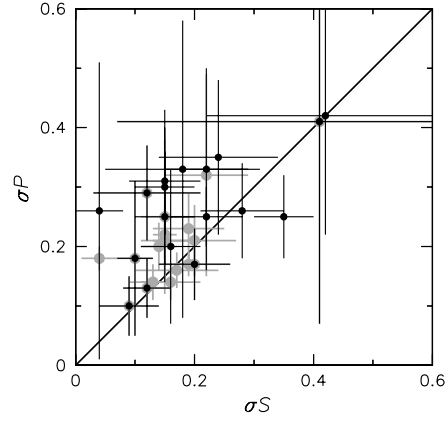}

\caption{Correlation in the dispersion of entropy and pressure. Grey
points indicate measurements within $0.4r_{500}$ and solid points -- in full
range of radii. The solid line show the one-to-one ratio between the entropy
and pressure.}
\label{f:corr}%
\end{figure}

\begin{figure*}
\includegraphics[width=8cm]{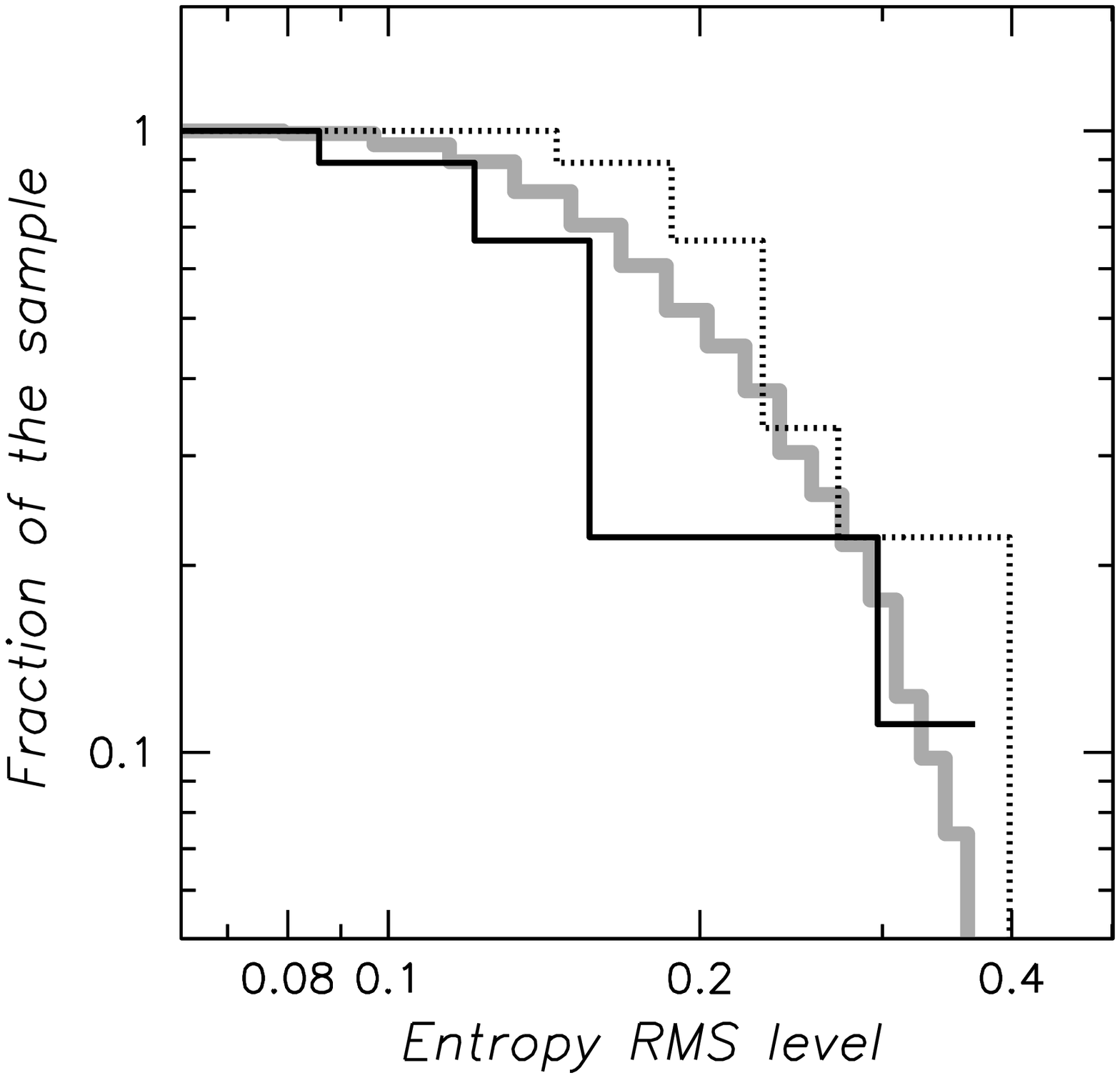}
\includegraphics[width=8cm]{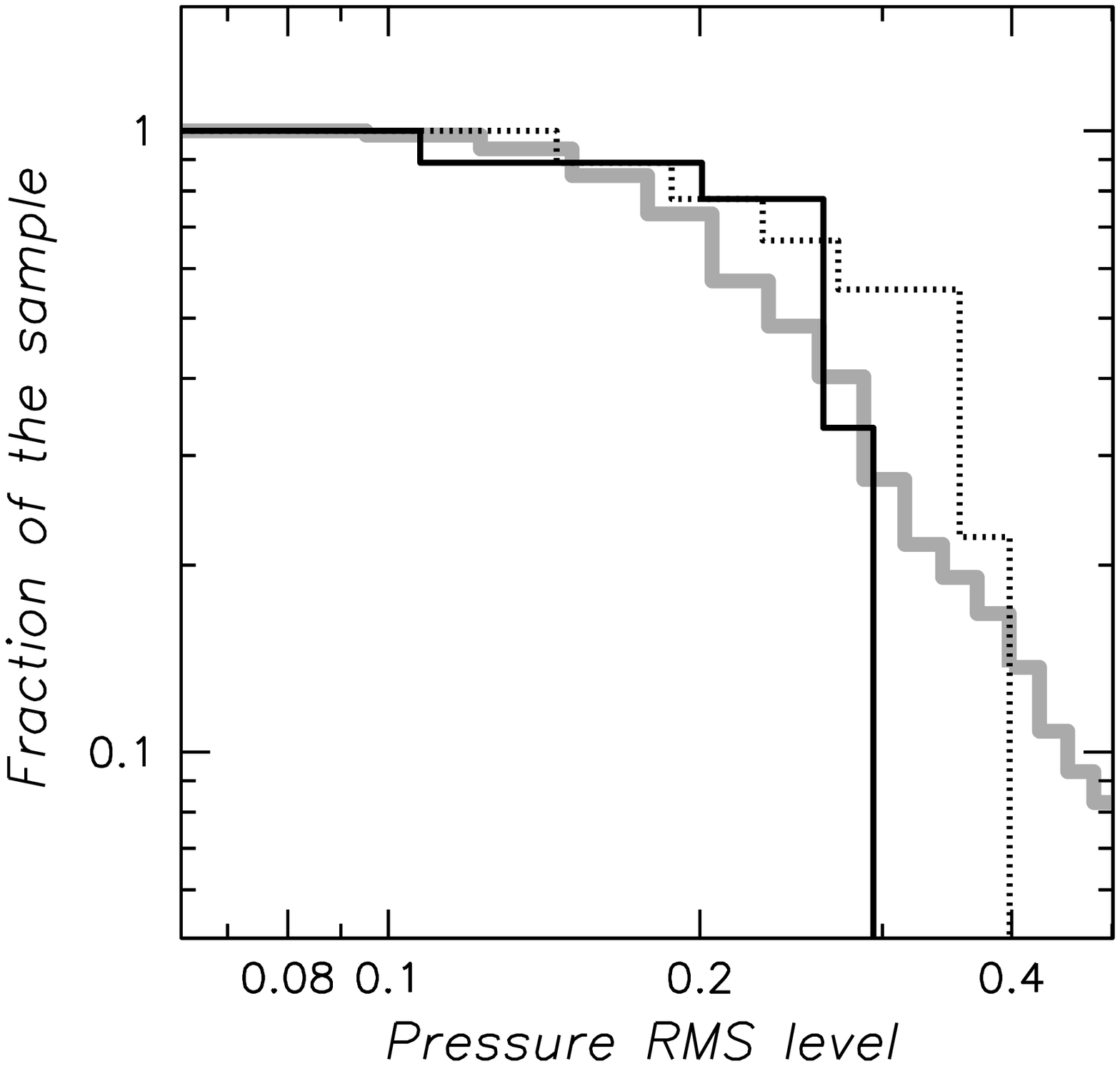} 
\caption{Fraction of
clusters with RMS of the entropy (left panel) and pressure (right panel)
parameter greater than the x-axis value. Black lines denote the results for
DXL cluster sample, obtained using two different masks (marked as solid and
dotted lines), correspondingly sampling entropy/pressure and
image/temperature. Grey line represents the results of a similar analysis
performed on a sample of 208 modeled clusters (Finoguenov et al. 2005).}
\label{f:c2s}%
\end{figure*}

\begin{figure*}
\centering
\includegraphics[width=8cm]{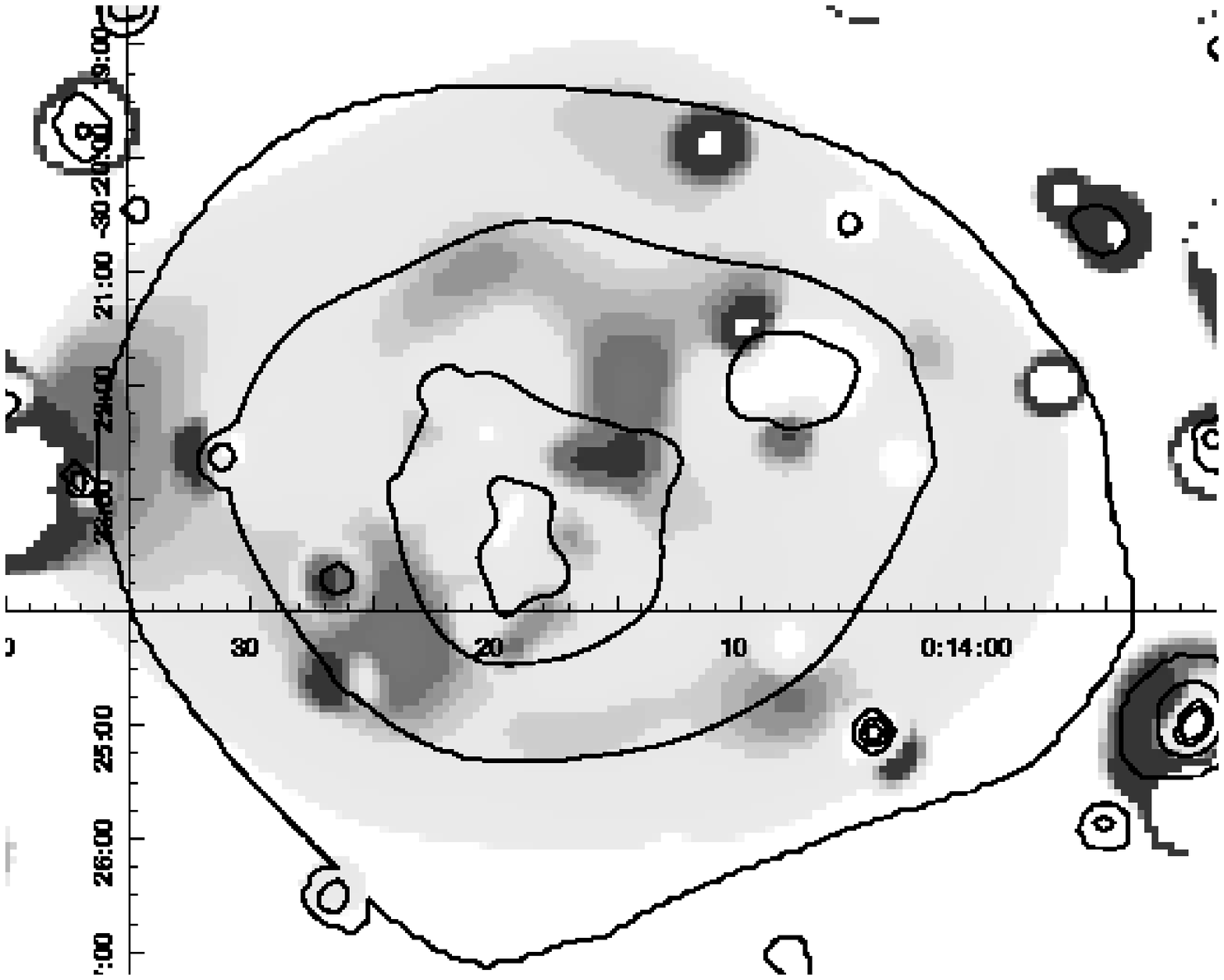}\hfill\includegraphics[width=8cm]{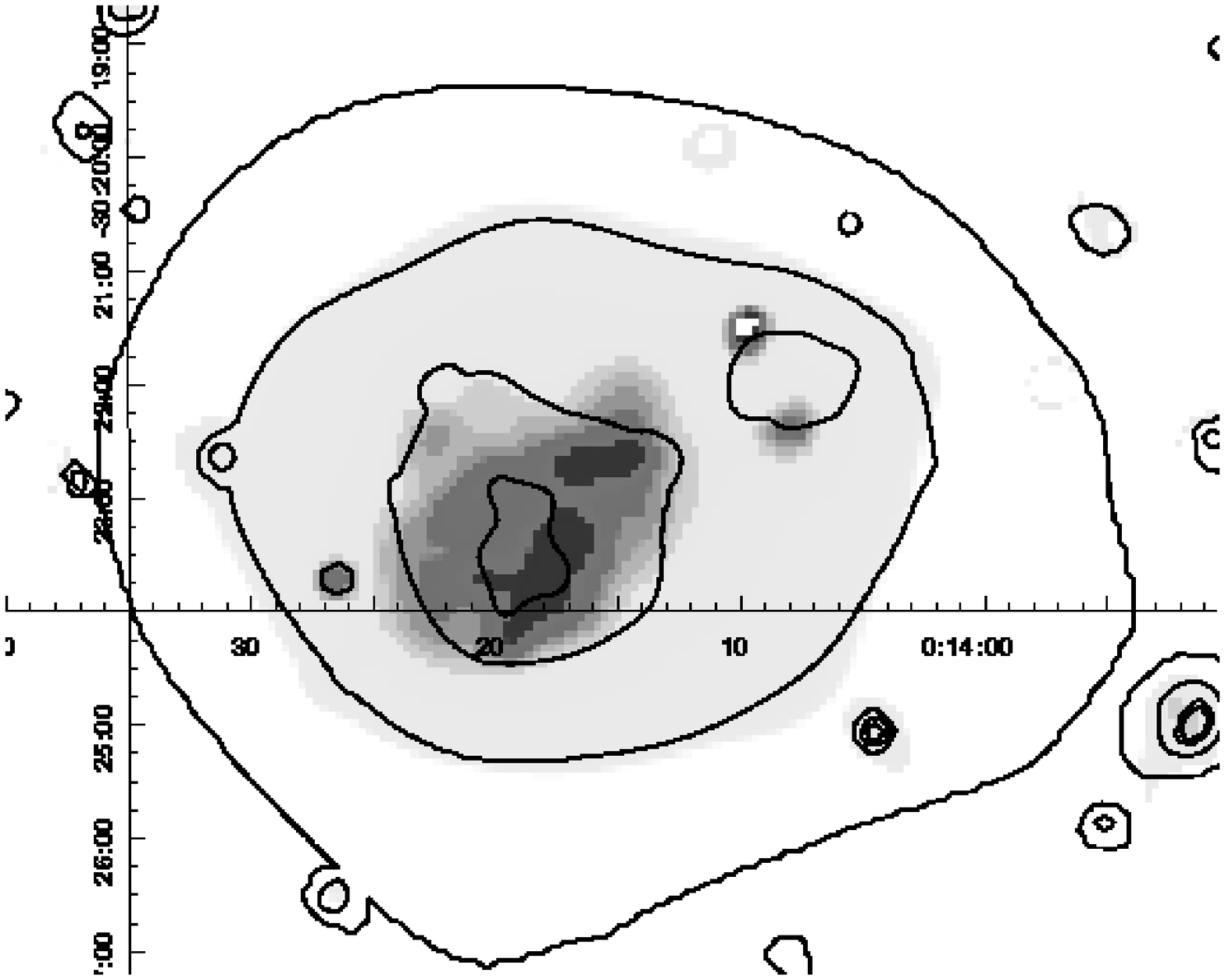}

\includegraphics[width=8cm]{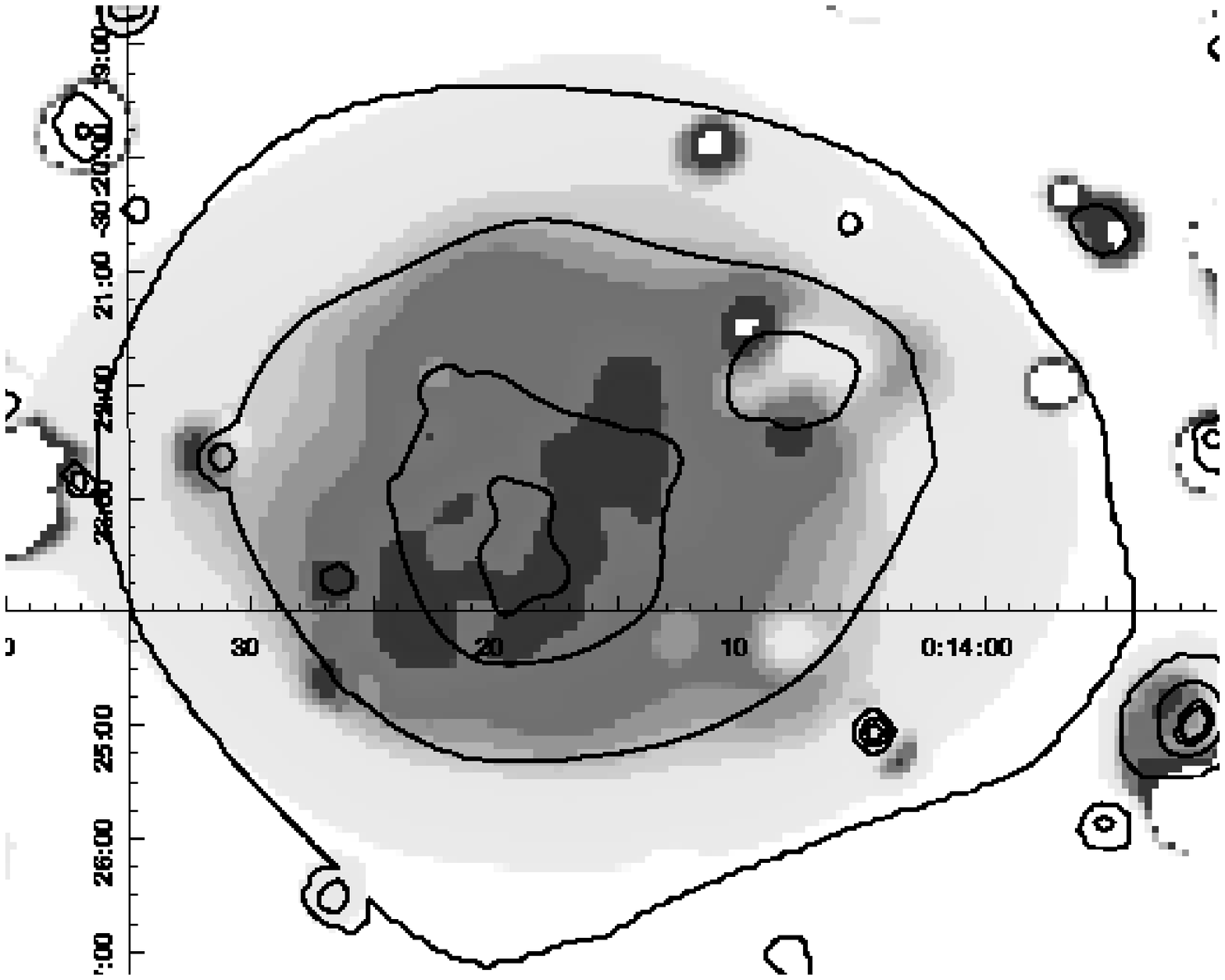}\hfill\includegraphics[width=8cm]{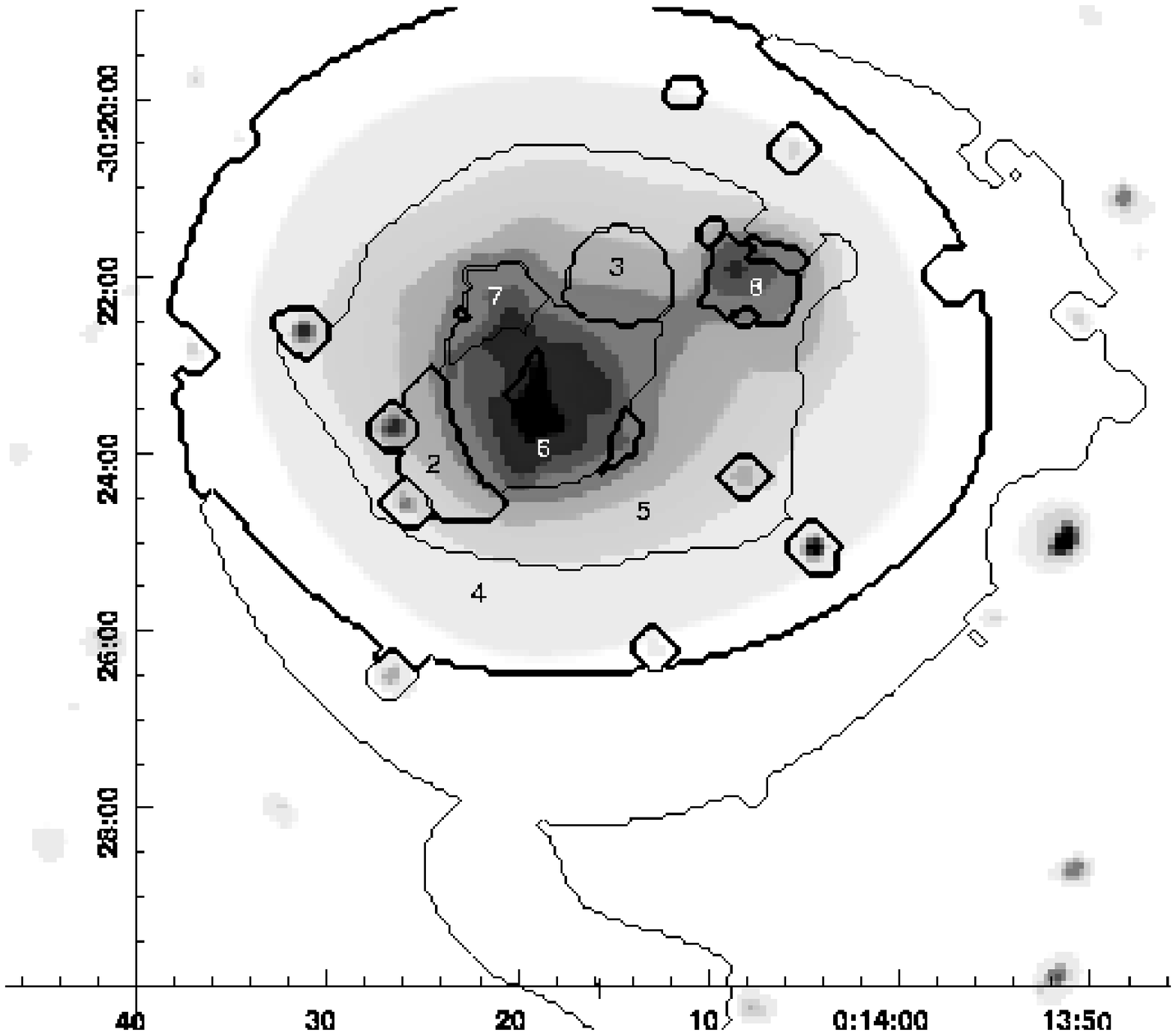}

\includegraphics[width=6cm]{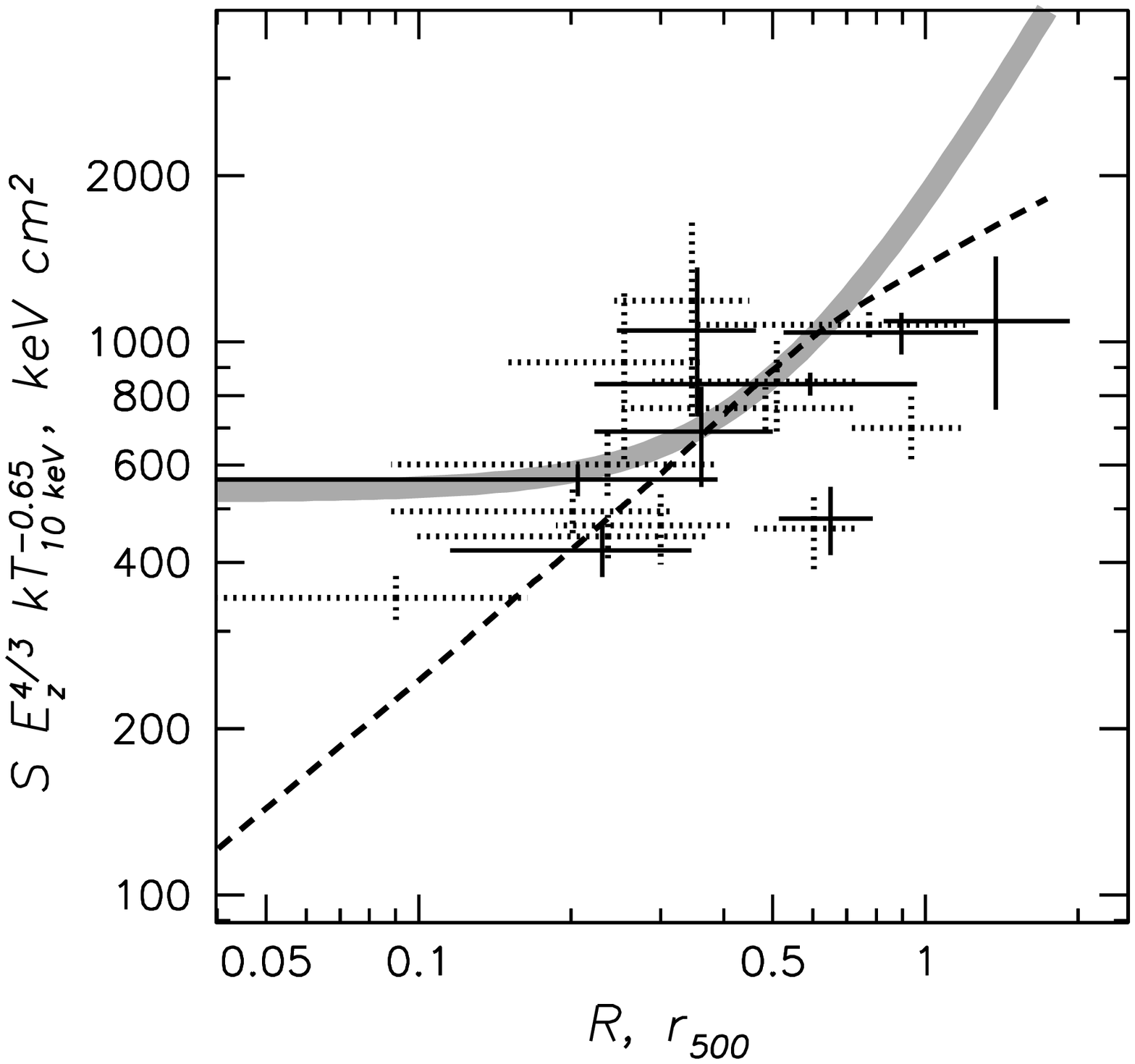}\hfill\includegraphics[width=6cm]{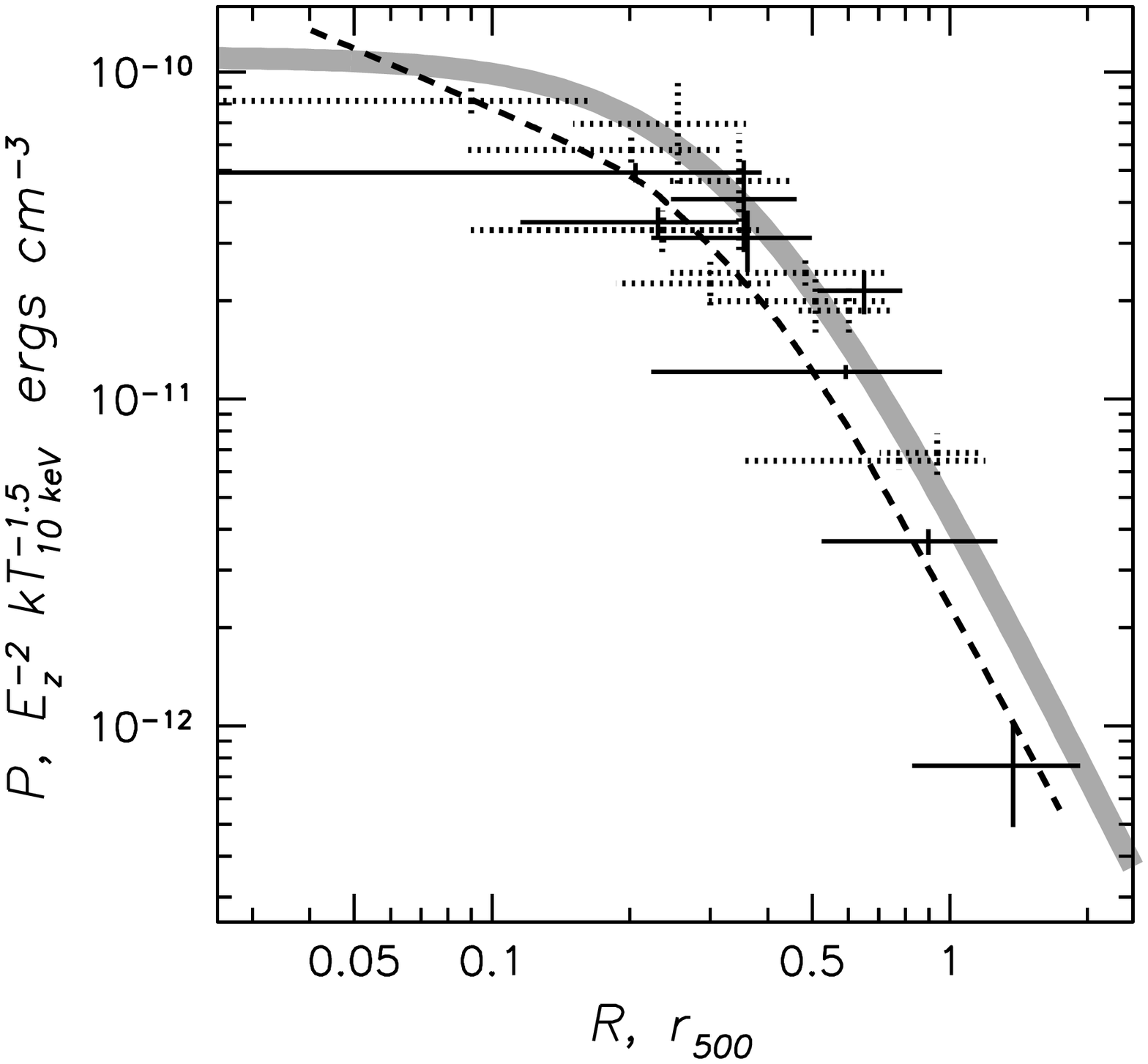}\hfill\includegraphics[width=6cm]{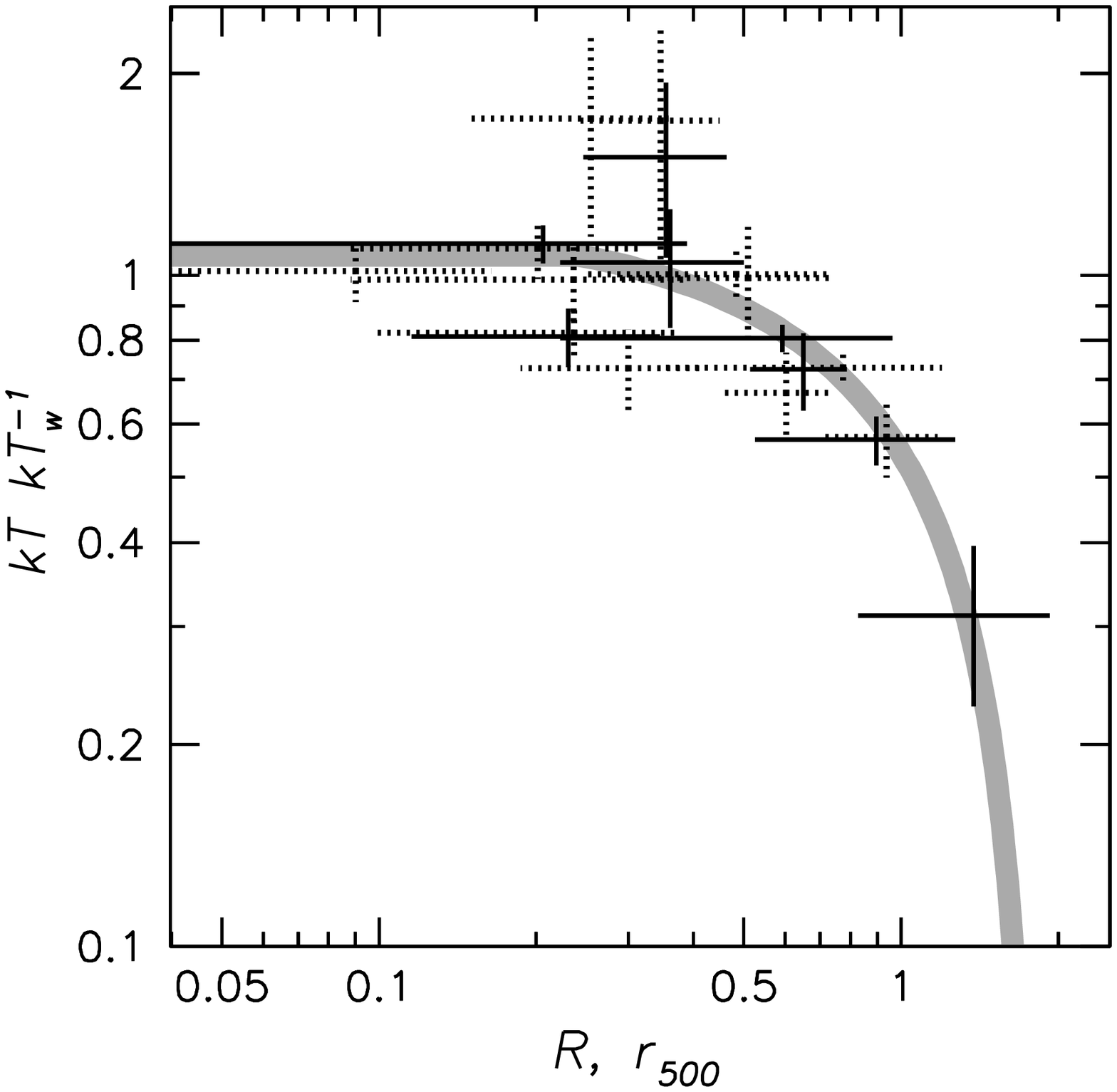}

\caption{RXCJ0014.3--3022. Top left: entropy map, top right: pressure
map, middle left: temperature map, middle right surface brightness. Entropy,
pressure and temperature maps are overlaid with the contours of equal
surface brightness in the 0.5--2 keV band. The surface brightness image is
overlaid with contours showing the spectral extraction regions with numbers
corresponding to those in Table \ref{t:cl01:t}. Coordinate grids are shown
for the epoch J2000. On the corresponding maps, zones of low entropy are
shown in white, zones of higher pressure, temperature are shown in
black. Lower left entropy profile, lower center: pressure profile, lower
right: temperature profile. The solid crosses denotes the tabulated data and
the dotted crosses show the rest of the results. The data is scaled for both
the cluster mass, using the relation of Ponman et al. (2003) and evolution
of the critical density with redshift. Dashed lines on both plots shows the
best fit to the whole sample, described in the text. Grey lines show the
results of the 1d analysis, using a beta-model and fits to the temperature
profile from Zhang et al. (2005, in prep.). The grey line in the temperature
panel is the universal temperature profile of Vikhlinin et al. (2005).}
\label{f:cl01}%
\end{figure*}

In Fig.\ref{f:corr} we study the scatter in the entropy and pressure,
reported in Table \ref{t:scatter}. The dispersion and normalization of the
profile are calculated weighting the points by the area of the corresponding
region. The renormalization is needed to remove a possible bias due to
assumption of a mean temperature in rescaling the profiles. We note that the
spread of values for the normalization is larger for the pressure, where
also the sensitivity to the assumed weighted temperature is larger. There is
also no clear correlation between the r.m.s. values and the normalization in
either entropy or pressure. The figure shows that the fluctuations in the
entropy and pressure are correlated in the magnitude, with the relation
favoring slightly higher pressure fluctuations as compared to the entropy,
but certainly not following the prescription for the shock heating, which at
the observed amplitudes predicts fluctuations only in the pressure and very
little in the entropy. This picture and a quantitative analysis below is very
similar to the results of Finoguenov et al. (2005), who analyzed a set of
cosmological simulations of clusters.  Apparently, azimuthal distortions,
related to gas and dark matter displacements during the merger and post
relaxation supersede the distortions associated with the shock heating. The
later accounts for about 15\% and its clear separation could be better
revealed in the future by comparison of pressure maps derived at X-rays and
thermal SZE observations.

In Fig.\ref{f:c2s} we summarize the fluctuation analysis and compare it with
the distribution for the cluster sample obtained in the cosmological
simulations (Finoguenov et al. 2005) sampling a similar range of
radii. Taking the comparison between the model and the results obtained with
the mask sampling pressure and entropy, we conclude that the clusters show a
similar degree of fluctuations and the probability of this to be randomly
drawn from the same distribution, calculated using the KS-test, is 60\% and
80\% for the pressure and entropy, respectively. We combined the results of
the two masks in the test.

\subsection{RXCJ0014.3--3022 }

\begin{table*}[ht]
{
\begin{center}
\footnotesize
{\renewcommand{\arraystretch}{0.9}\renewcommand{\tabcolsep}{0.09cm}
\caption{\footnotesize
Properties of main regions of RXCJ0014.3--3022.} 
\label{t:cl01:t}%
%c '0.45*sqrt(6.5)/(0.7+0.3*(1+0.2779)**3)**0.5'
\begin{tabular}{cccccccccccc}
 \hline
 \hline
N  &$kT$ &  $\rho_e$ & S & P, $10^{-12}$ & $M_{\rm gas}$ & $r_{\rm min}$ &
$r_{\rm max}$ & Remarks\\
    & keV     &  $10^{-4}$ cm$^{-3}$& keV cm$^2$ & ergs cm$^{-3}$ &
  $10^{12} M_\odot$ & Mpc & Mpc  \\
\hline
 1 &$ 2.6\pm0.7$&$  1.9\pm0.4$&$ 785\pm 242$&$  0.8\pm 0.3$&$25.9\pm5.9$&0.92&2.14& main-3    \\
 2 &$12.5\pm3.6$&$ 21.2\pm2.3$&$ 754\pm 226$&$ 42.4\pm13.2$&$ 2.7\pm0.3$&0.27&0.51& ridge S   \\
 3 &$ 8.7\pm1.7$&$ 23.2\pm1.6$&$ 495\pm 102$&$ 32.2\pm 6.8$&$ 3.4\pm0.2$&0.25&0.56& ridge N   \\
 4 &$ 4.7\pm0.4$&$  5.0\pm0.2$&$ 748\pm  64$&$  3.8\pm 0.3$&$58.8\pm1.8$&0.58&1.41& main-2    \\
 5 &$ 6.7\pm0.3$&$ 11.7\pm0.2$&$ 604\pm  29$&$ 12.5\pm 0.6$&$49.1\pm0.7$&0.25&1.07& main-1    \\
 6 &$ 9.2\pm0.6$&$ 34.5\pm0.6$&$ 405\pm  27$&$ 51.1\pm 3.4$&$14.9\pm0.2$&0.03&0.43& ridge     \\
 7 &$ 6.7\pm0.7$&$ 33.3\pm1.6$&$ 302\pm  32$&$ 35.9\pm 4.0$&$ 2.7\pm0.1$&0.13&0.38&disrupted core  \\
 8 &$ 6.0\pm0.8$&$ 23.0\pm1.6$&$ 345\pm  49$&$ 22.2\pm 3.3$&$ 3.2\pm0.2$&0.57&0.87&subcluster \\
\hline
\end{tabular}
}
\end{center}
}
%\vspace*{0.2cm}
\end{table*}

The cluster consists of two clumps. The large-scale emission is centered on
one of the peaks, that is therefore taken for the center of the main cluster
and its position is listed in Tab.\ref{t:ol}. In order to tabulate the basic
properties of the cluster we extracted the spectra from zones that according
to the hardness ratio have the same temperature to within 1 keV. Table
\ref{t:cl01:t} contains three zones with temperatures exceeding 7 keV. These
zones form a ridge passing between the two subclusters and surrounding the
main cluster center. The elongated shape of the ridge suggests its origin in
the interaction between the clusters. The temperature enhancement compared
to region 5 is a factor of $1.4\pm0.1$, which corresponds to a Mach number
of $1.4\pm0.1$. Although, higher temperatures are also observed, they are
not statistically different from 9 keV. The pressure peak is not at the
cluster center, but is within the ridge.

The temperature as well as the pressure of the subcluster (region 8) are
similar to that of the disrupted core of the main cluster (region 7, as well
as 5). Assuming that the mass scales with the central pressure as $M \sim
P^{1.5}$, the initial mass ratio of the two clusters is $1.7\pm0.2$. An
expected final temperature of the cluster after the merger from the $M-T$
relation (Finoguenov et al. 2001) is $9.1\pm0.5$ keV, similar to the
observed temperature of the ridge. This comparison suggests that the grown
mass of the new cluster will be able to cope with the increased pressure of
the gas.

\begin{figure*}
\includegraphics[width=17cm]{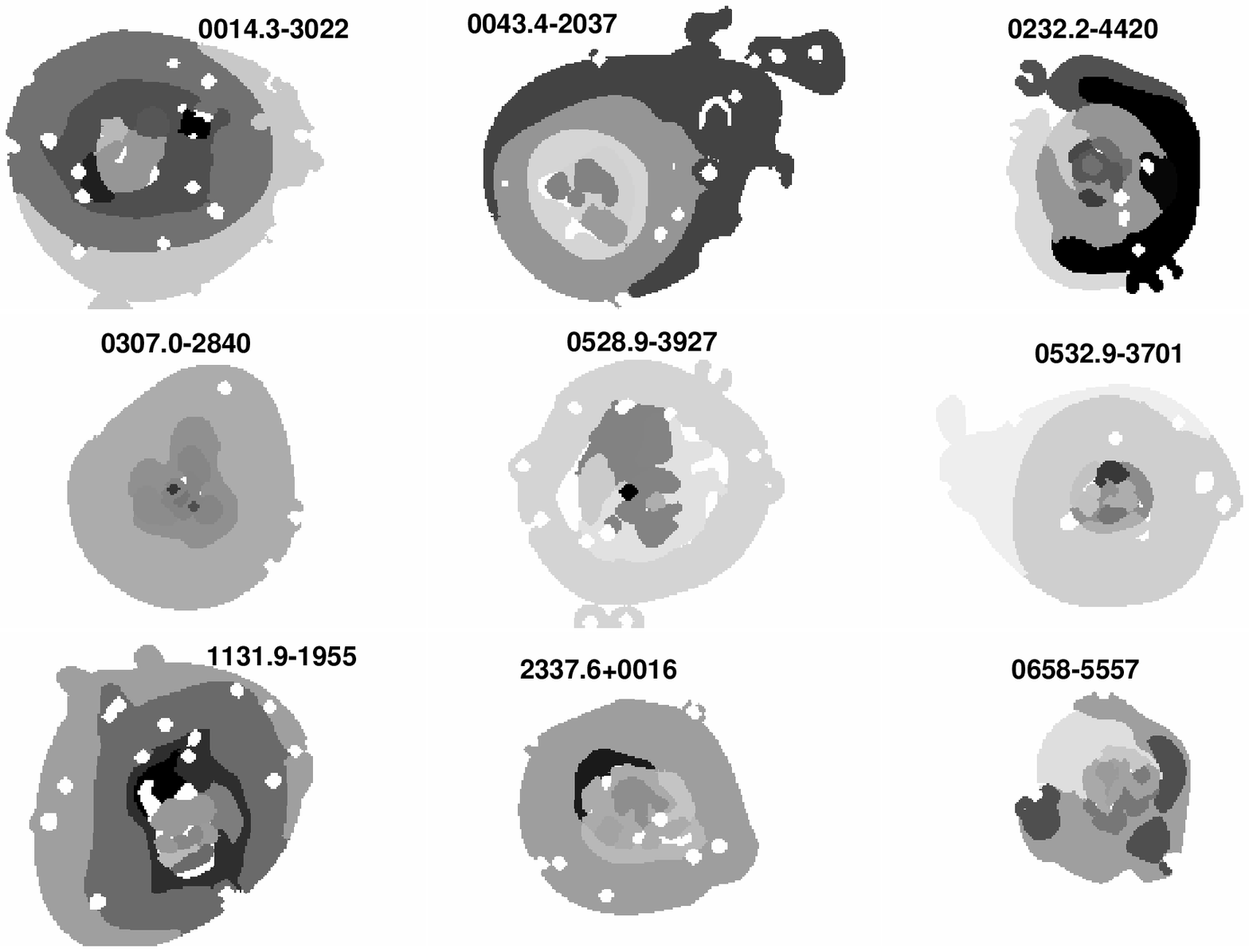}

\caption{Maps of the ratio between between the observed pressure and the
average trends measured in the DXL sample. The different shades of grey
stand for a different value of the ratio, light -- 0.5 (e.g. tail of RXCJ
0532.9-3701), grey -- 1.1 (a dominant color of RXCJ 0532.9-3701), dark grey
-- 1.4 (a tail of RXCJ 0043.4-2037), black -- $>2$.}
\label{f:mapp}%
\end{figure*}

In Fig.\ref{f:cl01} we also compare the derived entropies and pressures with
the average trend for the DXL sample and further illustrate their ratio in
Figs. \ref{f:maps}-\ref{f:mapp}. It can be seen that the entropy of the core
region is higher by a factor of $1.3\pm0.05$. Low entropy clump is seen at
the position of the second core. The pressure of the bridge appears to match
well the adopted scaling temperature of 8.3 keV. The pressure of the main
cluster at the distance of the subcluster (region 5) is by a factor of
$1.5\pm0.05$ higher than the model, despite of a relatively low temperature
of 6.7 keV.

The subcluster is characterized by higher pressure (best seen in the Table
\ref{t:cl01:t} and as deviation in the pressure profile) as well as lower
entropy. This implies that the low entropy gas of the subcluster is retained
within its own dark matter halo. This picture is different from the case of
A3667 (Briel et al. 2003), where low entropy gas is displaced from the
pressure peak. The subcluster in A2744 is a cool core, rather than a cold
front.  Our identification of the core of the main cluster is further
confirmed by the entropy plot, as it lies on the prediction for the
self-similar scaling. It is located 0.2 Mpc apart from the cluster center
and may even present the still unperturbed part of an initially large
cluster core extending to that radii.

In Fig.\ref{f:cl01} we also show the results produced by the standard
analysis using the beta model (e.g. Jones \& Forman 1984). We find an
overall agreement between the two methods.

For many of the described features there is a qualitative agreement with
Chandra results of Kempner \& David (2004). Similar high temperature ridges
have been claimed for A1644 (Reiprich et al. 2004) and A3921 (Belsole et
al. 2005), yet none of them are found to dominate the pressure of the
cluster, as in the case of A2744.

\begin{figure*}
\includegraphics[width=8cm]{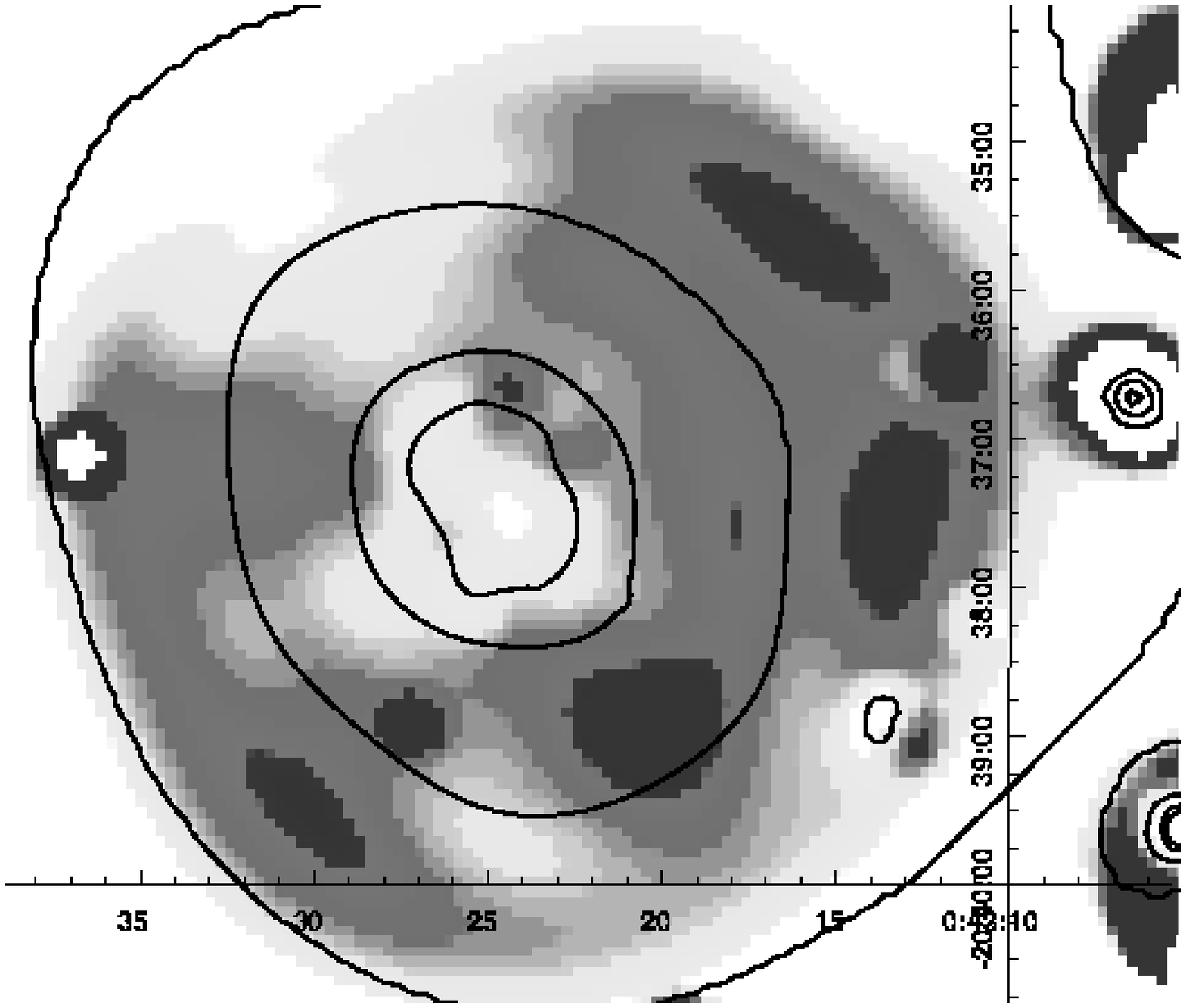}\hfill\includegraphics[width=8cm]{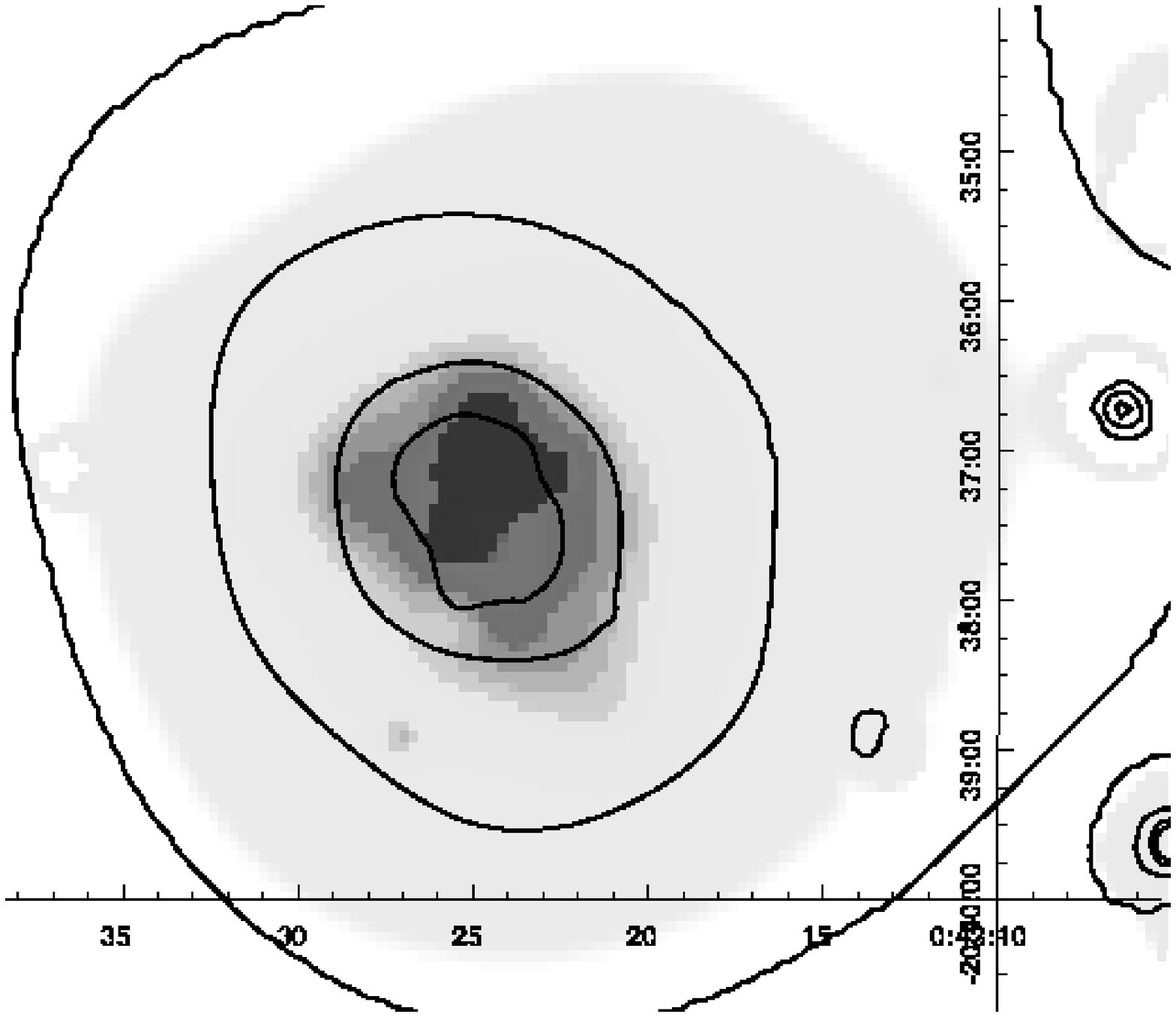}

\includegraphics[width=8cm]{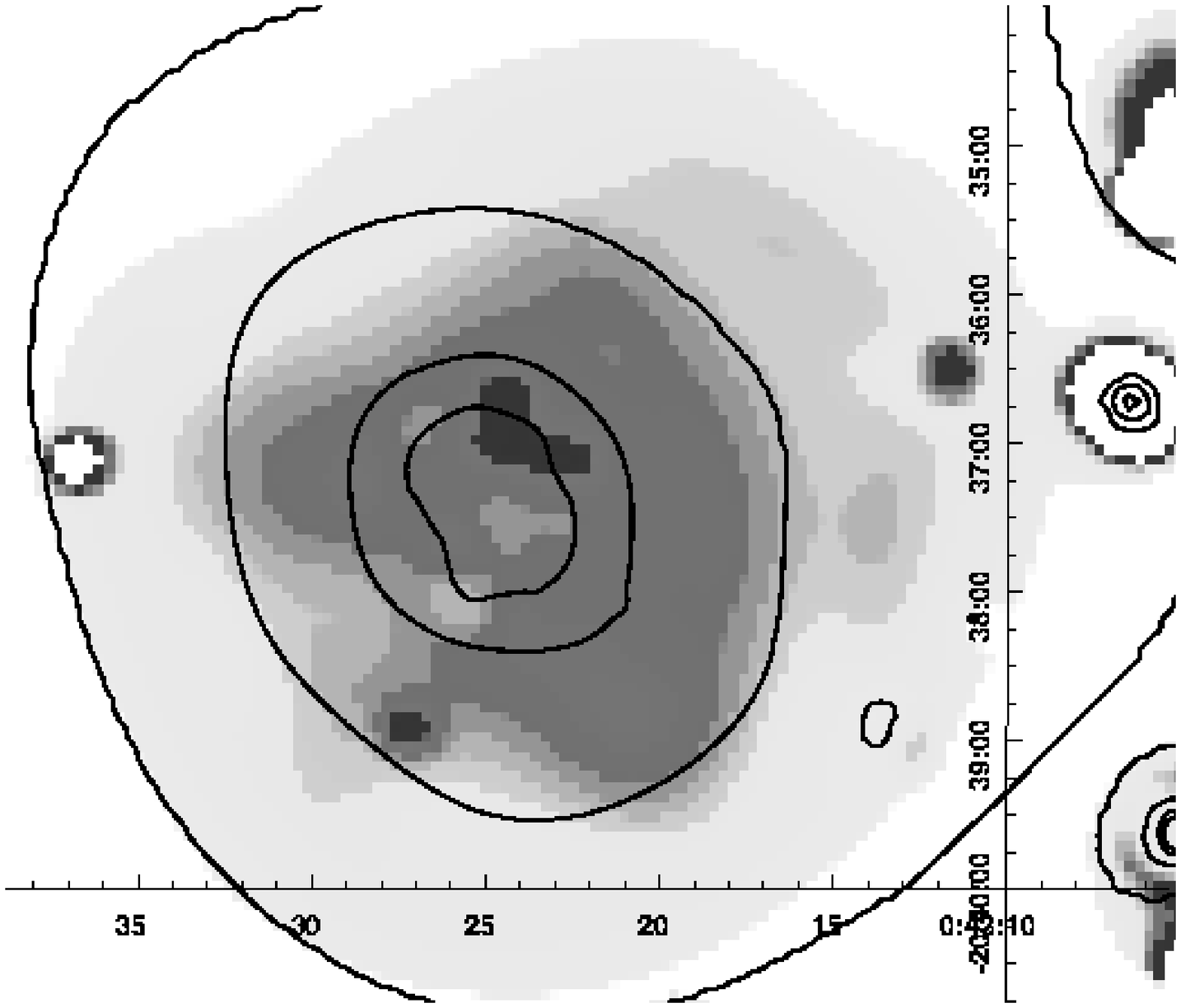}\hfill\includegraphics[width=8cm]{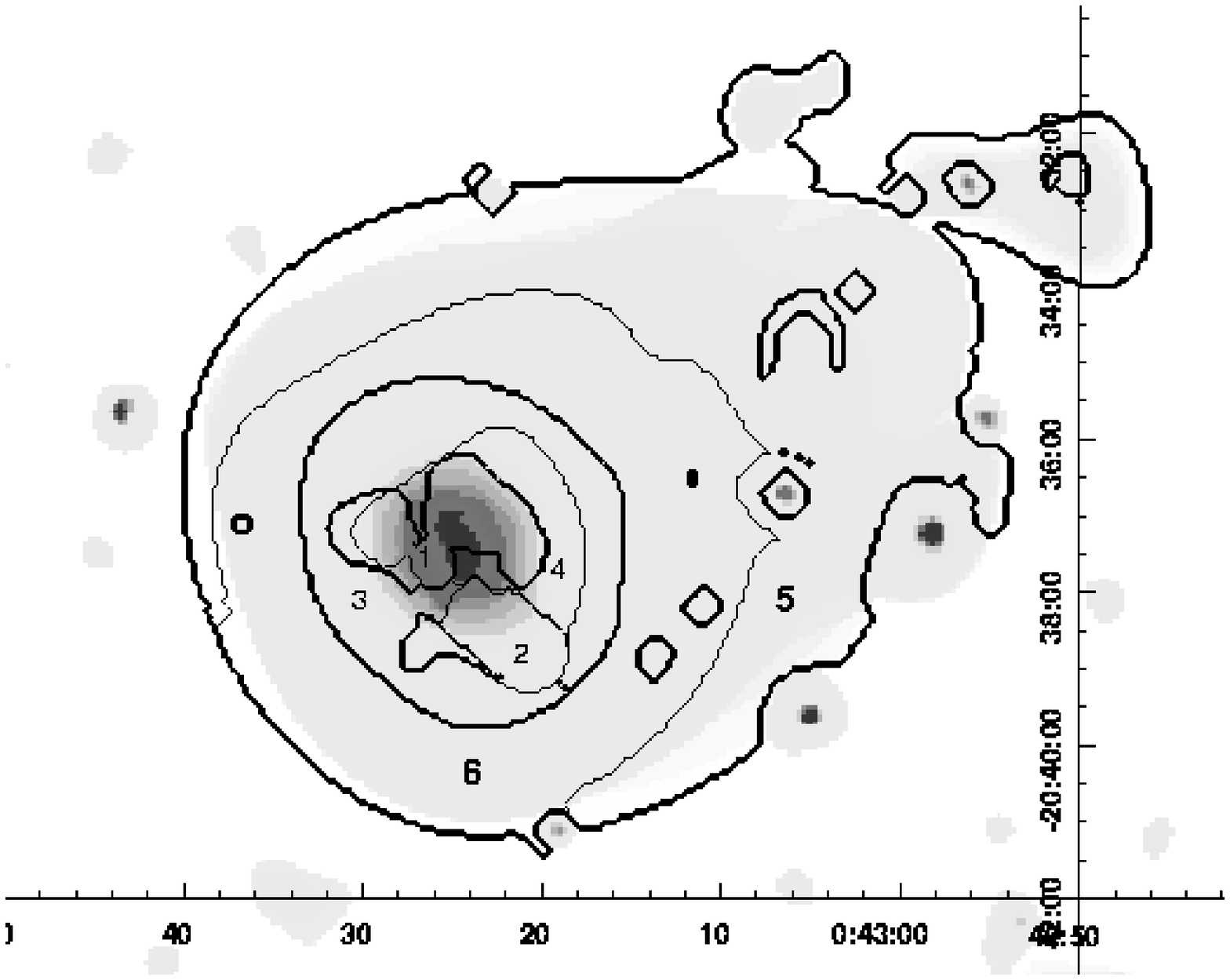}

\includegraphics[width=6cm]{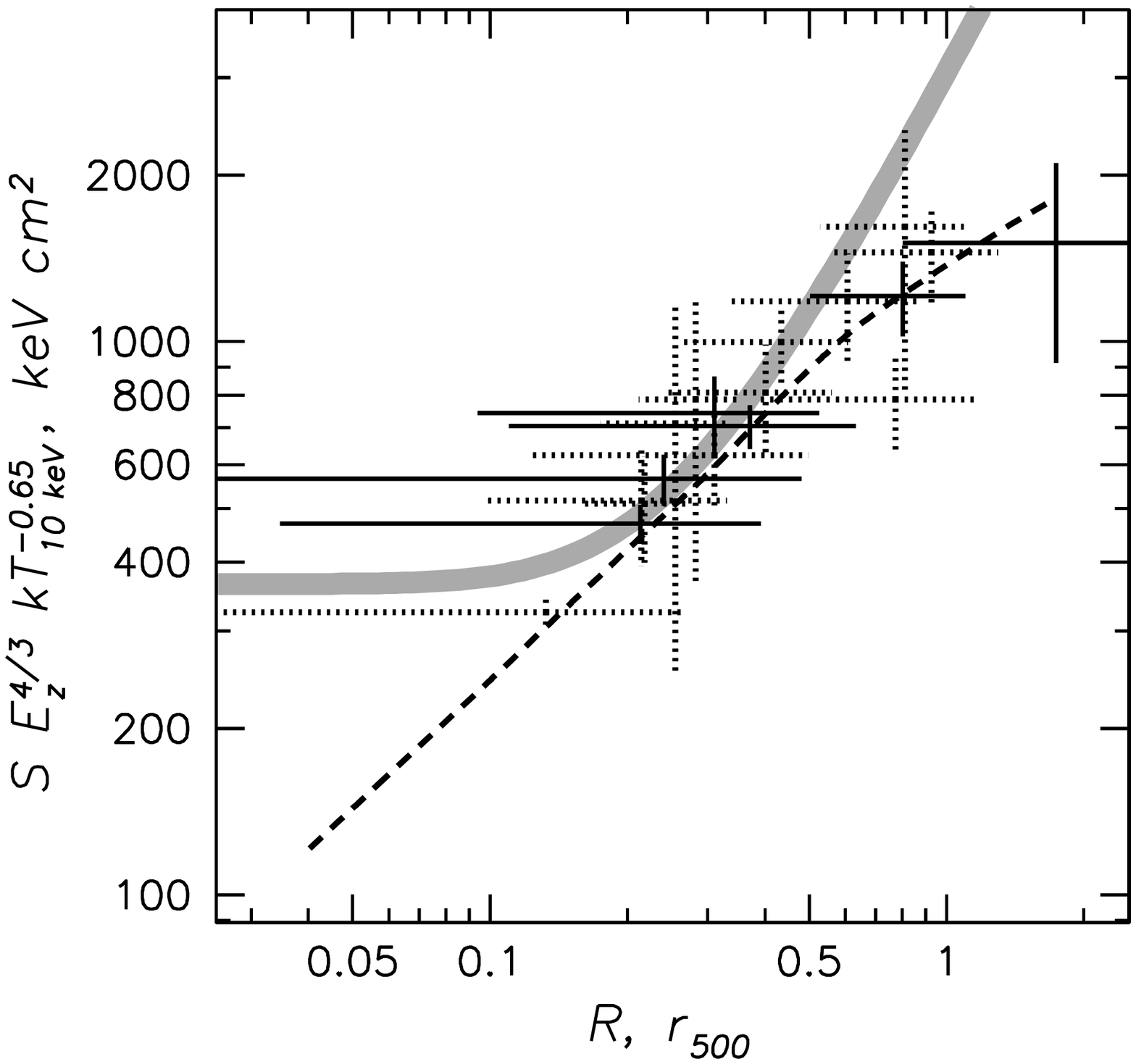}\hfill\includegraphics[width=6cm]{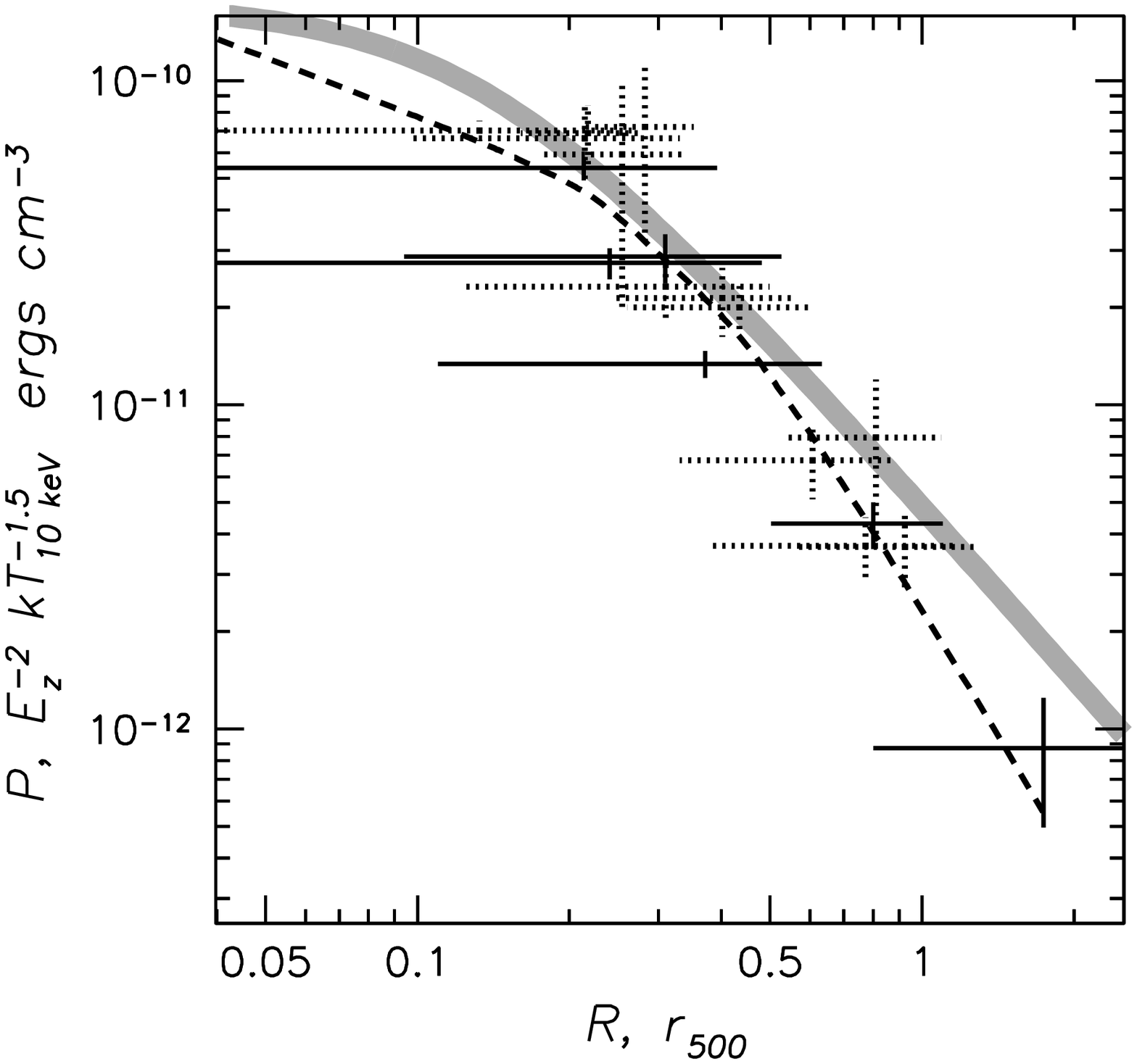}\hfill\includegraphics[width=6cm]{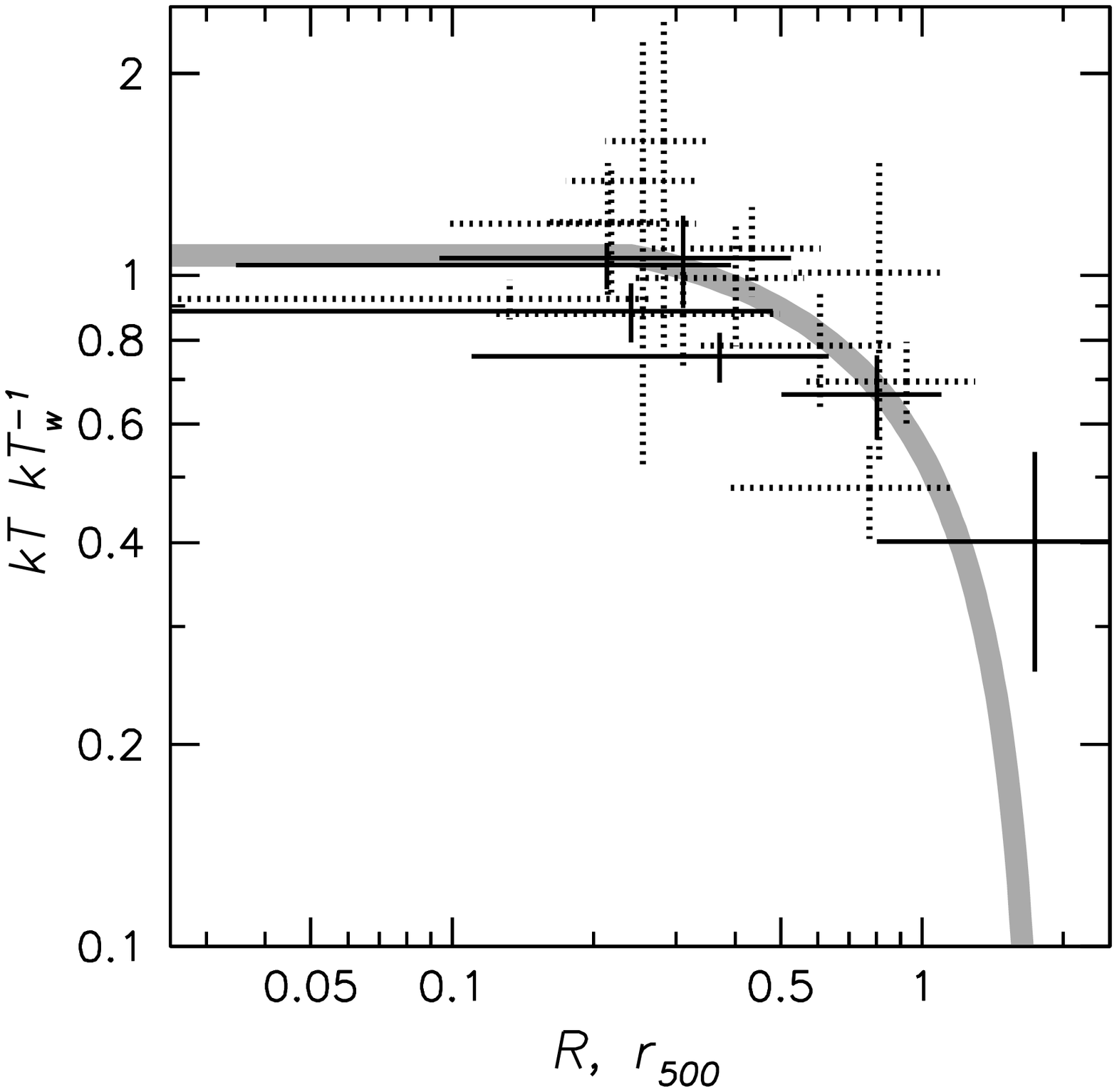}

\caption{RXCJ0043.4--2037. Figure caption is similar to that of
Fig.\ref{f:cl01}.  The surface brightness image is overlaid with contours
showing the spectral extraction regions with numbers corresponding to those
in Table \ref{t:cl02:t}.}
\label{f:cl02}%
\end{figure*}

\begin{table*}[ht]
{
\begin{center}
\footnotesize
{\renewcommand{\arraystretch}{0.9}\renewcommand{\tabcolsep}{0.09cm}
\caption{\footnotesize
Properties of main regions of RXCJ0043.4--2037.}
\label{t:cl02:t}%

\begin{tabular}{cccccccccccc}
 \hline
 \hline
N  &$kT$ &  $\rho_e$ & S & P, $10^{-12}$ & $M_{\rm gas}$ & $r_{\rm min}$ &
$r_{\rm max}$ & Remarks\\
   & keV     &  $10^{-4}$ cm$^{-3}$& keV cm$^2$ & ergs cm$^{-3}$ & $10^{12}
   M_\odot$  & Mpc & Mpc&  \\
\hline
 1 &$ 7.0\pm0.6$&$ 36.0\pm0.8$&$ 300\pm  24$&$ 40.6\pm 3.4$&$ 9.7\pm0.2$&0.04&0.40 & core\\
 2 &$ 7.2\pm1.1$&$ 18.7\pm1.3$&$ 475\pm  78$&$ 21.6\pm 3.7$&$ 3.9\pm0.3$&0.10&0.53\\
 3 &$ 5.1\pm0.4$&$ 12.2\pm0.5$&$ 450\pm  40$&$ 10.1\pm 1.0$&$16.0\pm0.7$&0.11&0.64\\
 4 &$ 6.0\pm0.6$&$ 21.5\pm1.0$&$ 361\pm  38$&$ 20.7\pm 2.3$&$ 7.0\pm0.3$&0.00&0.49\\
 5 &$ 2.7\pm1.0$&$  1.5\pm0.4$&$ 963\pm 379$&$  0.7\pm 0.3$&$37.6\pm9.0$&0.81&2.71 & filament?\\
 6 &$ 4.5\pm0.7$&$  4.5\pm0.4$&$ 772\pm 119$&$  3.2\pm 0.5$&$27.7\pm2.2$&0.51&1.11 & outskirts\\
\hline
\end{tabular}
}
\end{center}
}
%\vspace*{0.2cm}
\end{table*}

\subsection{RXCJ0043.4--2037 }

If one judges from the image, this is one of the most relaxed clusters in
the sample. The image has a single center, which is located at the center of
the large-scale emission. However, there is an indication of small pressure
and entropy distortions and the statistical analysis reveals that the
fluctuations are quite large. The pressure shape follows essentially the
temperature shape. Some small-scale fluctuations in the temperature map are
still seen. The image is extended to the north-west.

The results of the spectral analysis are reported in Table \ref{t:cl02:t}
and shown in Fig.\ref{f:cl02}. The most significant feature is the presence
of a region with extremely low entropy, seen as dotted cross in the entropy
profile in Fig.\ref{f:cl02} at the $0.7r_{500}$ distance to the center to
the south. This gas is in pressure equilibrium with the cluster, leading to
a suggestion that we observe the debris of accreted group. A comparison with
simulations would be useful here to shed more light on the stage of this
suggested disruption. Given the statistical uncertainty it is difficult to
conclude on the origin of the extent towards the north-west in the lowest
levels of the X-ray surface brightness (region 5). Presumably, this
extension, located outside of $r_{500}$, is due to an accretion of a
filament.

\begin{figure*}
\includegraphics[width=8cm]{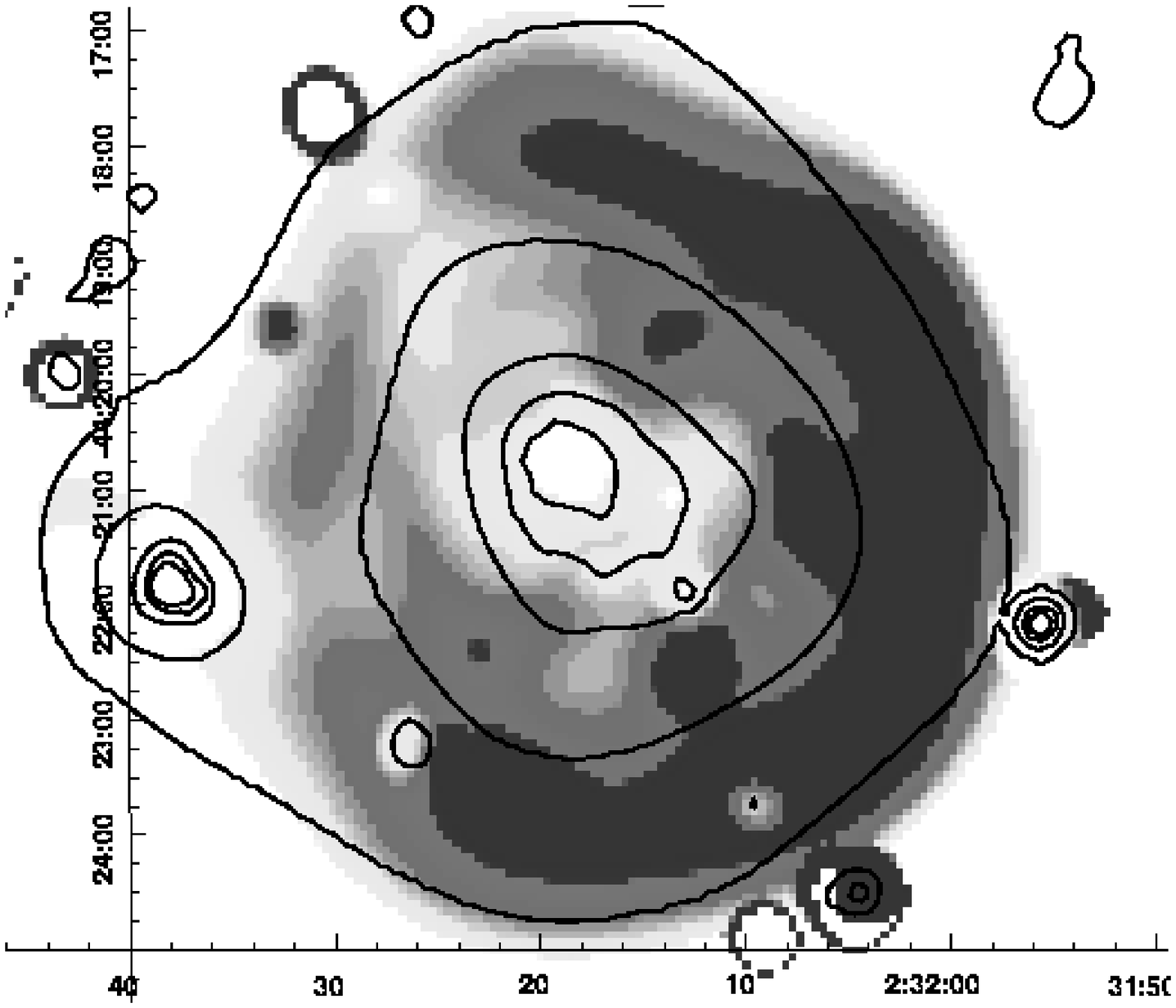}\hfill\includegraphics[width=8cm]{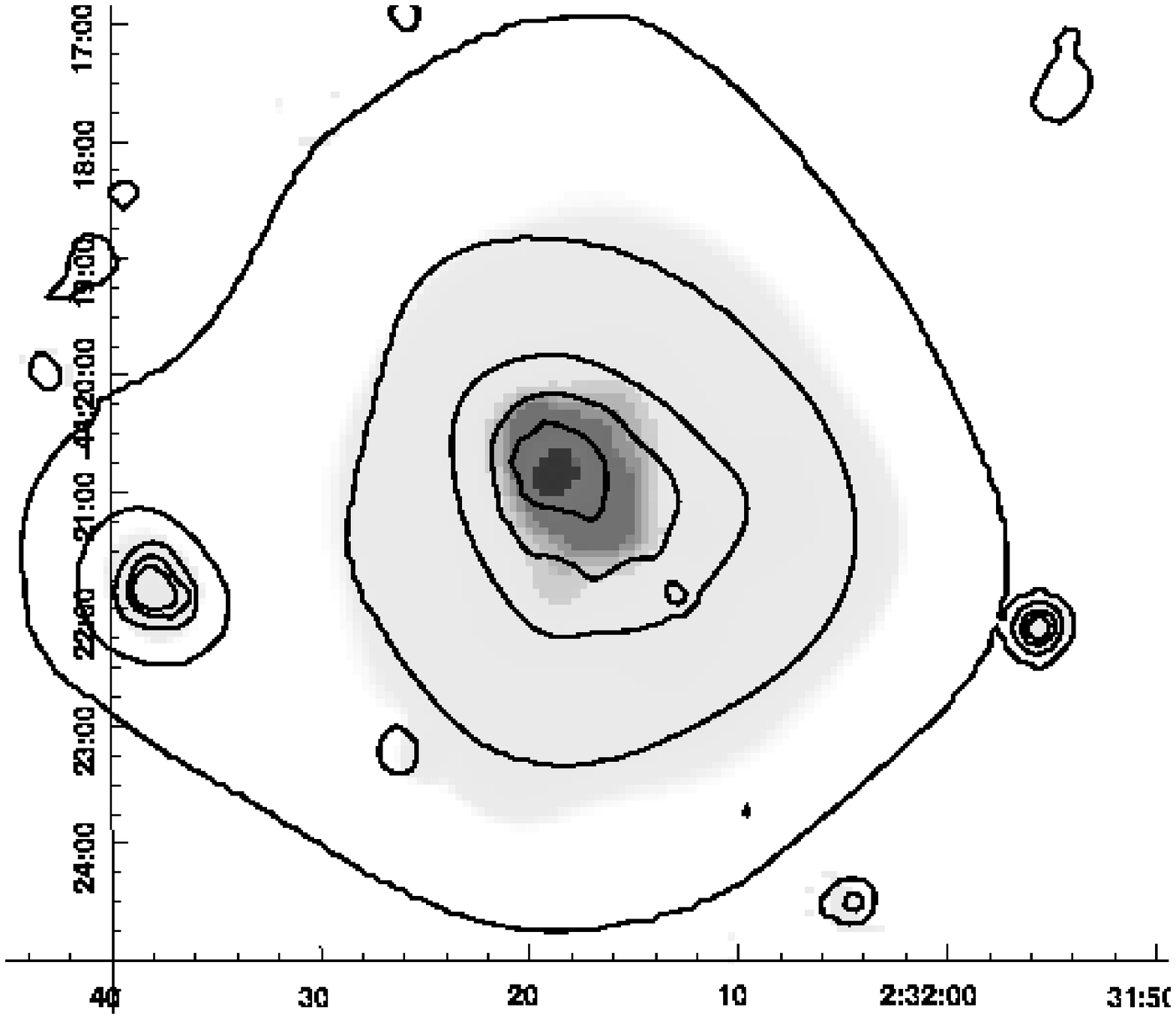}

\includegraphics[width=8cm]{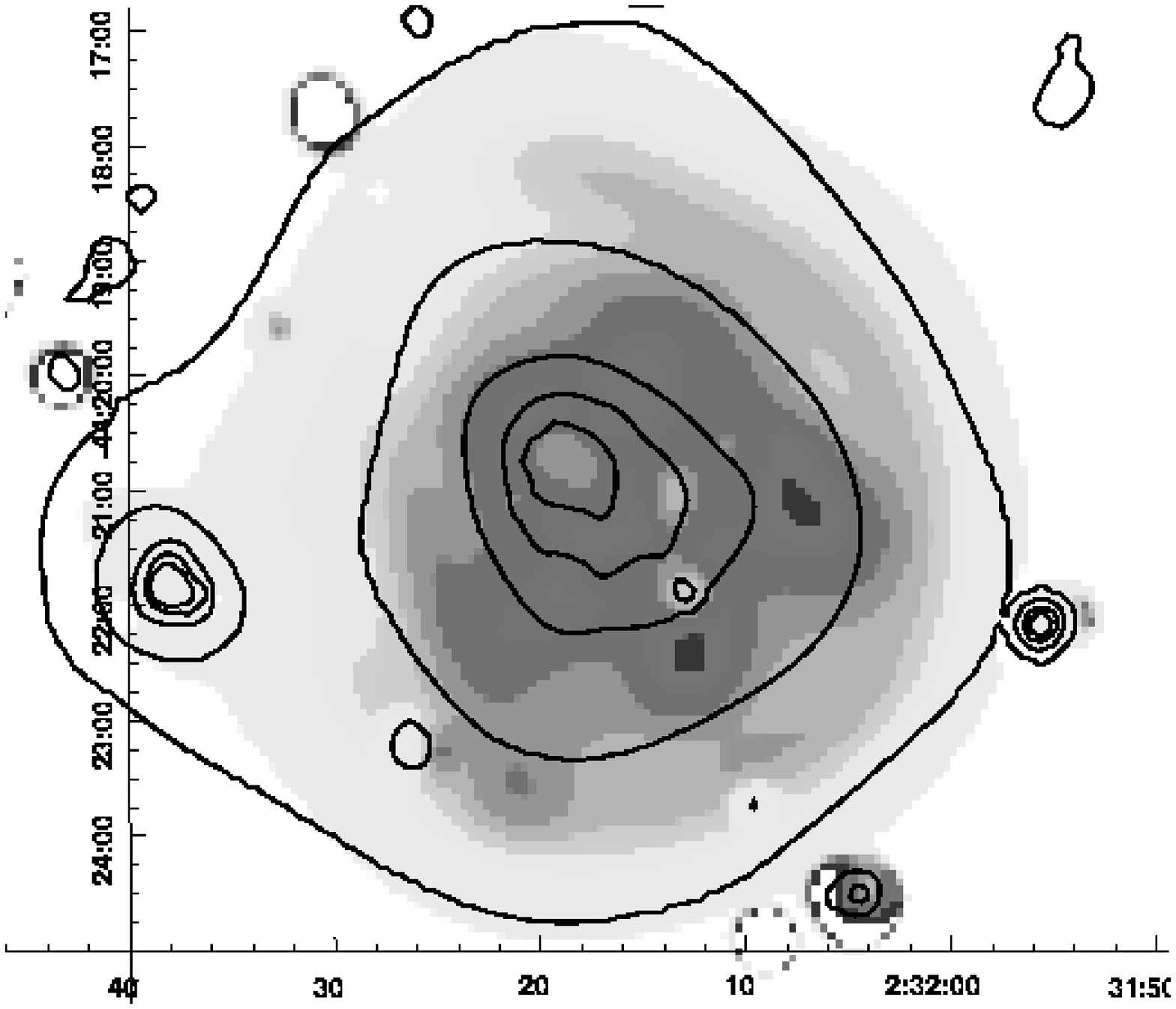}\hfill\includegraphics[width=8cm]{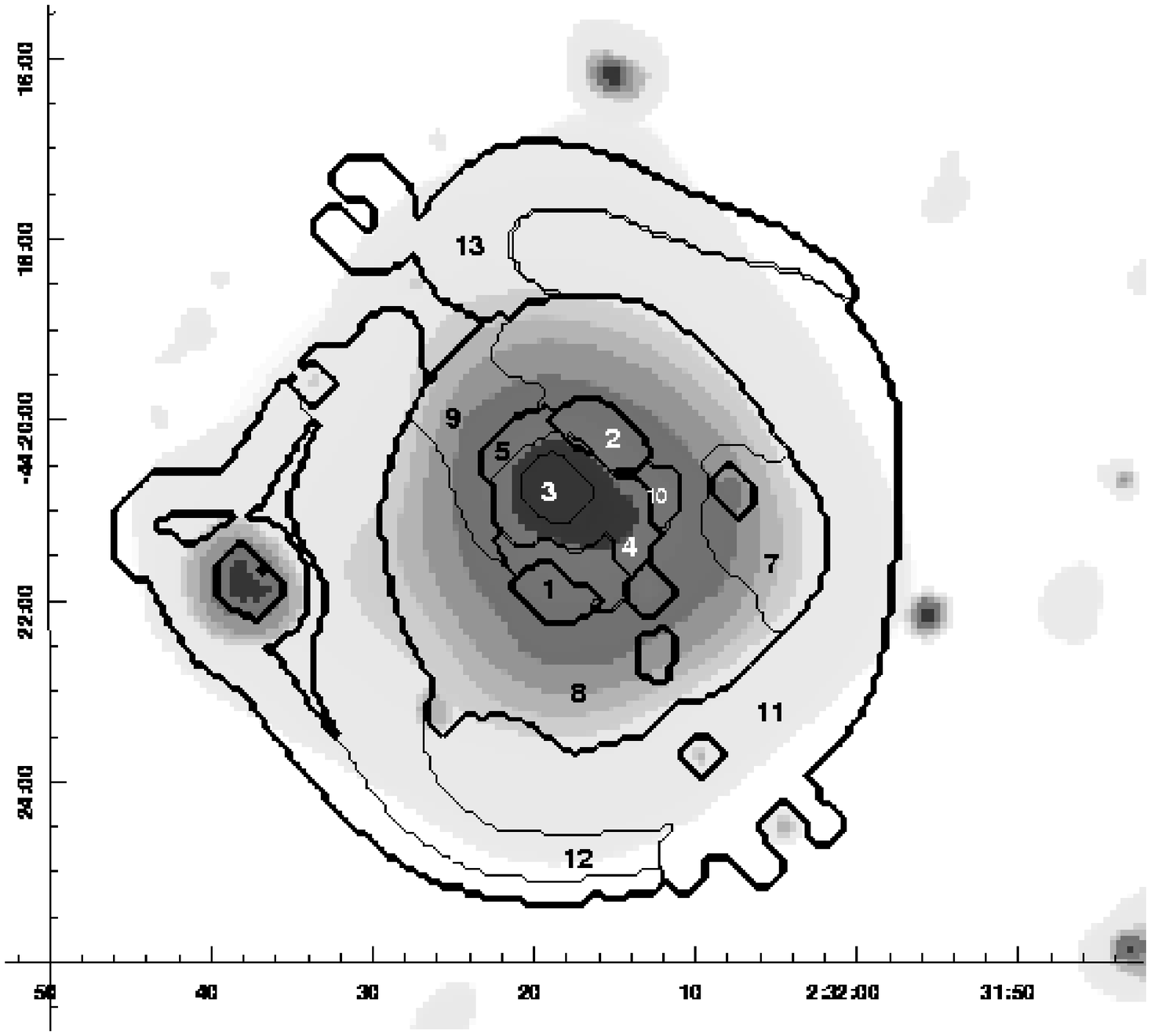}

\includegraphics[width=6cm]{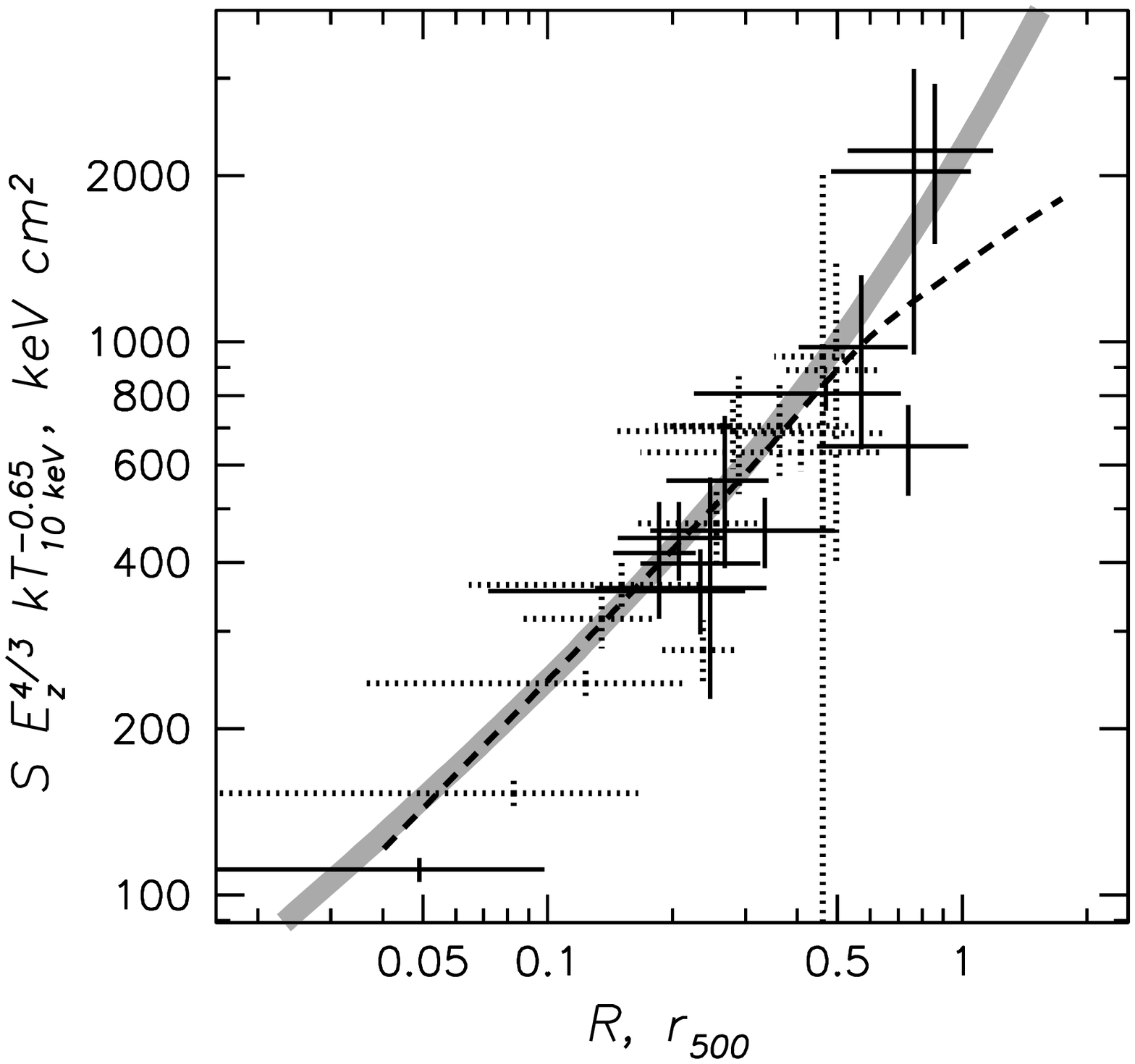}\hfill\includegraphics[width=6cm]{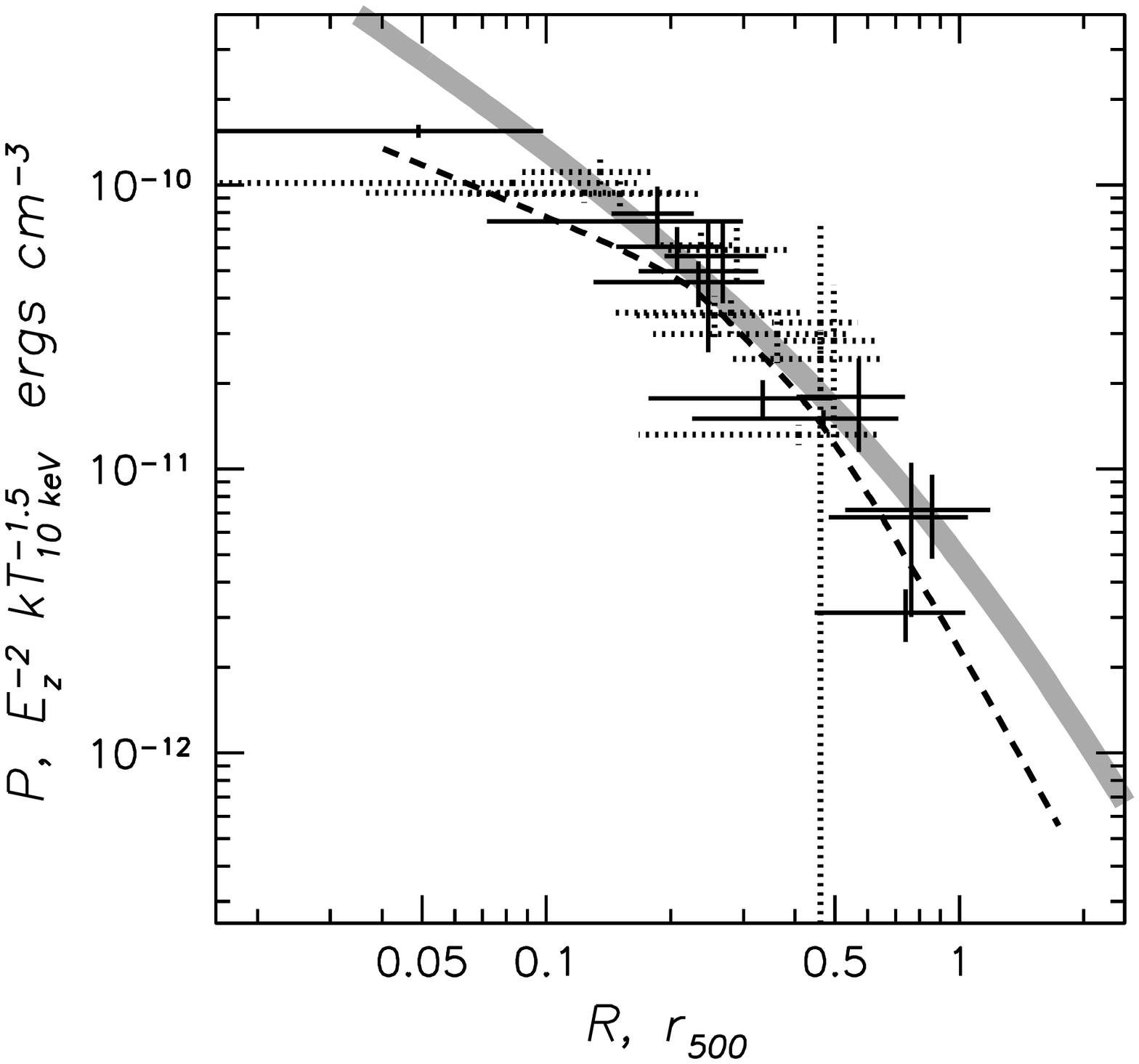}\hfill\includegraphics[width=6cm]{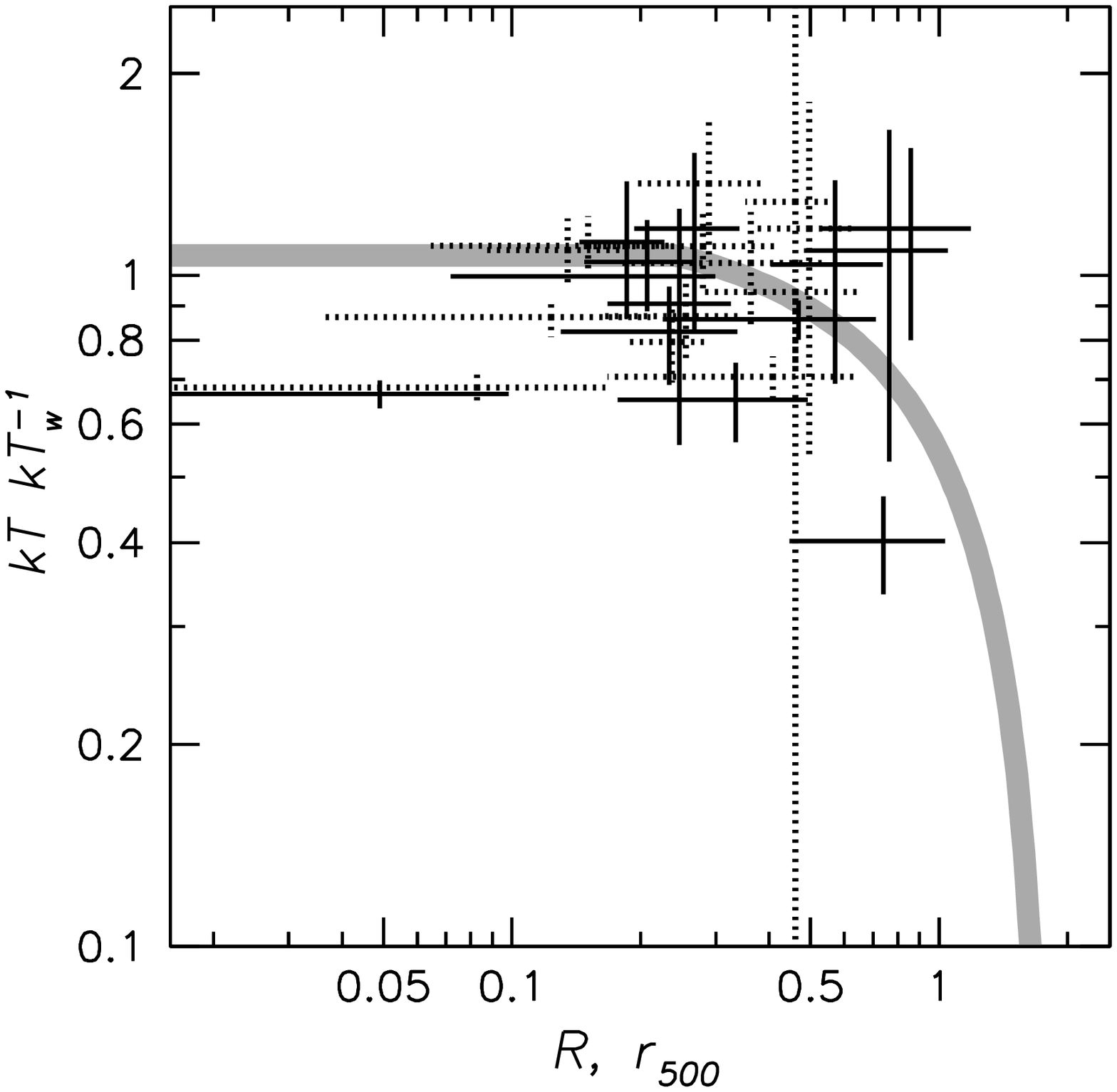}

\caption{RXCJ0232.2--4420. Figure caption is similar to that of
Fig.\ref{f:cl01}.  The surface brightness image is overlaid with contours
showing the spectral extraction regions with numbers corresponding to those
in Table \ref{t:cl03:t}.}
\label{f:cl03}%

\end{figure*}

\begin{table*}[ht]
{
\begin{center}
\footnotesize
{\renewcommand{\arraystretch}{0.9}\renewcommand{\tabcolsep}{0.09cm}
\caption{\footnotesize
Properties of main regions of RXCJ0232.2--4420.}
\label{t:cl03:t}%

\begin{tabular}{cccccccccccc}
 \hline
 \hline
N  &$kT$ &  $\rho_e$ & S & P, $10^{-12}$ & $M_{\rm gas}$ & $r_{\rm min}$ & $r_{\rm max}$& Remarks\\
  & keV     &  $10^{-4}$ cm$^{-3}$& keV cm$^2$ & ergs cm$^{-3}$ & $10^{12} M_\odot$  & Mpc & Mpc  & \\
\hline
 1 &$ 8.9\pm2.6$&$ 34.8\pm4.3$&$ 388\pm 119$&$ 49.8\pm15.9$&$ 1.1\pm0.1$&0.21&0.36\\
 2 &$ 8.0\pm1.2$&$ 42.1\pm3.0$&$ 305\pm  50$&$ 53.7\pm 9.2$&$ 1.3\pm0.1$&0.16&0.29\\
 3 &$ 5.1\pm0.2$&$169.0\pm4.3$&$  77\pm   4$&$136.9\pm 7.3$&$
 1.7\pm0.0$&0.00&0.10 & cool core\\
 4 &$ 7.6\pm0.5$&$ 54.3\pm1.3$&$ 245\pm  17$&$ 65.9\pm 4.7$&$ 5.7\pm0.1$&0.08&0.32\\
 5 &$ 8.5\pm2.0$&$ 51.3\pm4.8$&$ 287\pm  68$&$ 70.1\pm17.4$&$ 0.8\pm0.1$&0.15&0.24\\
 6 &$ 6.3\pm1.0$&$ 40.1\pm3.2$&$ 248\pm  43$&$ 40.3\pm 7.4$&$ 2.0\pm0.2$&0.14&0.36\\
 7 &$ 7.9\pm2.6$&$ 12.6\pm1.7$&$ 677\pm 235$&$ 15.9\pm 5.7$&$ 3.6\pm0.5$&0.43&0.79\\
 8 &$ 6.5\pm0.4$&$ 12.7\pm0.3$&$ 557\pm  39$&$ 13.3\pm 1.0$&$24.5\pm0.6$&0.24&0.76\\
 9 &$ 5.0\pm0.7$&$ 19.8\pm1.5$&$ 315\pm  46$&$ 15.7\pm 2.4$&$
 4.2\pm0.3$&0.19&0.53& entropy tail\\
10 &$ 6.9\pm2.7$&$ 39.8\pm11.$&$ 275\pm 118$&$ 44.0\pm21.3$&$ 0.8\pm0.2$&0.18&0.35\\
11 &$ 8.9\pm2.8$&$  4.4\pm0.3$&$1532\pm 494$&$  6.4\pm
2.1$&$17.7\pm1.3$&0.57&1.27& forward shock\\
12 &$ 3.1\pm0.5$&$  5.6\pm0.7$&$ 448\pm  83$&$  2.8\pm
0.6$&$10.0\pm1.3$&0.48&1.11& outskirts\\
13 &$ 8.3\pm4.3$&$  4.5\pm0.9$&$1405\pm 749$&$  6.0\pm 3.3$&$ 5.7\pm1.2$&0.52&1.12\\
\hline
\end{tabular}
}
\end{center}
}
%\vspace*{0.2cm}
\end{table*}

\subsection{RXCJ0232.2--4420 } 

The hardness ratio map reveals a soft compact source in the center. In the
image we see some elongation to the north. In the inner region the surface
brightness distribution shows an indication of a triangular shape with one
tip of the triangle pointing towards the north and the other to the west. As
a result what we see is that there is colder material in the east compared
to the west.  The pressure map looks rather symmetric, but we have an
asymmetric entropy structure. While a possible scenario could be a slow
infall of the material from the north-east, the west part of the cluster
appears to be systematically hotter as compared to the eastern side, which
could be an indication of a strong shock.

The spectroscopic analysis is reported in Table \ref{t:cl03:t} and
Fig.\ref{f:cl03}. The point-like source appears to be the low-entropy core
of the cluster, with one of the lowest entropies ($77\pm4$ keV cm$^2$) in
our sample. The hot region on the west is characterized by both an
enhancement in the pressure and entropy, yet of high statistical
uncertainty. We attribute the appearance of this region to a forward shock
of the Mach number of $3\pm1$. The low entropy gas in the north east (region
9) is confirmed in the spectral analysis. The behavior of the entropy in
the outskirts of this cluster led us to a conclusion, supported by other
clusters in this sample: the clusters in the advanced stage of interaction
have systematically higher entropy at $r_{500}$ compared to the average
trend. We ascribe this result to the heating by the forward shock, induced
during the merger. 

\begin{figure*}
\includegraphics[width=8cm]{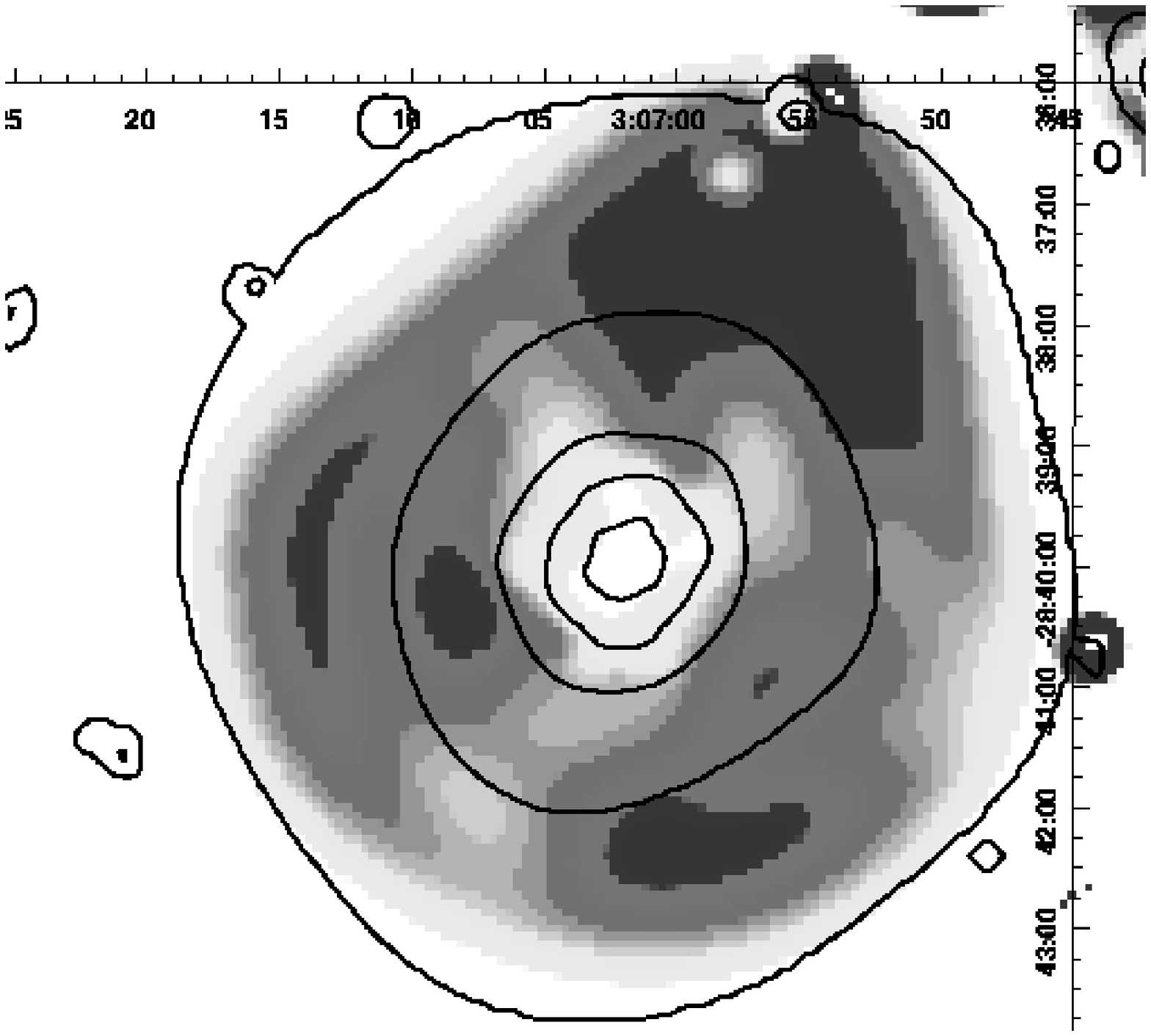}\hfill\includegraphics[width=8cm]{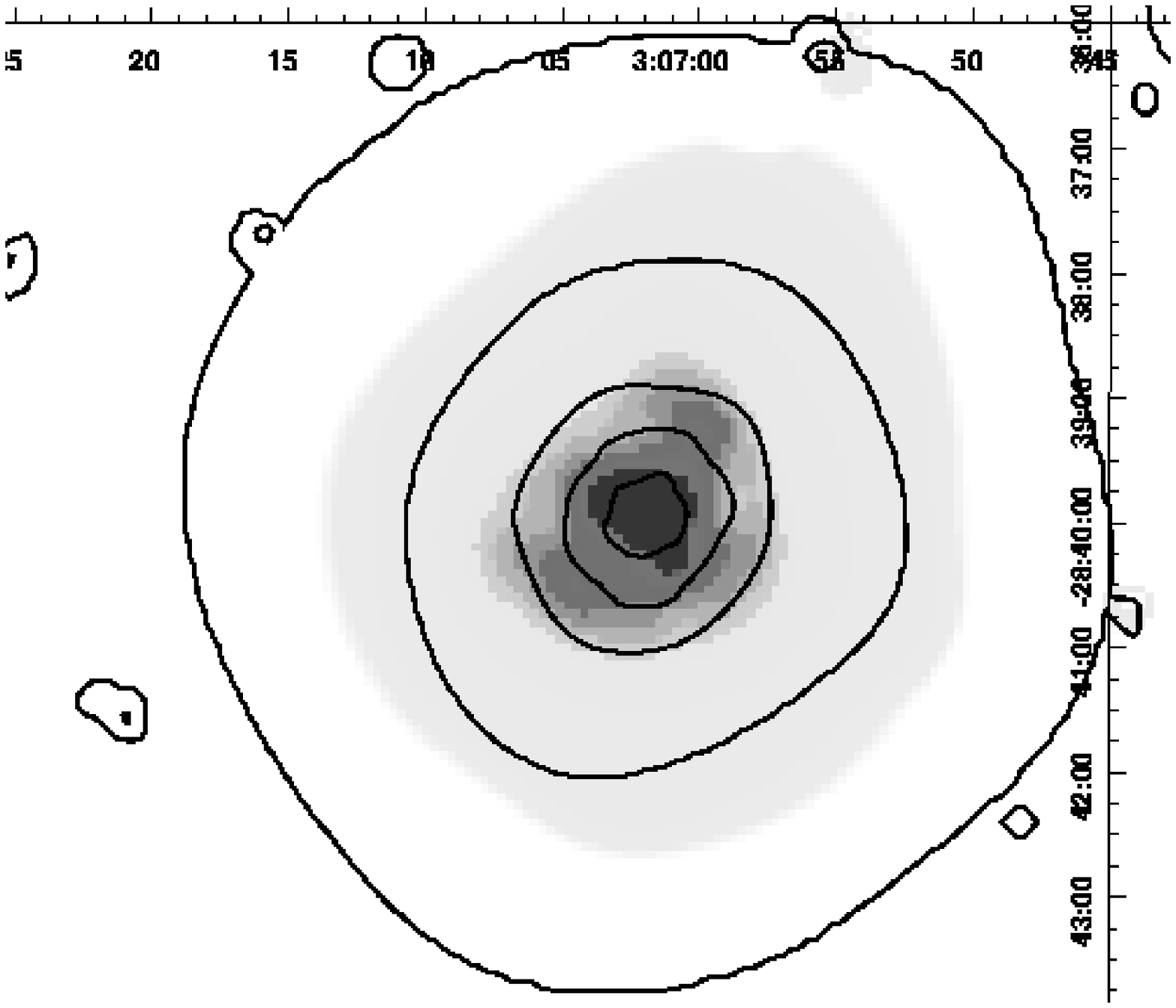}

\includegraphics[width=8cm]{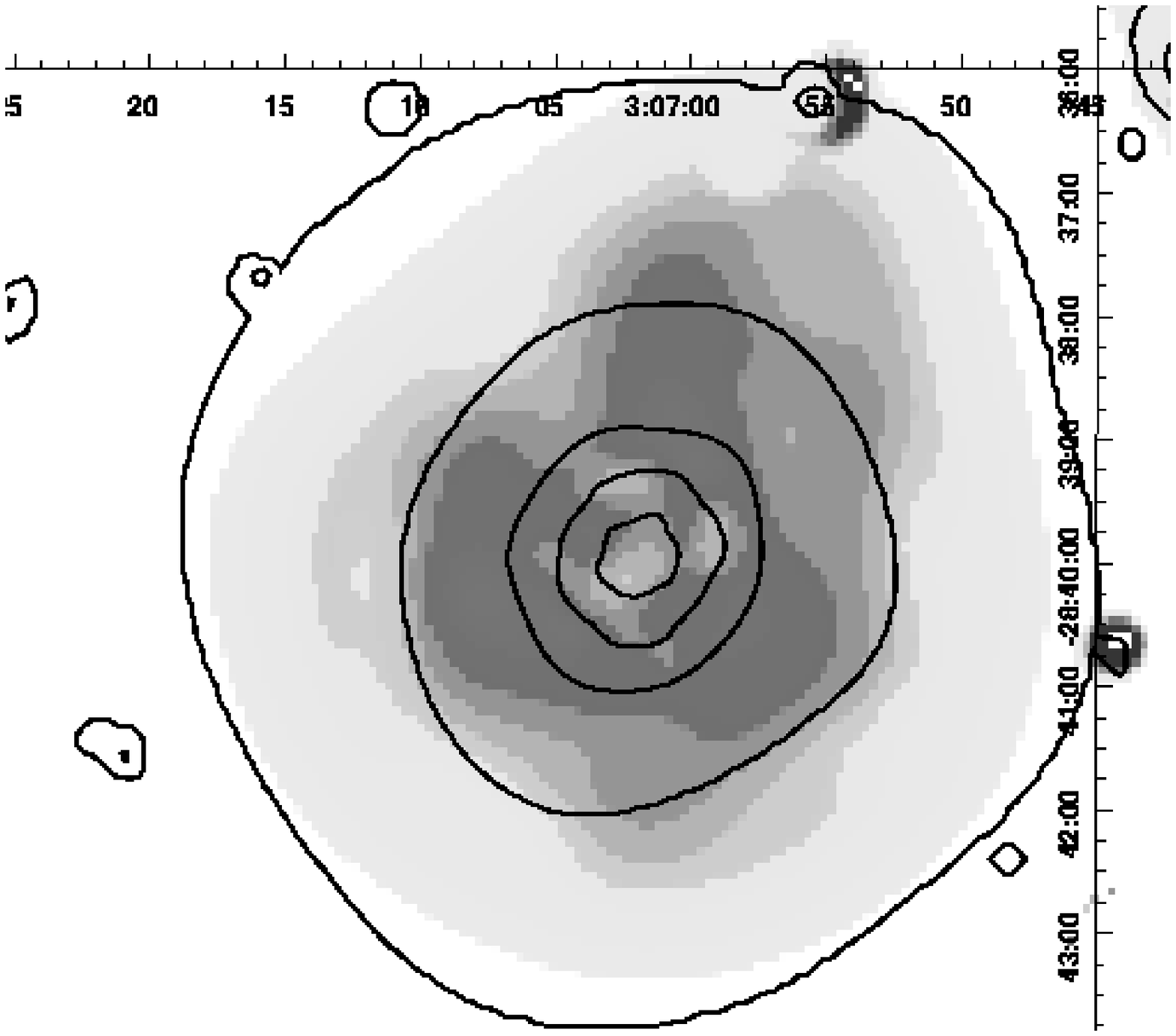}\hfill\includegraphics[width=8cm]{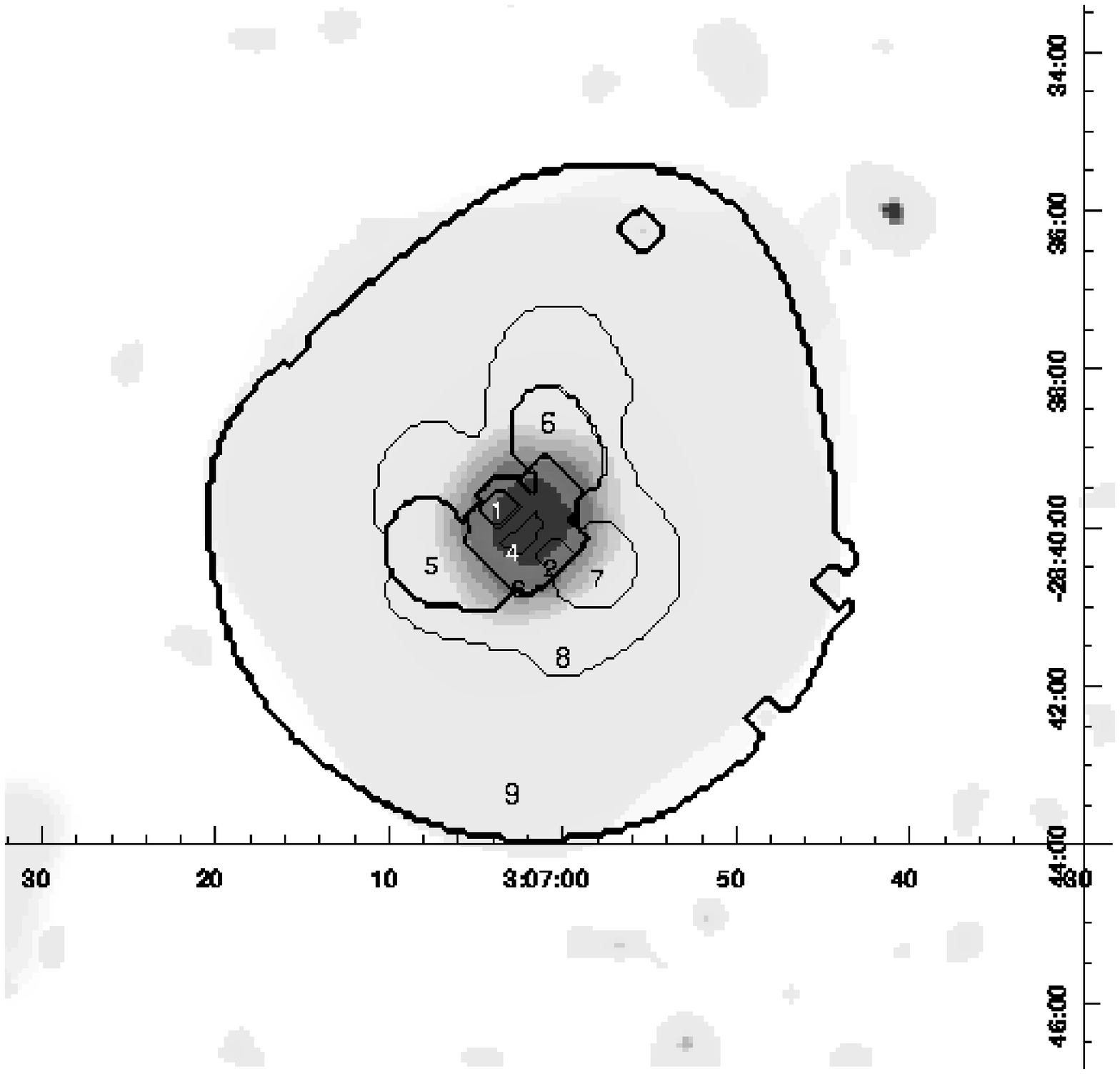}

\includegraphics[width=6cm]{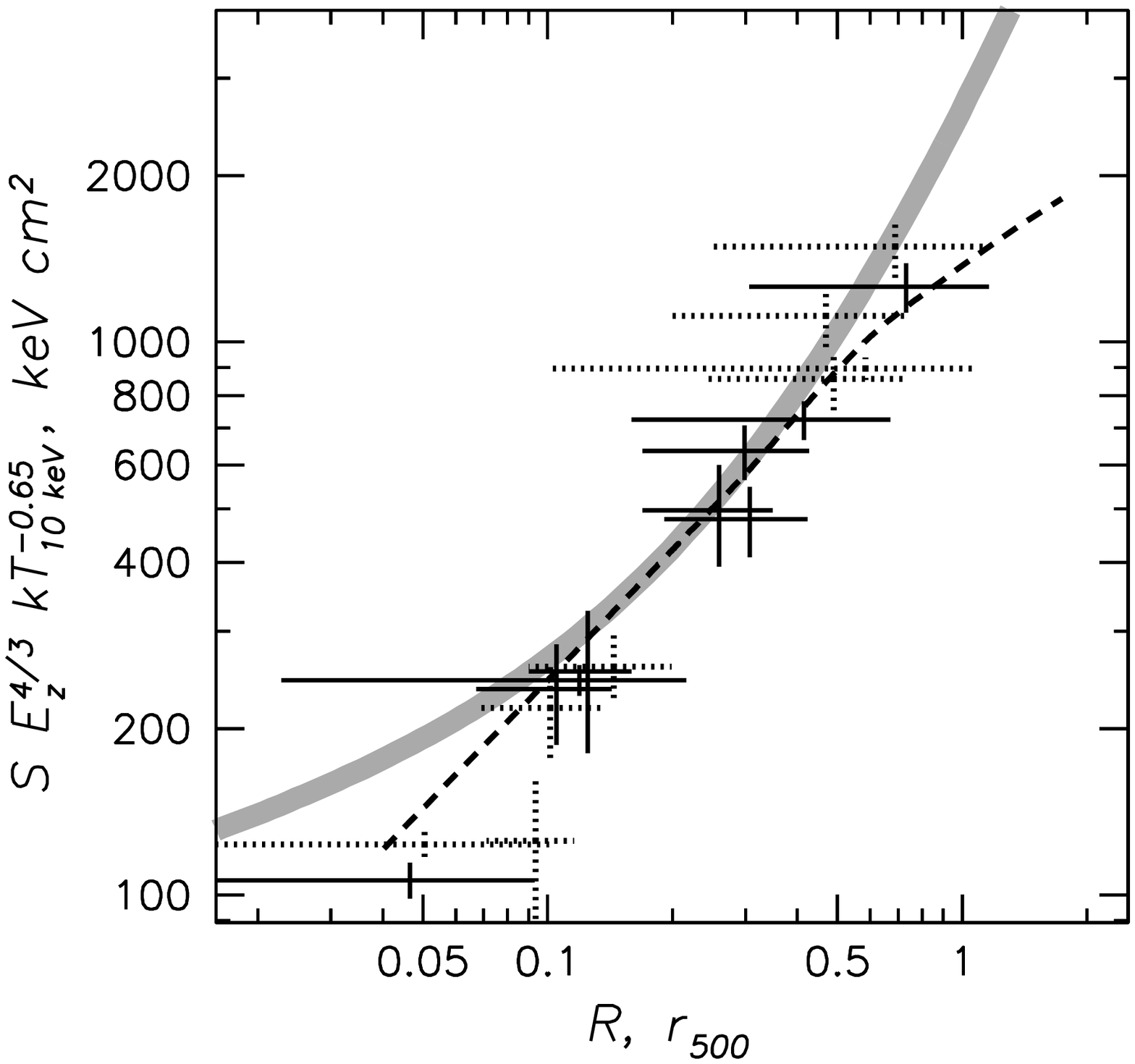}\hfill\includegraphics[width=6cm]{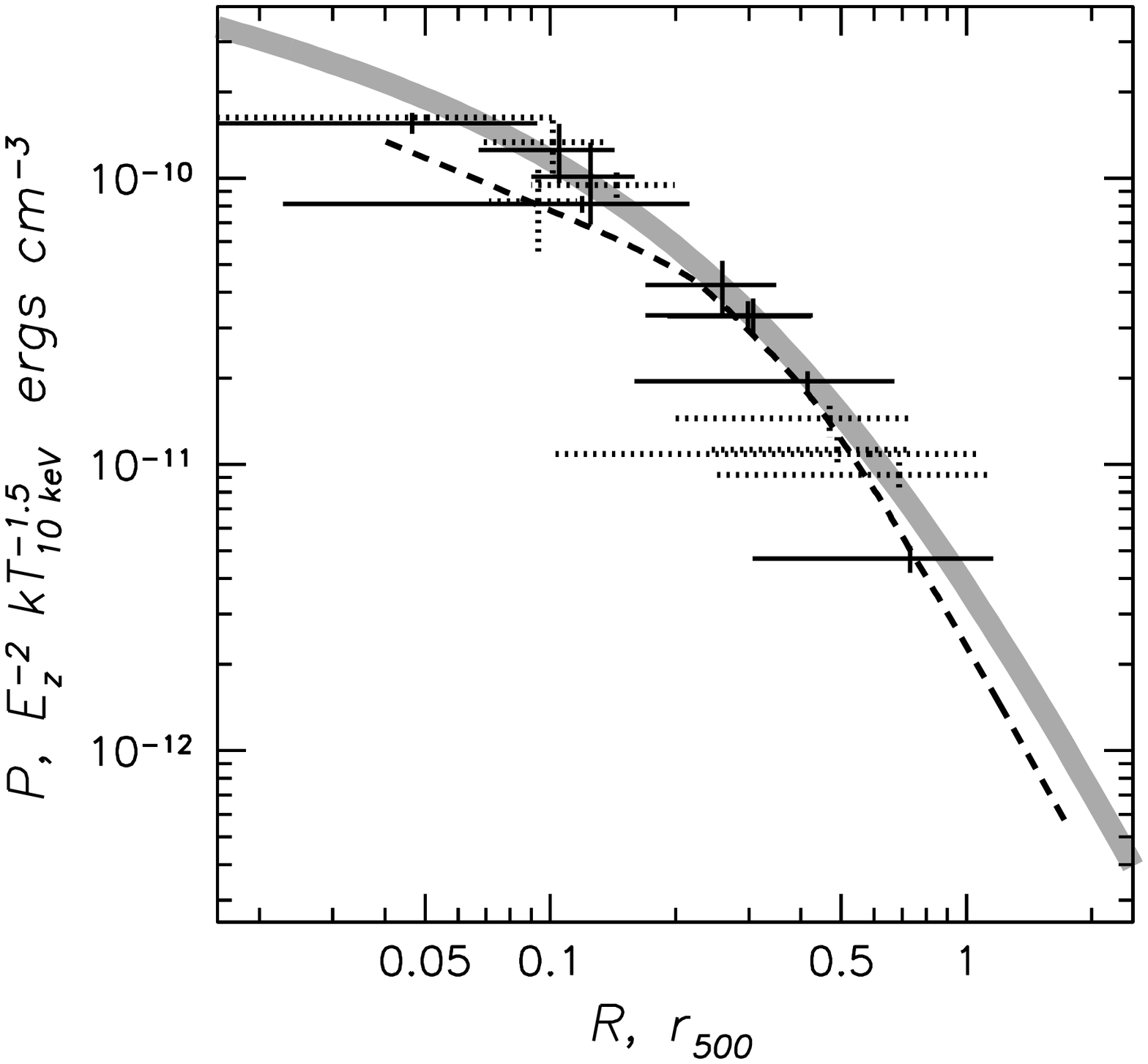}\hfill\includegraphics[width=6cm]{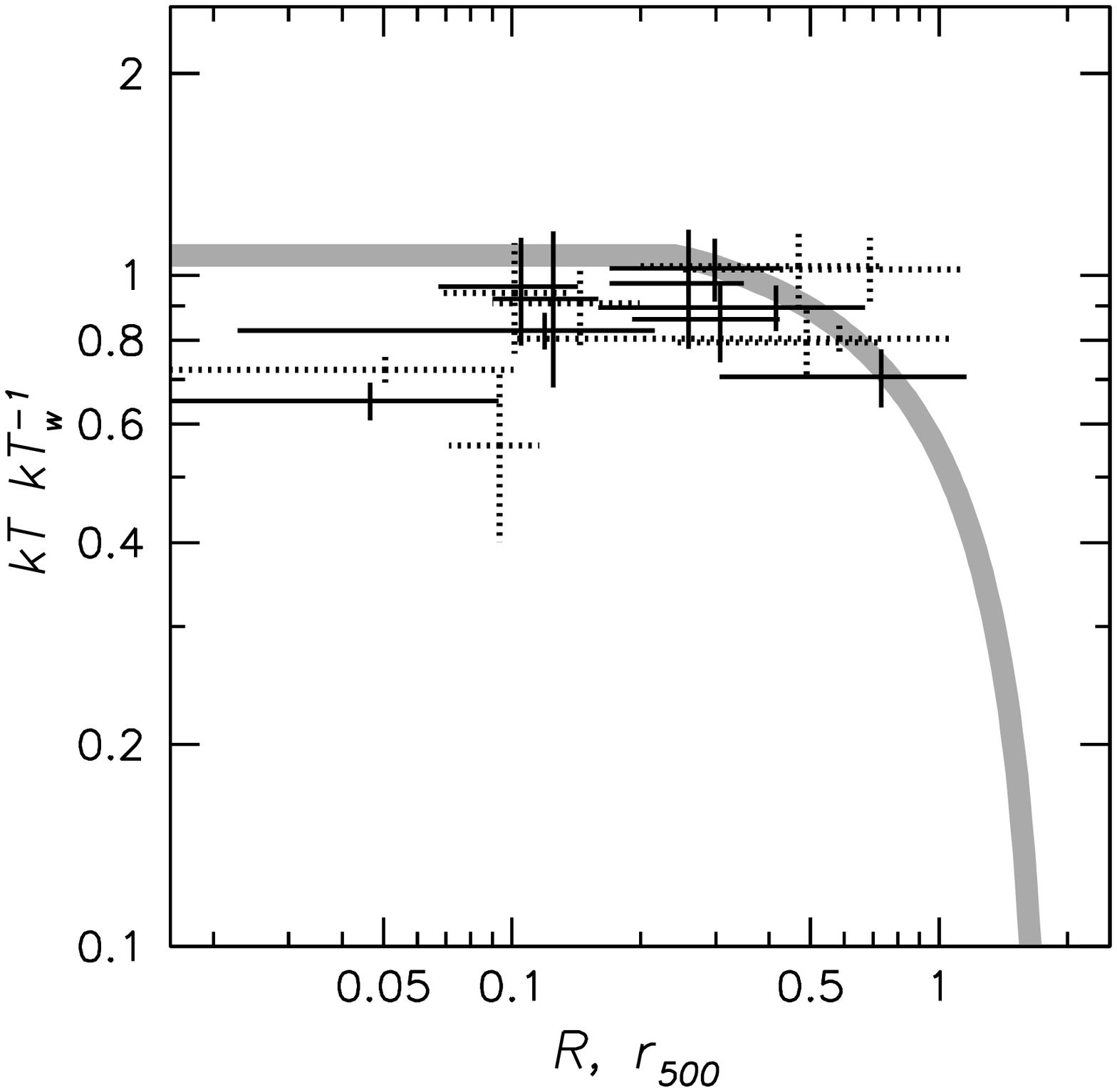}

\caption{RXCJ0307.0--2840. Figure caption is similar to that of
Fig.\ref{f:cl01}.  The surface brightness image is overlaid with contours
showing the spectral extraction regions with numbers corresponding to those
in Table \ref{t:cl05:t}.}
\label{f:cl05}%

\end{figure*}

\begin{table*}[ht]
{
\begin{center}
\footnotesize
{\renewcommand{\arraystretch}{0.9}\renewcommand{\tabcolsep}{0.09cm}
\caption{\footnotesize
Properties of main regions of RXCJ0307.0--2840.}
\label{t:cl05:t}%

\begin{tabular}{cccccccccccc}
 \hline
 \hline
N &$kT$ &  $\rho_e$ & S & P, $10^{-12}$ & $M_{\rm gas}$ & $r_{\rm min}$ &
$r_{\rm max}$ & Remarks\\
 & keV     &  $10^{-4}$ cm$^{-3}$& keV cm$^2$ & ergs cm$^{-3}$ &
  $10^{12} M_\odot$  & Mpc & Mpc &   \\
\hline
 1&$ 6.3\pm1.2$&$ 85.9\pm12.2$&$ 151\pm  31$&$ 87.4\pm20.3$&$ 0.3\pm0.0$&0.07&0.14& \\
 2&$ 6.1\pm1.6$&$ 72.2\pm12.9$&$ 163\pm  47$&$ 70.4\pm22.3$&$ 0.2\pm0.0$&0.09&0.16& \\
 3&$ 5.5\pm0.3$&$ 64.7\pm 1.6$&$ 157\pm  10$&$ 56.6\pm 3.8$&$ 4.1\pm0.1$&0.02&0.22& \\
 4&$ 4.3\pm0.3$&$157.7\pm 8.6$&$  68\pm   5$&$108.5\pm 9.2$&$
 0.6\pm0.0$&0.00&0.09&cool core \\
 5&$ 6.8\pm0.7$&$ 21.3\pm 1.2$&$ 408\pm  46$&$ 23.1\pm 2.8$&$ 2.7\pm0.2$&0.17&0.43& \\
 6&$ 5.7\pm0.8$&$ 25.2\pm 1.9$&$ 307\pm  45$&$ 22.9\pm 3.6$&$ 2.1\pm0.2$&0.19&0.43& \\
 7&$ 6.4\pm1.3$&$ 28.6\pm 2.2$&$ 318\pm  66$&$ 29.5\pm 6.4$&$ 1.4\pm0.1$&0.17&0.35& \\
 8&$ 5.9\pm0.5$&$ 14.3\pm 0.4$&$ 465\pm  37$&$ 13.6\pm 1.1$&$14.1\pm0.4$&0.16&0.68& \\
 9&$ 4.7\pm0.5$&$  4.4\pm 0.2$&$ 808\pm  83$&$  3.3\pm 0.3$&$44.5\pm1.8$&0.31&1.17& \\
\hline
\end{tabular}
}
\end{center}
}
%\vspace*{0.2cm}
\end{table*}

\subsection{RXCJ0307.0--2840 } 

The cluster seems to be quite relaxed judging from the symmetric appearance
of the image and the pressure map. The temperature map shows small amplitude
fluctuations. According to the spectroscopic analysis, reported in Table
\ref{t:cl05:t} and shown in Fig.\ref{f:cl05}, the cluster has a cool core,
but otherwise exhibits no significant deviations from the modified scaling
relation, bearing a slightly higher entropy and showing no fluctuations in
the pressure profile. The 1d and 2d modeling agree well with each other.

\begin{figure*}
\includegraphics[width=8cm]{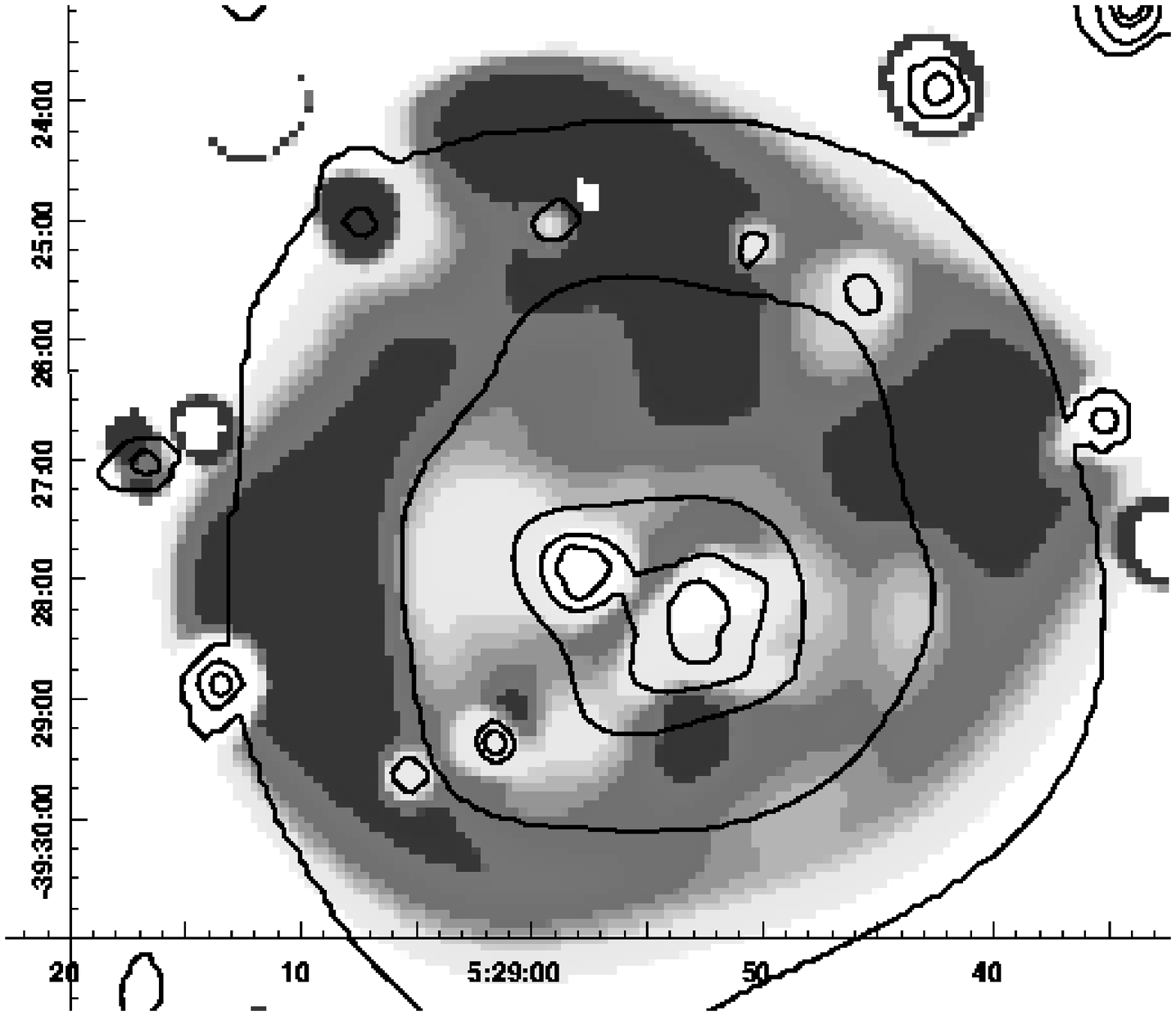}\hfill\includegraphics[width=8cm]{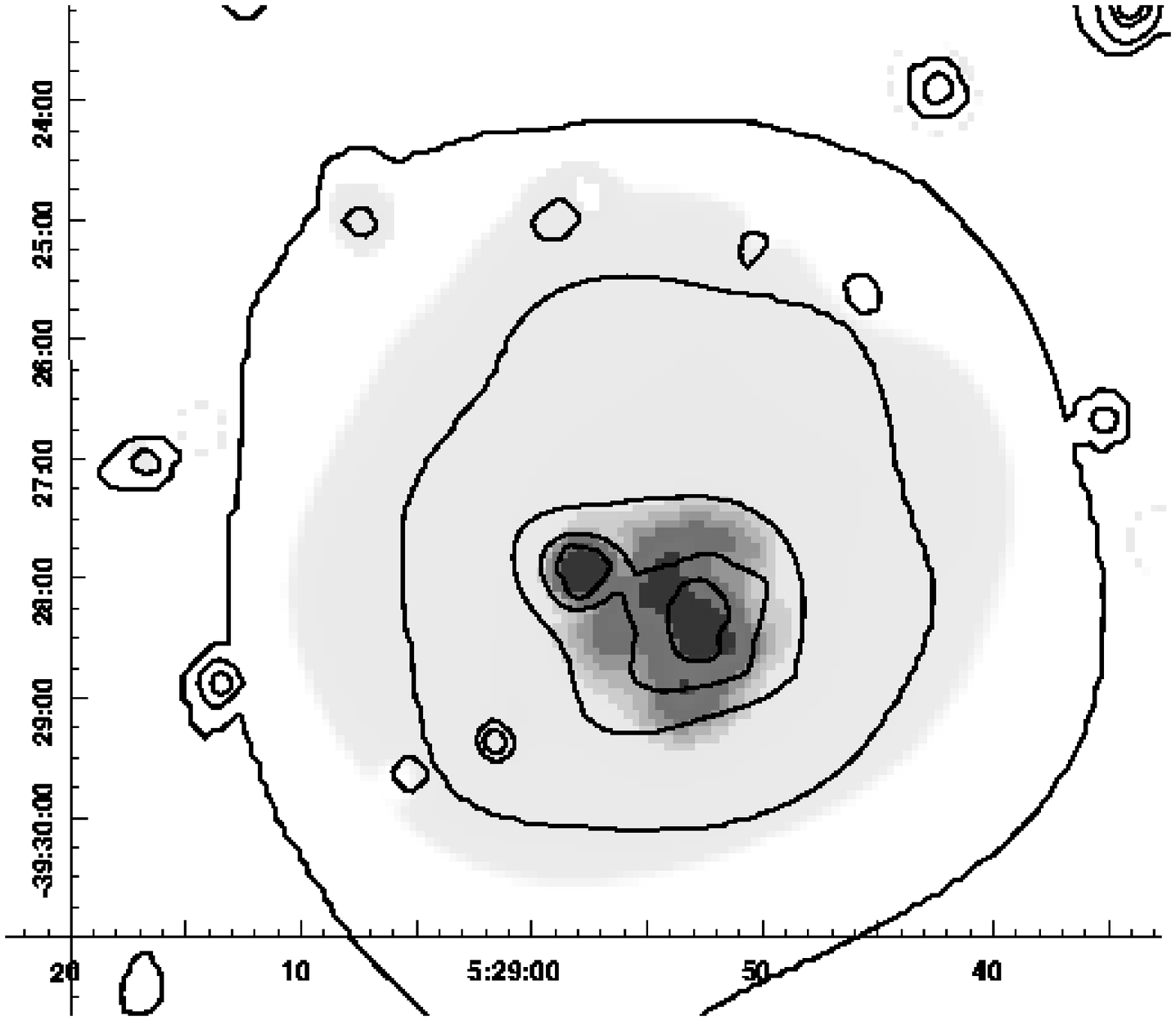}

\includegraphics[width=8cm]{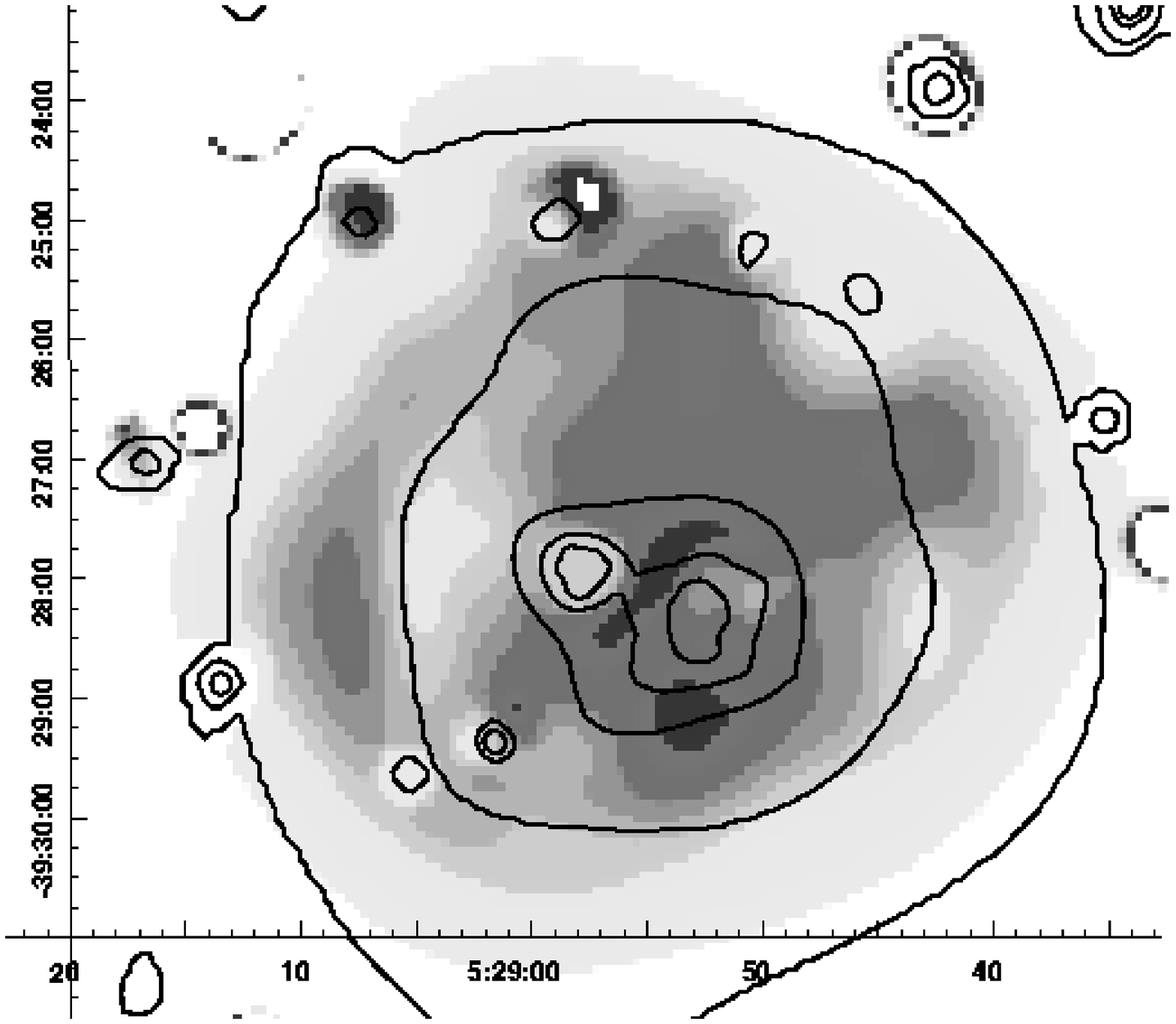}\hfill\includegraphics[width=8cm]{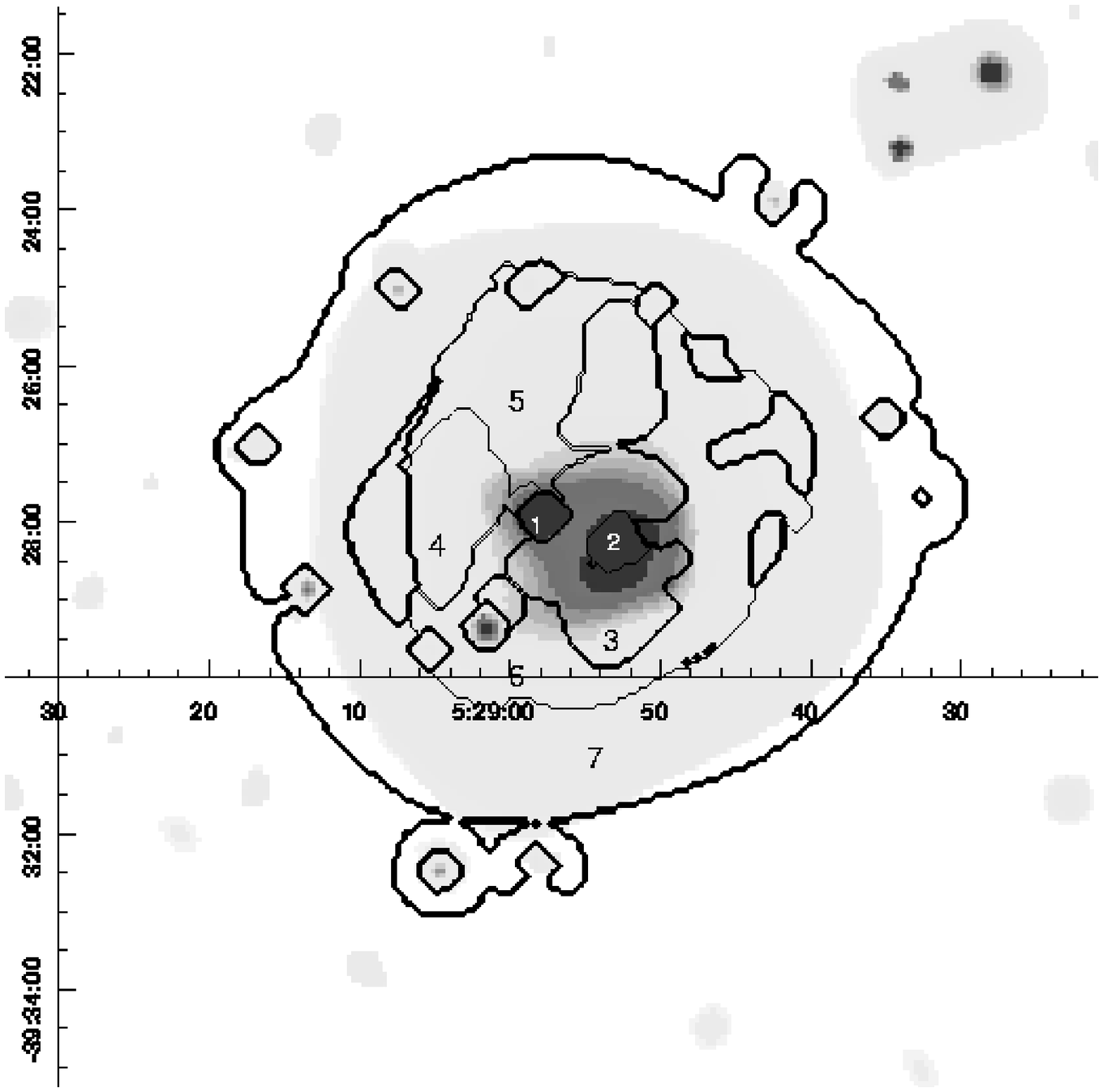}

\includegraphics[width=6cm]{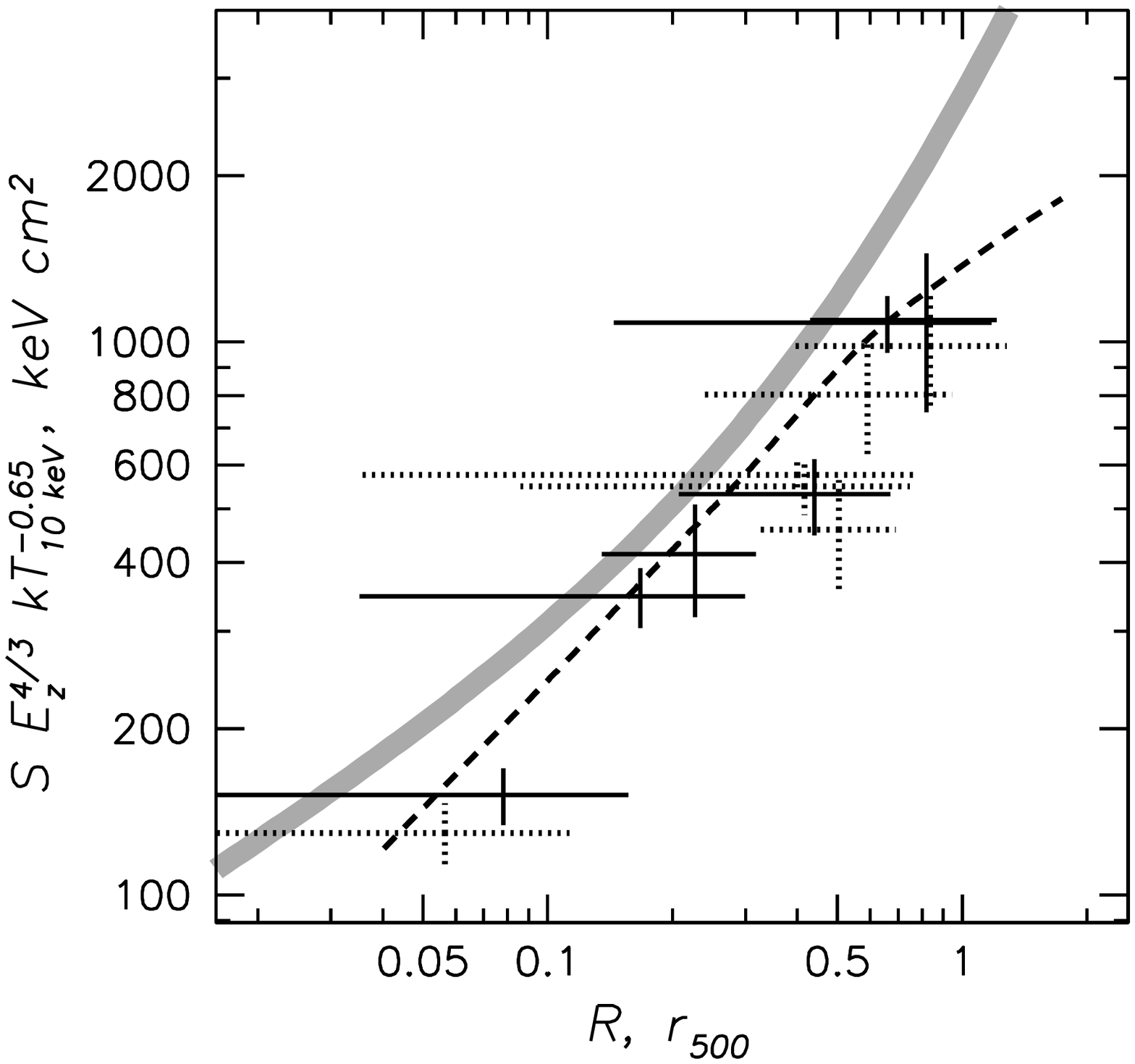}\hfill\includegraphics[width=6cm]{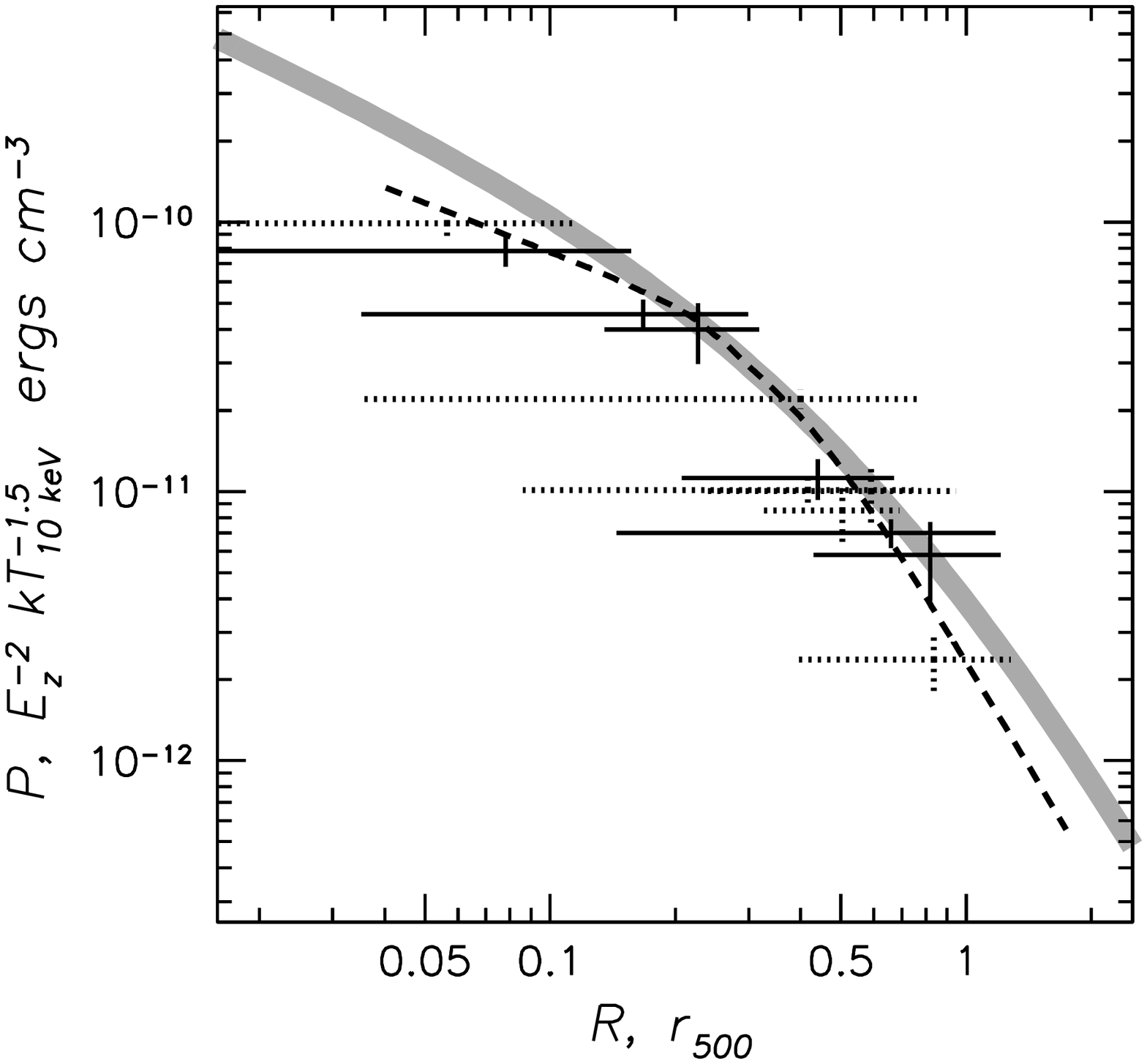}\hfill\includegraphics[width=6cm]{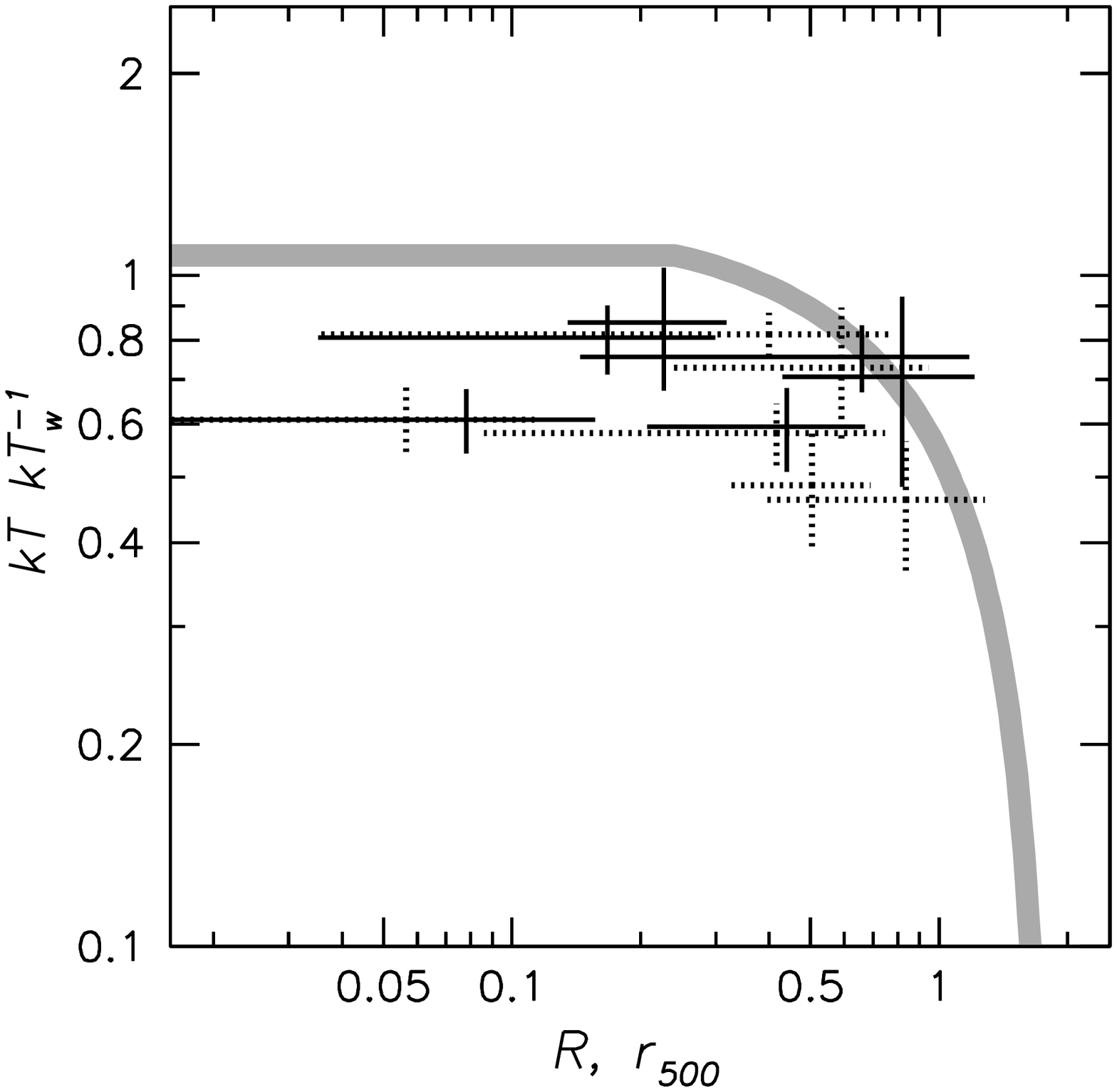}

\caption{RXCJ0528.9--3927. Figure caption is similar to that of
Fig.\ref{f:cl01}.  The surface brightness image is overlaid with contours
showing the spectral extraction regions with numbers corresponding to those
in Table \ref{t:cl08:t}.}
\label{f:cl08}%

\end{figure*}

\begin{table*}[ht]
{
\begin{center}
\footnotesize
{\renewcommand{\arraystretch}{0.9}\renewcommand{\tabcolsep}{0.09cm}
\caption{\footnotesize
Properties of main regions of RXCJ0528.9--3927 (the region 1 turned to be a
QSO and is omitted).}
\label{t:cl08:t}%

\begin{tabular}{cccccccccccc}
 \hline
 \hline
N  &$kT$ &  $\rho_e$ & S & P, $10^{-12}$ & $M_{\rm gas}$ & $r_{\rm min}$ &
$r_{\rm max}$ & Remarks\\
  & keV     &  $10^{-4}$ cm$^{-3}$& keV cm$^2$ & ergs cm$^{-3}$ & $10^{12} M_\odot$  & Mpc & Mpc  & \\
\hline
 2&$ 4.7\pm0.5$&$ 93.8\pm 6.0$&$ 106\pm  12$&$ 70.6\pm 9.0$&$
 2.0\pm0.1$&0.00&0.17& core region\\
 3&$ 6.2\pm0.7$&$ 41.2\pm 2.5$&$ 242\pm  30$&$ 41.1\pm 5.4$&$ 6.1\pm0.4$&0.04&0.32& \\
 4&$ 6.5\pm1.4$&$ 34.3\pm 5.0$&$ 288\pm  66$&$ 35.9\pm 9.1$&$ 2.1\pm0.3$&0.15&0.34& \\
 5&$ 5.4\pm1.7$&$  6.0\pm 0.4$&$ 764\pm 244$&$  5.2\pm 1.7$&$ 9.1\pm0.7$&0.46&1.31& \\
 6&$ 4.6\pm0.7$&$ 13.8\pm 1.4$&$ 370\pm  58$&$ 10.1\pm 1.8$&$10.2\pm1.0$&0.22&0.72& \\
 7&$ 5.8\pm0.7$&$  6.8\pm 0.3$&$ 755\pm  89$&$  6.3\pm 0.8$&$54.6\pm2.1$&0.16&1.27& \\
\hline
\end{tabular}
}
\end{center}
}
%\vspace*{0.2cm}
\end{table*}

\subsection{RXCJ0528.9--3927} 

A strong soft point source near the center ($<20$\% of the flux) prevents us
from a more detailed analysis of the central region.  The pressure peak is
distorted on small scales, while on large scales it appears quite
relaxed. There is some entropy and temperature structure in the north-west.

The spectroscopic analysis reported in Table \ref{t:cl08:t} and
Fig.\ref{f:cl08} does not include the region 1, which is associated with the
point-like source and reveals a clearly non-thermal spectrum. The suggested
fluctuations in the north-west are confirmed to be a zone of lower entropy
and also in general the entropy of this cluster is lower at almost every
radii, compared to the general trend.

\begin{figure*}
\includegraphics[width=8cm]{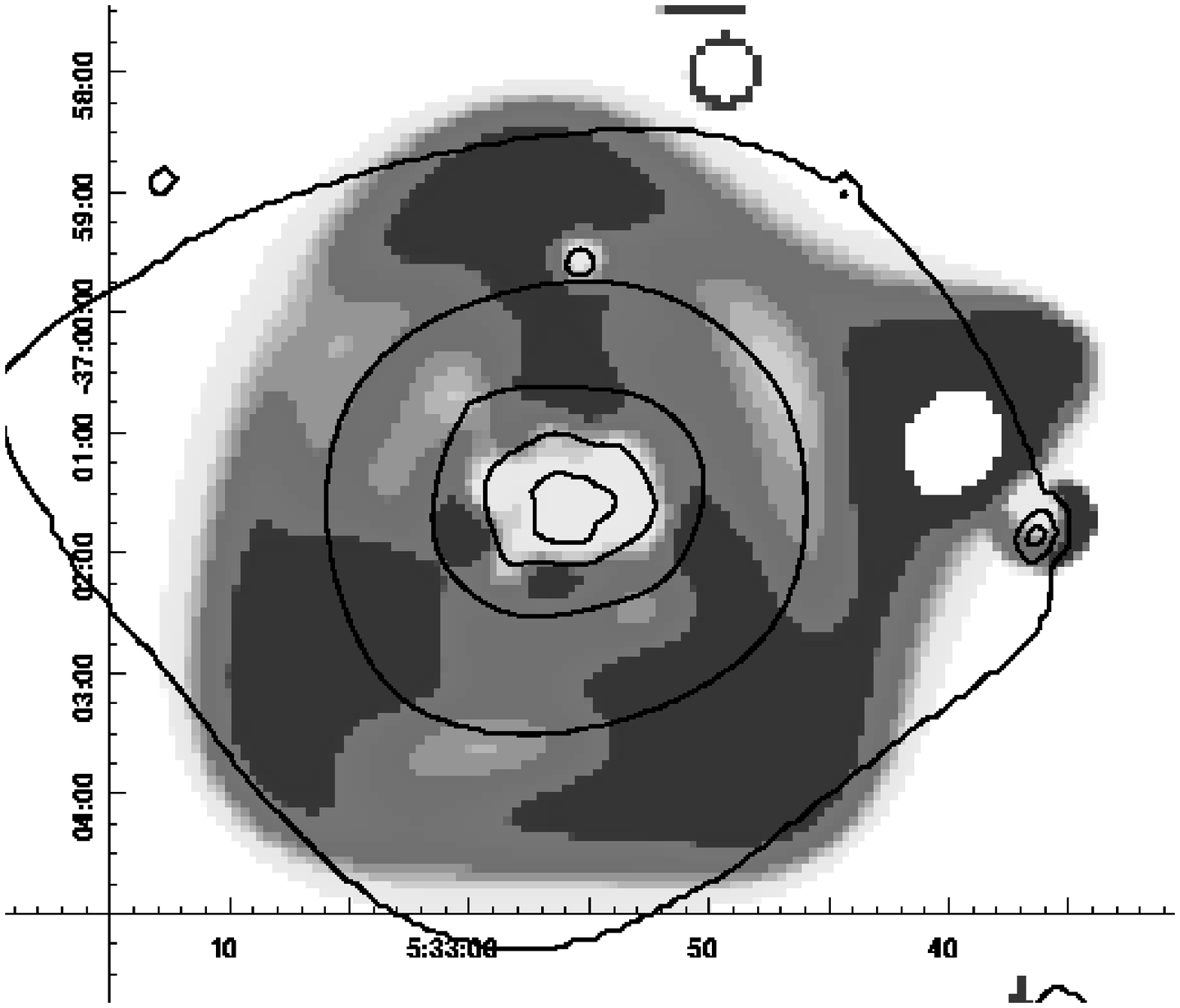}\hfill\includegraphics[width=8cm]{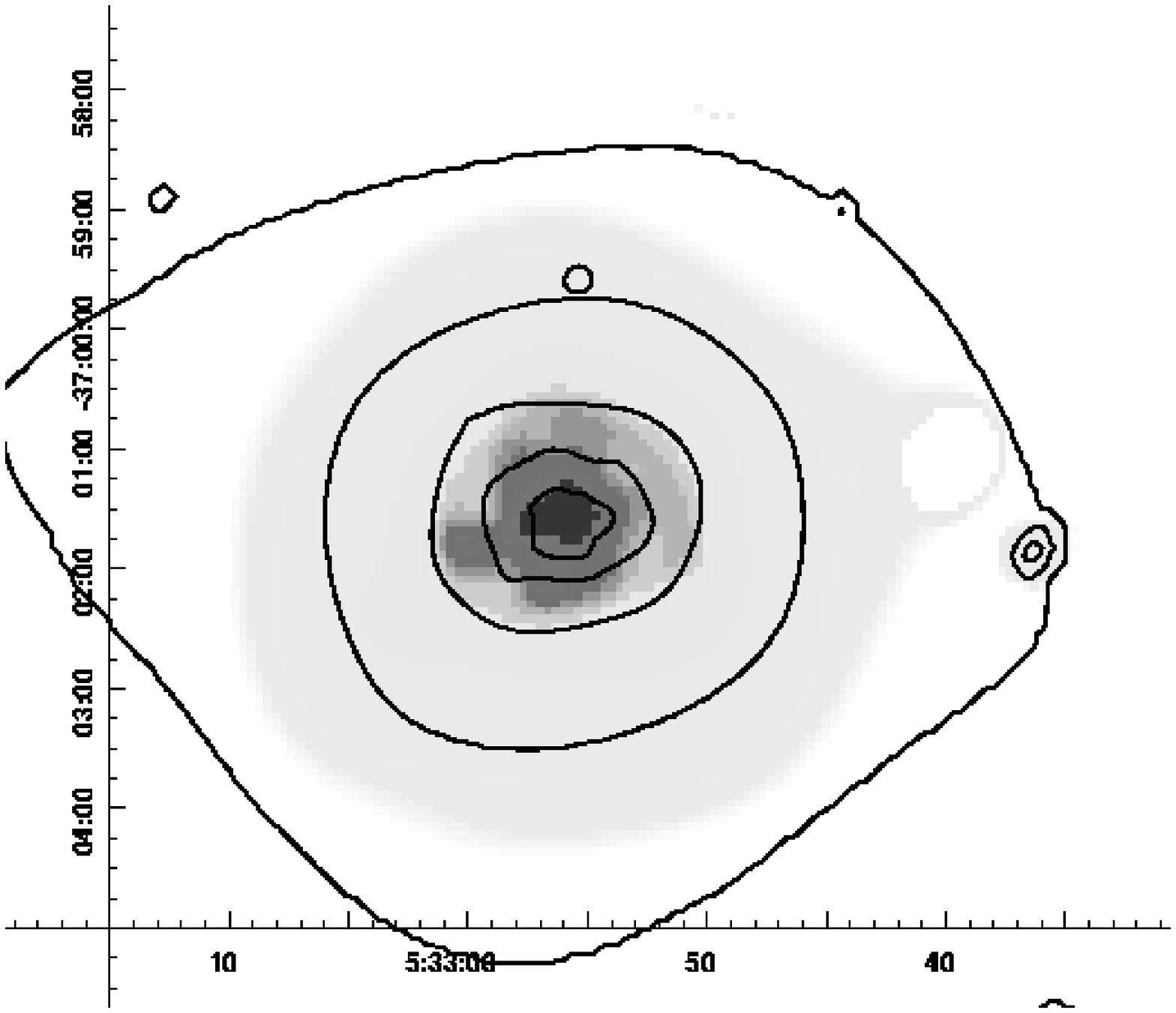}

\includegraphics[width=8cm]{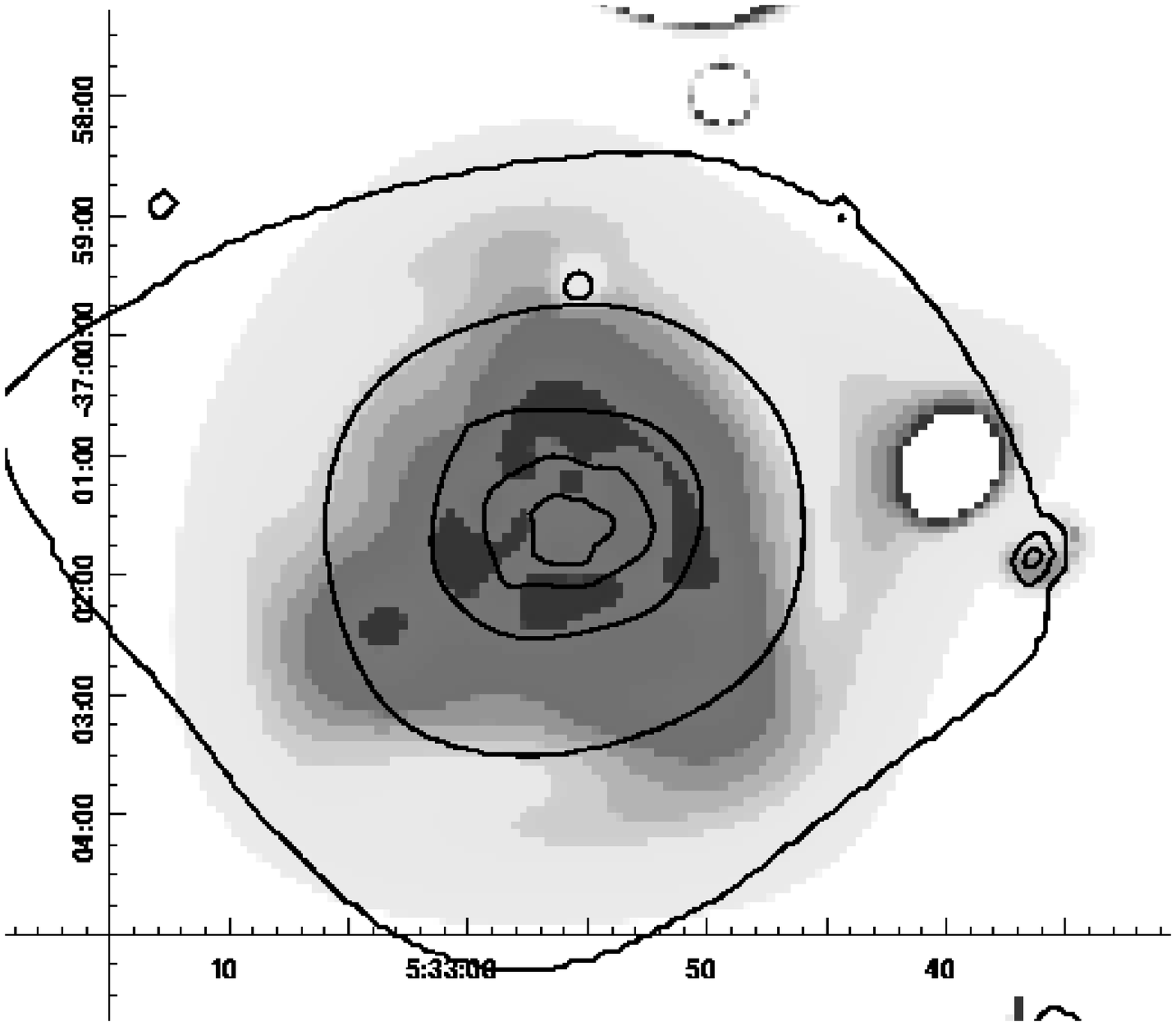}\hfill\includegraphics[width=8cm]{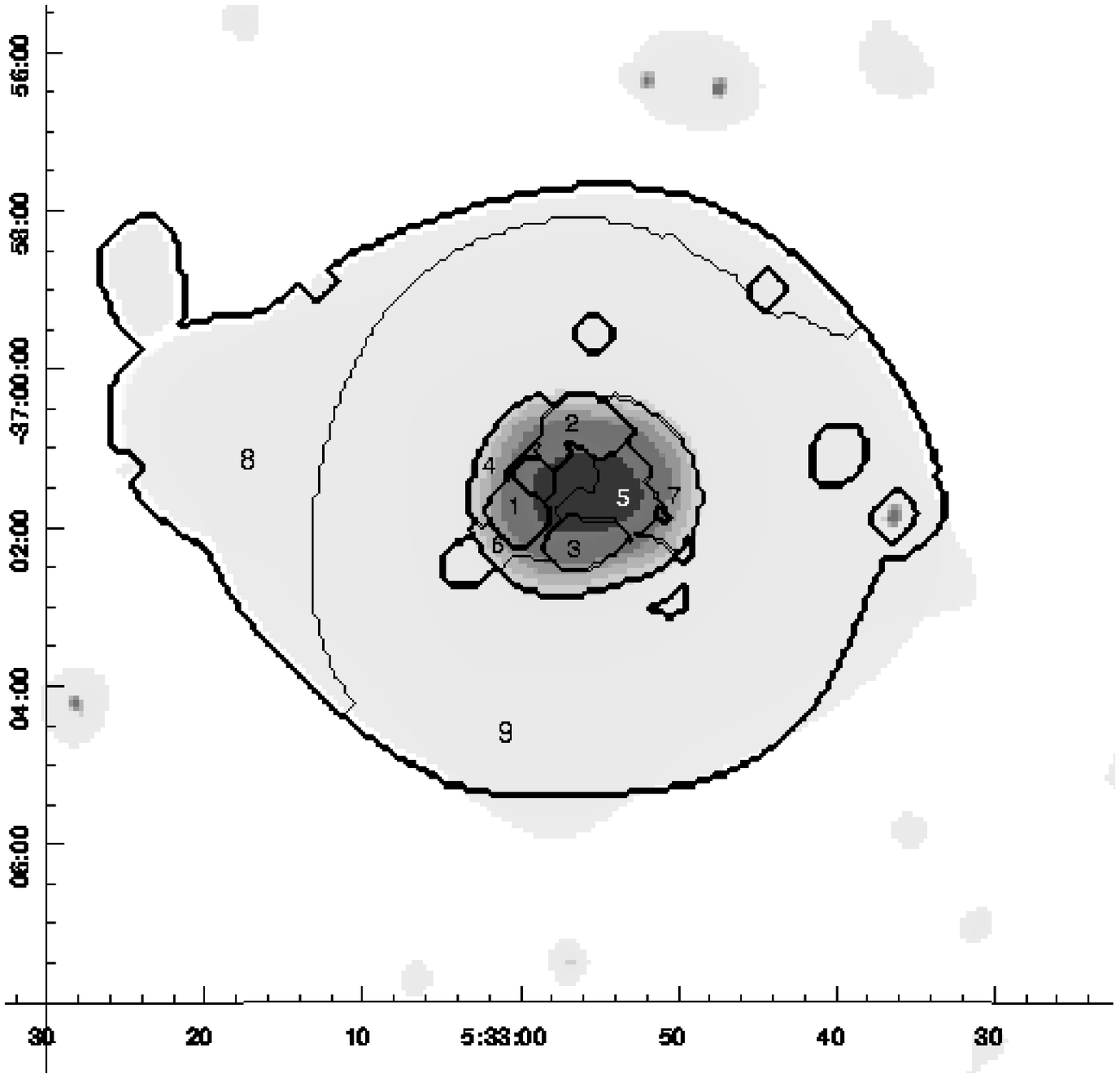}

\includegraphics[width=6cm]{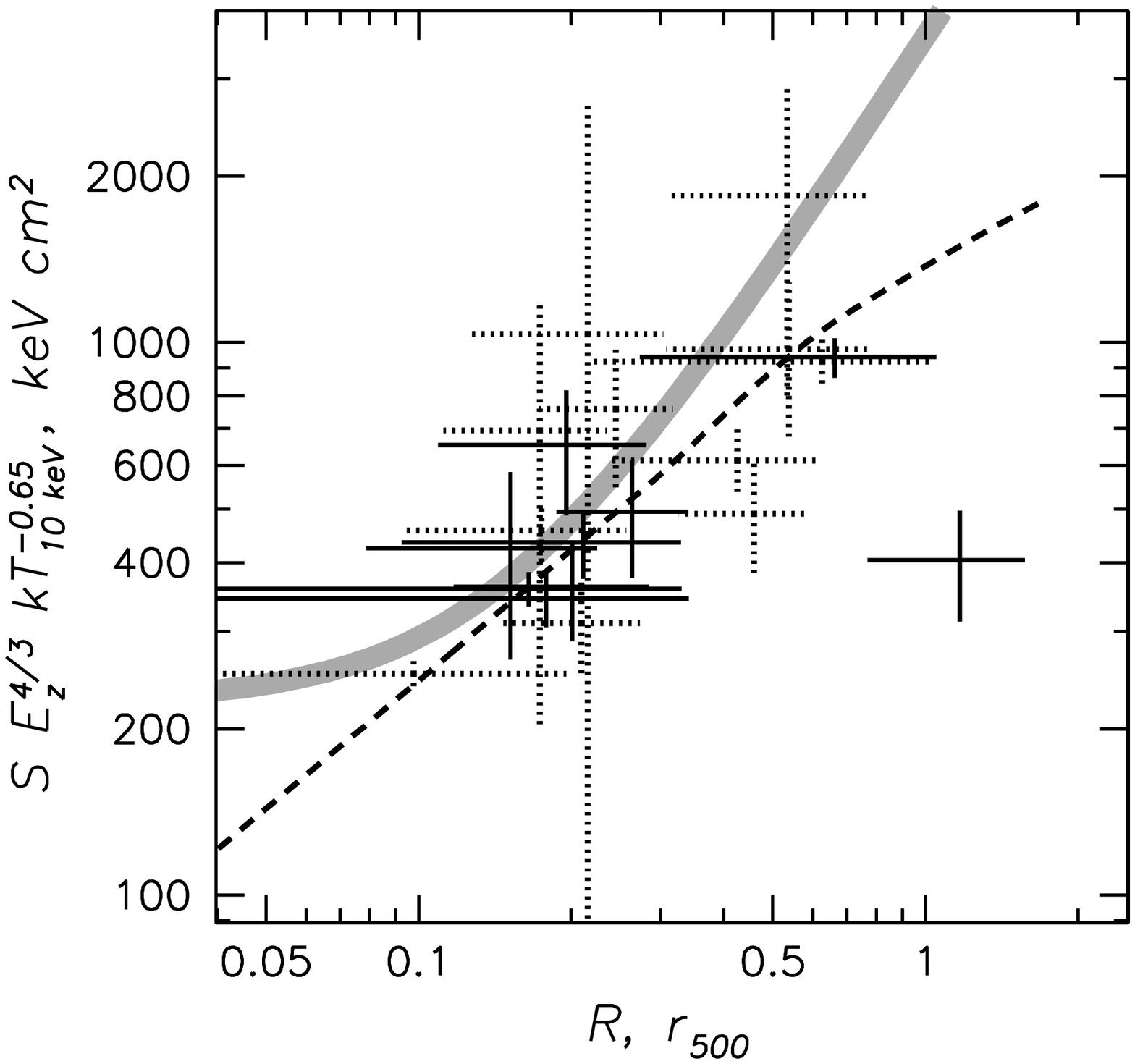}\hfill\includegraphics[width=6cm]{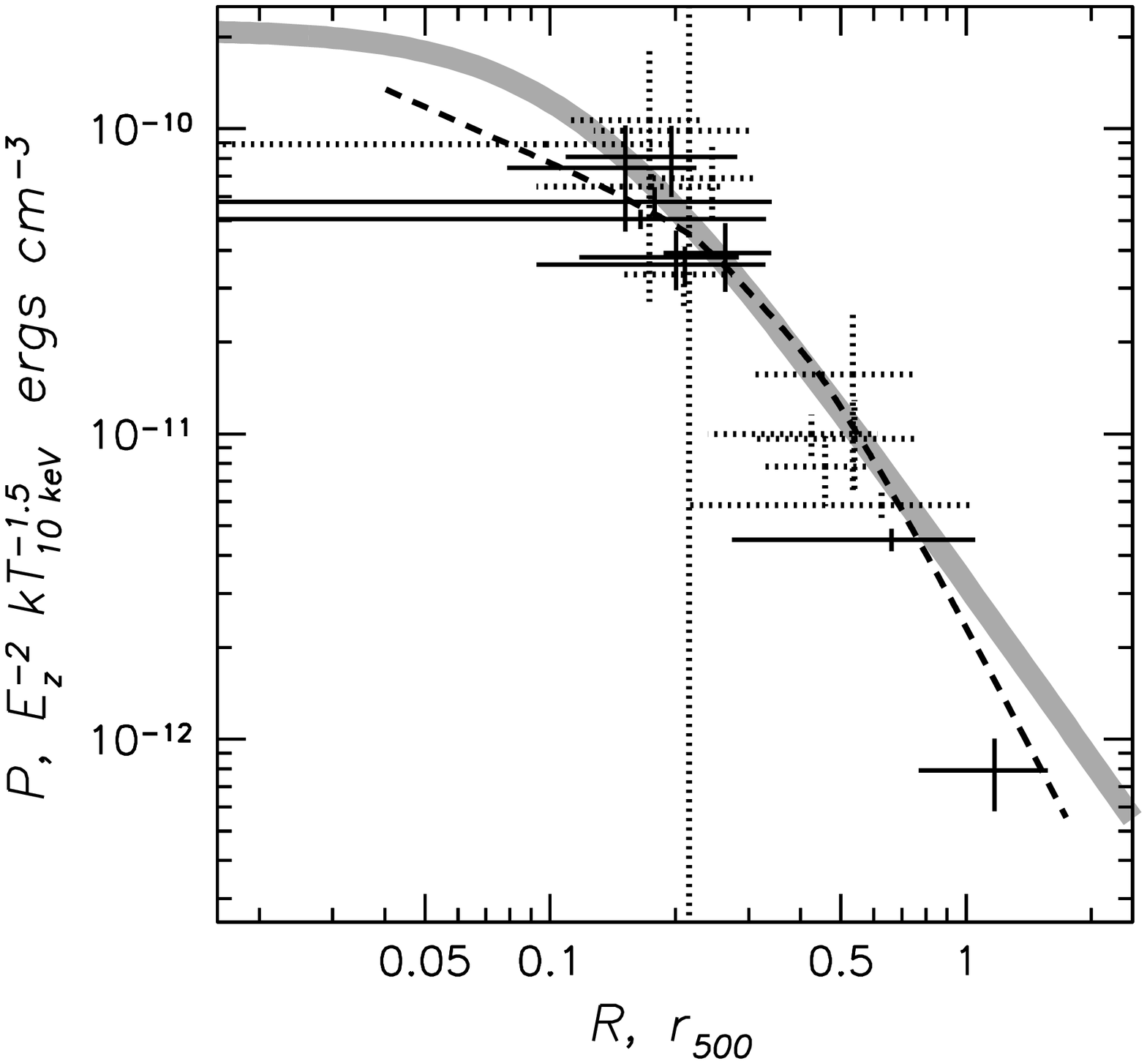}\hfill\includegraphics[width=6cm]{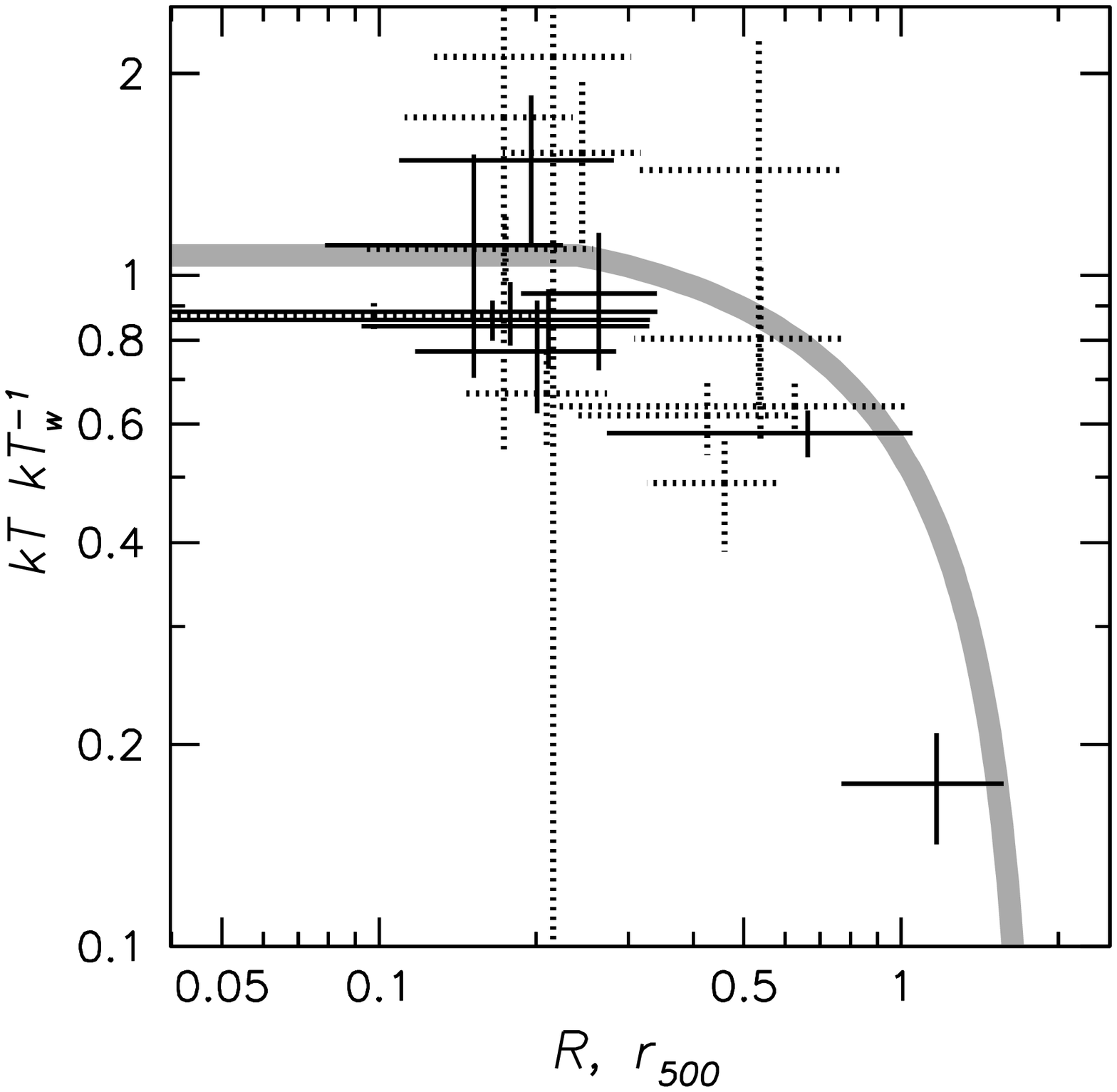}

\caption{RXCJ0532.9--3701. Figure caption is similar to that of
Fig.\ref{f:cl01}.  The surface brightness image is overlaid with contours
showing the spectral extraction regions with numbers corresponding to those
in Table \ref{t:cl09:t}.}
\label{f:cl09}%

\end{figure*}

\begin{table*}[ht]
{
\begin{center}
\footnotesize
{\renewcommand{\arraystretch}{0.9}\renewcommand{\tabcolsep}{0.09cm}
\caption{\footnotesize
Properties of main regions of RXCJ0532.9--3701.
\label{t:cl09:t}}

\begin{tabular}{cccccccccccc}
 \hline
 \hline
N  &$kT$ &  $\rho_e$ & S & P, $10^{-12}$ & $M_{\rm gas}$ & $r_{\rm min}$ &
$r_{\rm max}$& Remarks\\
  & keV     &  $10^{-4}$ cm$^{-3}$& keV cm$^2$ & ergs cm$^{-3}$ & $10^{12} M_\odot$  & Mpc & Mpc  & \\
\hline
 1&$ 6.0\pm1.1$&$ 36.0\pm 3.9$&$ 256\pm  52$&$ 34.6\pm 7.6$&$ 1.1\pm0.1$&0.13&0.31& \\
 2&$11.6\pm2.9$&$ 39.7\pm 3.1$&$ 461\pm 118$&$ 73.6\pm19.3$&$ 2.0\pm0.2$&0.12&0.31& \\
 3&$ 8.6\pm3.2$&$ 48.9\pm 5.3$&$ 300\pm 112$&$ 67.7\pm25.8$&$ 1.4\pm0.1$&0.09&0.24& \\
 4&$ 6.5\pm0.9$&$ 31.1\pm 1.8$&$ 307\pm  43$&$ 32.6\pm 4.8$&$ 2.4\pm0.1$&0.10&0.36& \\
 5&$ 6.7\pm0.5$&$ 43.0\pm 1.2$&$ 253\pm  18$&$ 46.1\pm 3.4$&$ 5.9\pm0.2$&0.00&0.36& core\\
 6&$ 6.9\pm0.7$&$ 47.6\pm 2.0$&$ 243\pm  27$&$ 52.5\pm 6.1$&$ 2.9\pm0.1$&0.02&0.37& \\
 7&$ 7.3\pm1.7$&$ 30.4\pm 3.0$&$ 349\pm  84$&$ 35.6\pm 9.0$&$ 2.3\pm0.2$&0.20&0.37& \\
 8&$ 1.4\pm0.3$&$  3.3\pm 0.6$&$ 286\pm  65$&$  0.7\pm 0.2$&$17.0\pm3.2$&0.84&1.71& filament\\
 9&$ 4.5\pm0.4$&$  5.6\pm 0.2$&$ 665\pm  55$&$  4.1\pm 0.4$&$57.0\pm1.7$&0.30&1.15& \\
\hline
\end{tabular}
}
\end{center}
}
%\vspace*{0.2cm}
\end{table*}

\subsection{RXCJ0532.9--3701 } 

This cluster shows one of the most symmetrical images in the sample. It has
a low-entropy core, but not low enough for a cool core. The pressure map is
also quite symmetric. In the intensity we see a slightly boxy structure and
associated with it a clover leaf structure in the entropy. On large scales
the temperature is anti-correlated to the density in azimuthal direction,
resulting in a constant pressure.

The spectroscopic analysis is reported in Table \ref{t:cl09:t} and
Fig.\ref{f:cl09}. The cluster exhibits a low-entropy infalling zone to the
east (region 8). Its entropy is $286\pm 65$ keV cm$^2$, which is lower
compared to the $665\pm 55$ keV cm$^2$ value for the preceding zone (region
9). However, the low-entropy zone exhibits a pressure typical for its
distance to the center. The indicated pressure and entropy fluctuations are
rather marginal, $\sim 2$ sigma. The strong fluctuations in both entropy and
pressure, reported in Tab.\ref{t:scatter} are entirely due to the presence
of the filament.

\begin{figure*}
\includegraphics[width=8cm]{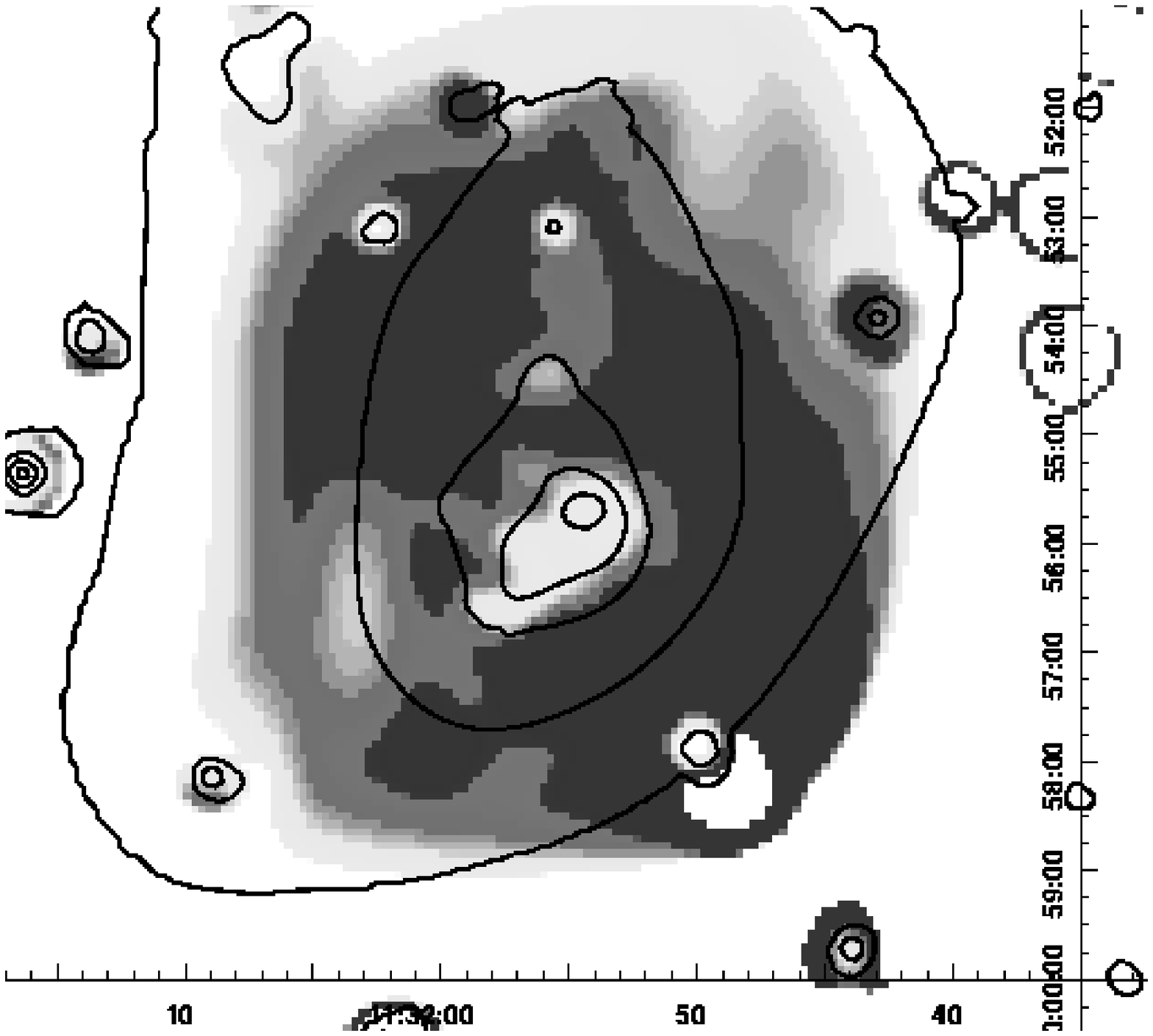}\hfill\includegraphics[width=8cm]{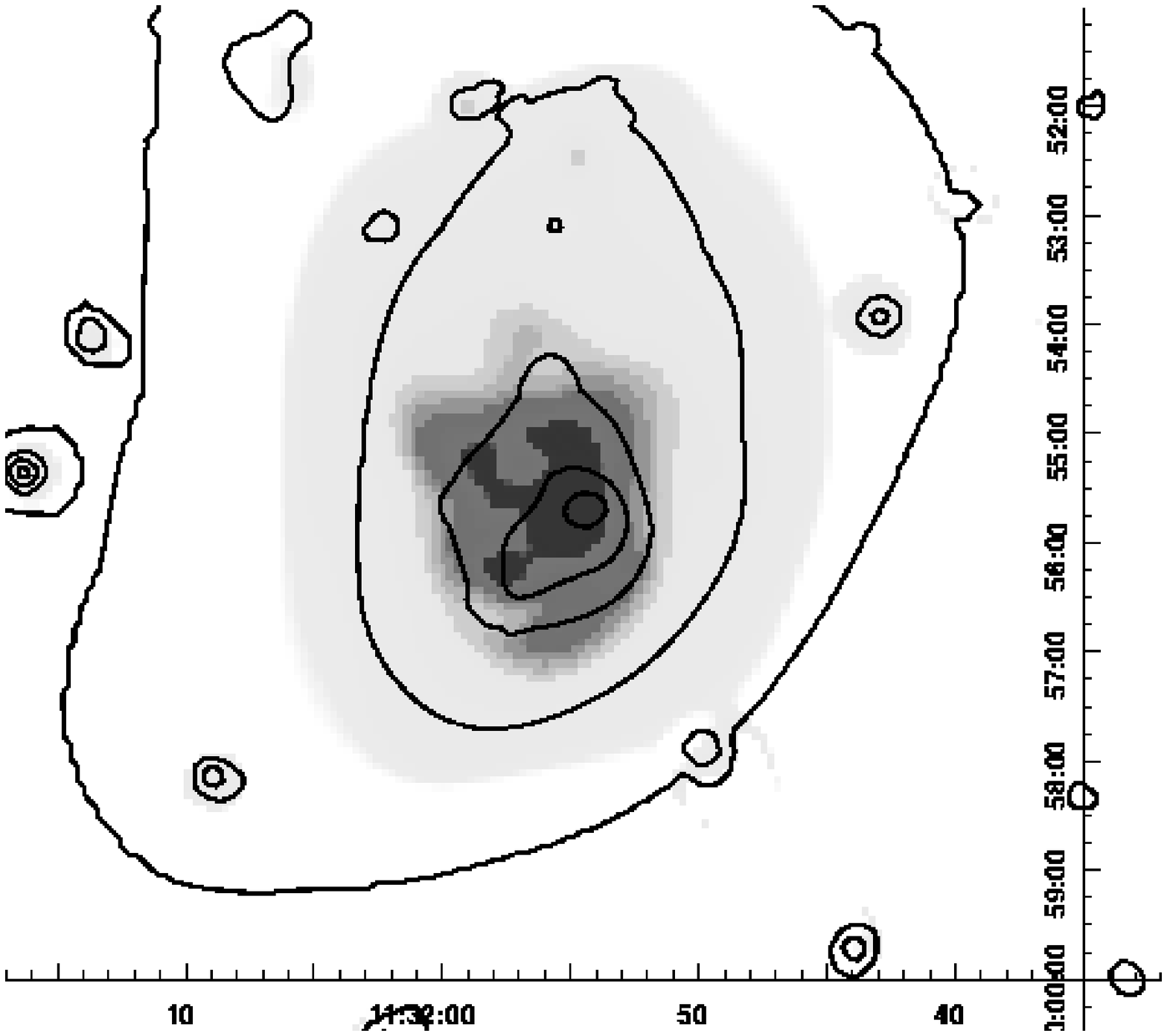}

\includegraphics[width=8cm]{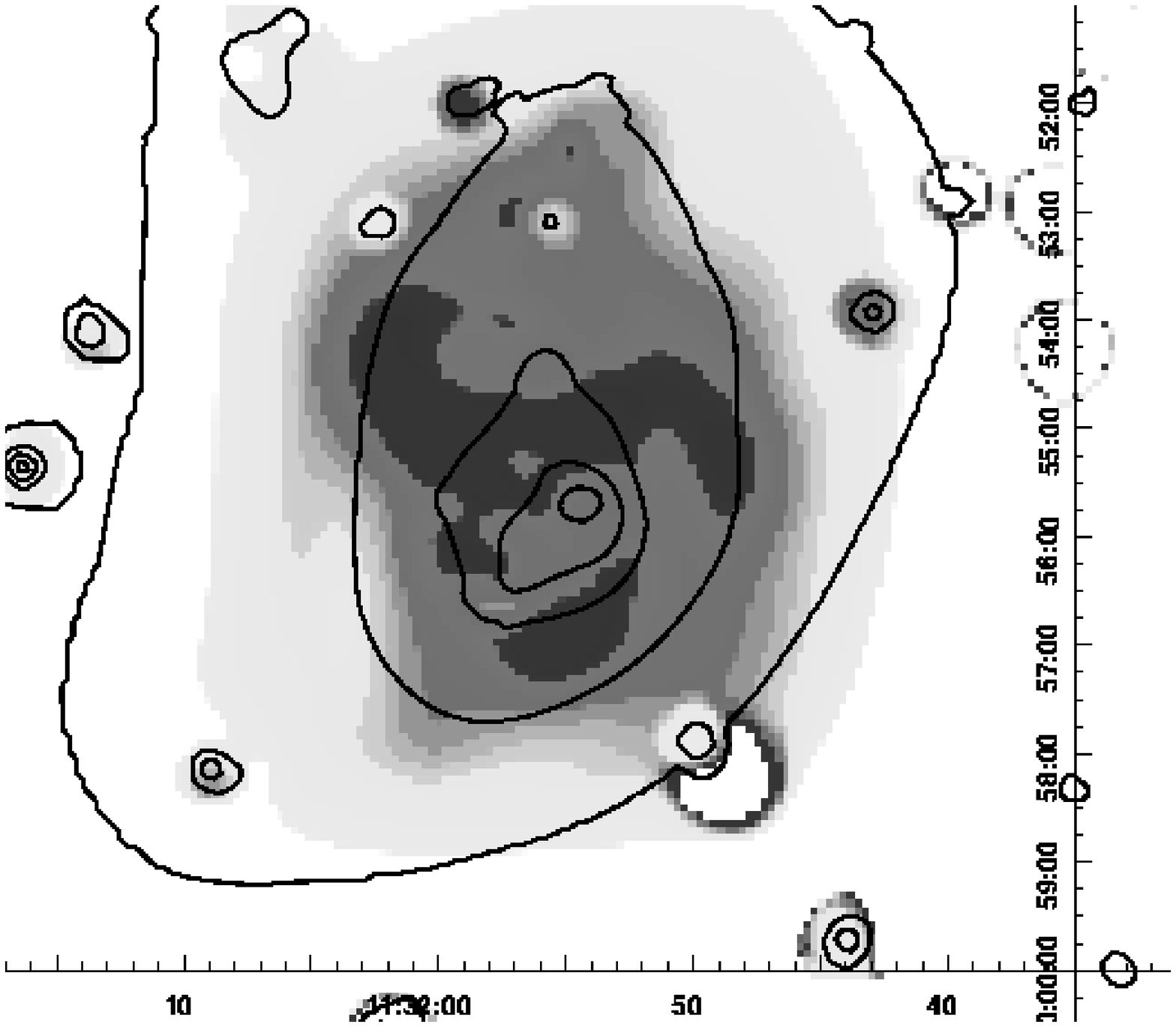}\hfill\includegraphics[width=8cm]{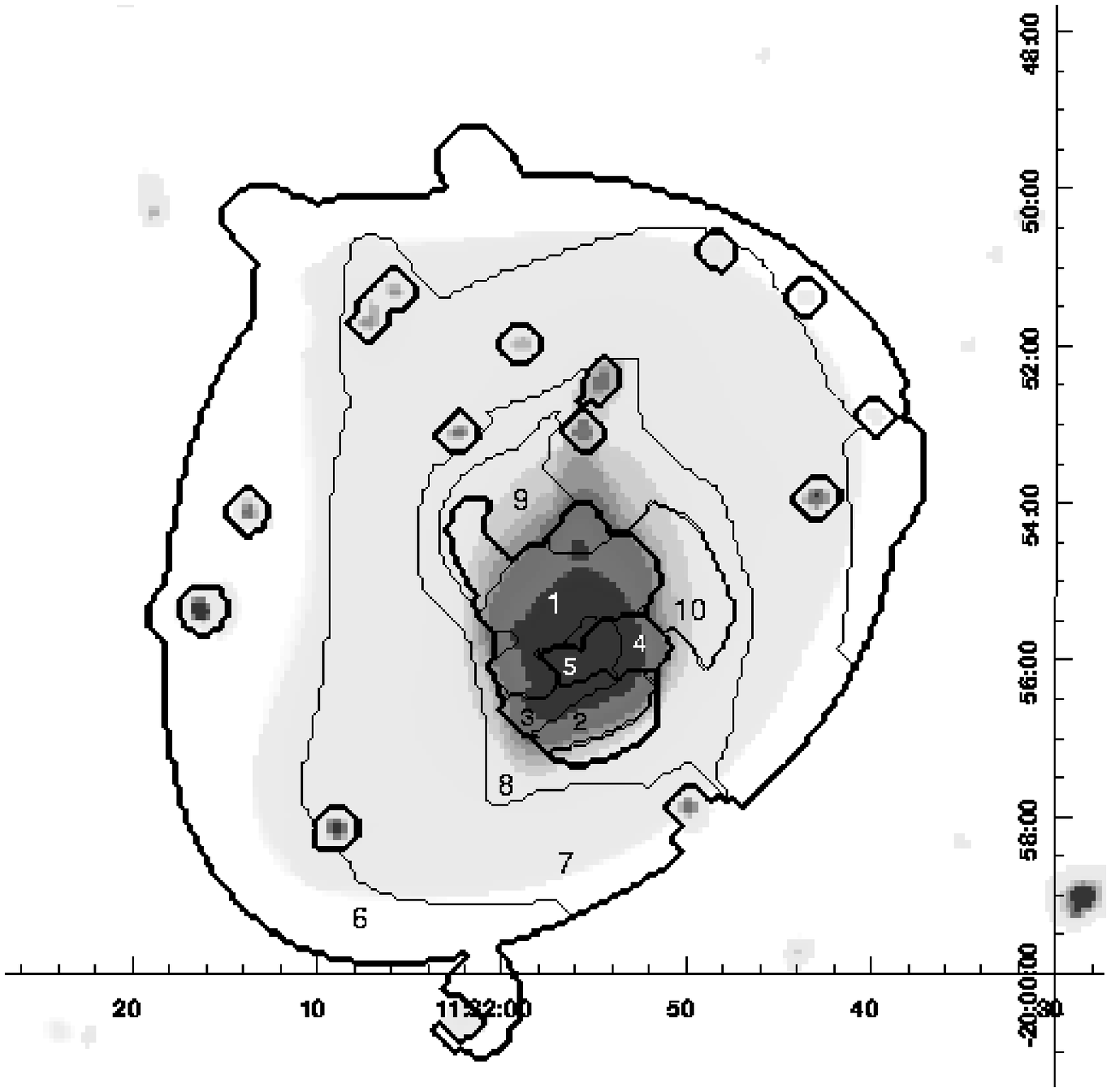}

\includegraphics[width=6cm]{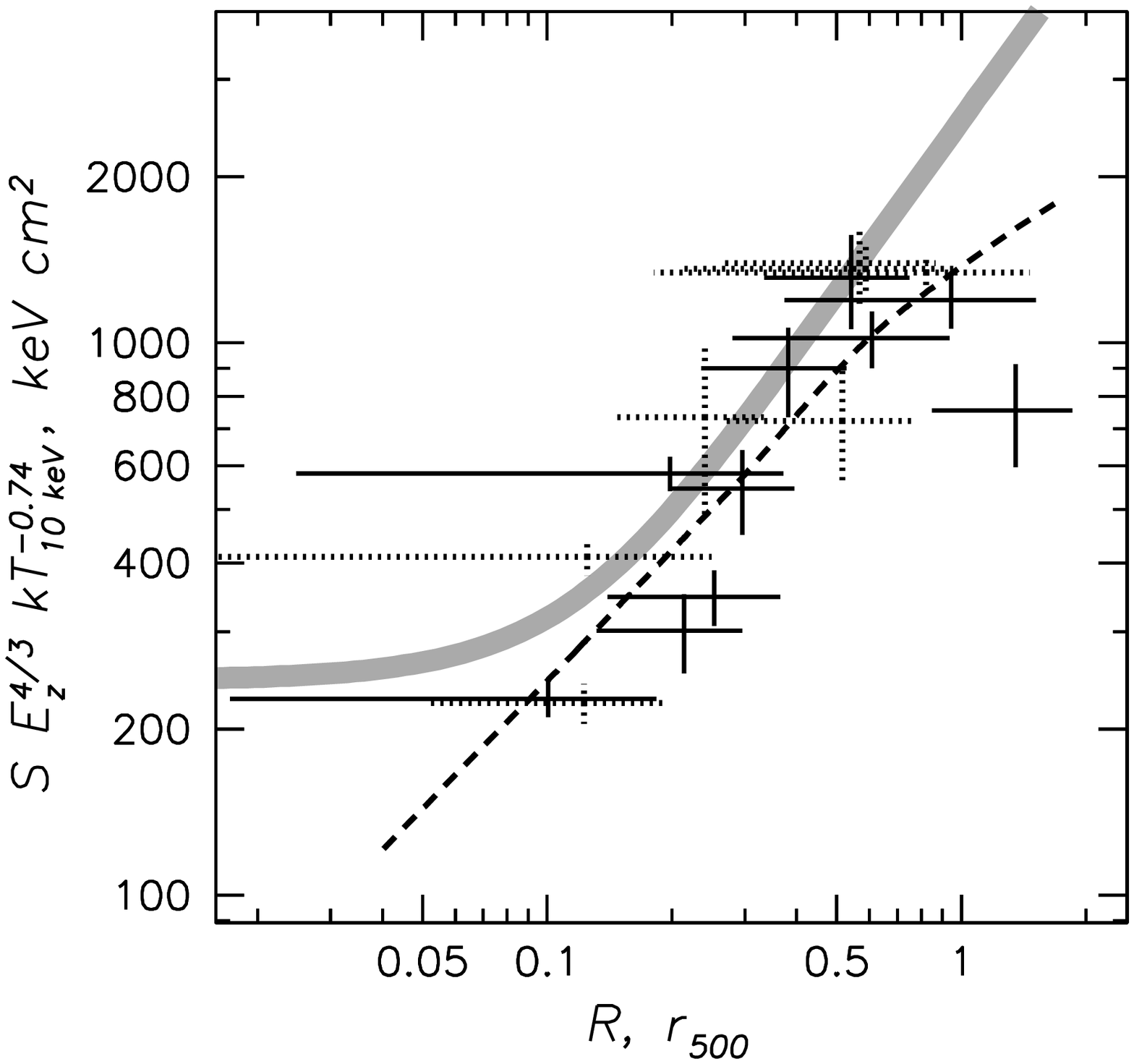}\hfill\includegraphics[width=6cm]{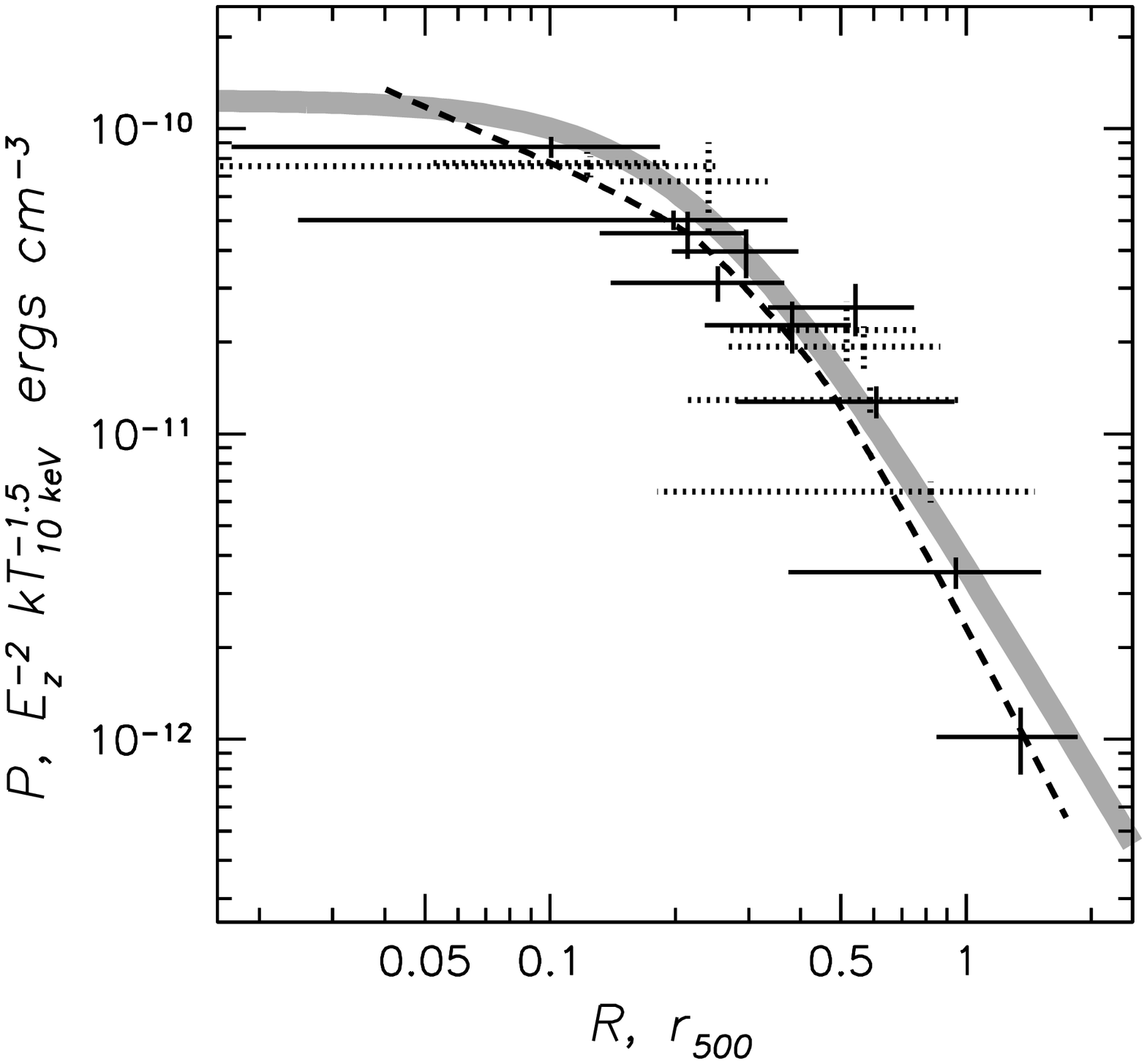}\hfill\includegraphics[width=6cm]{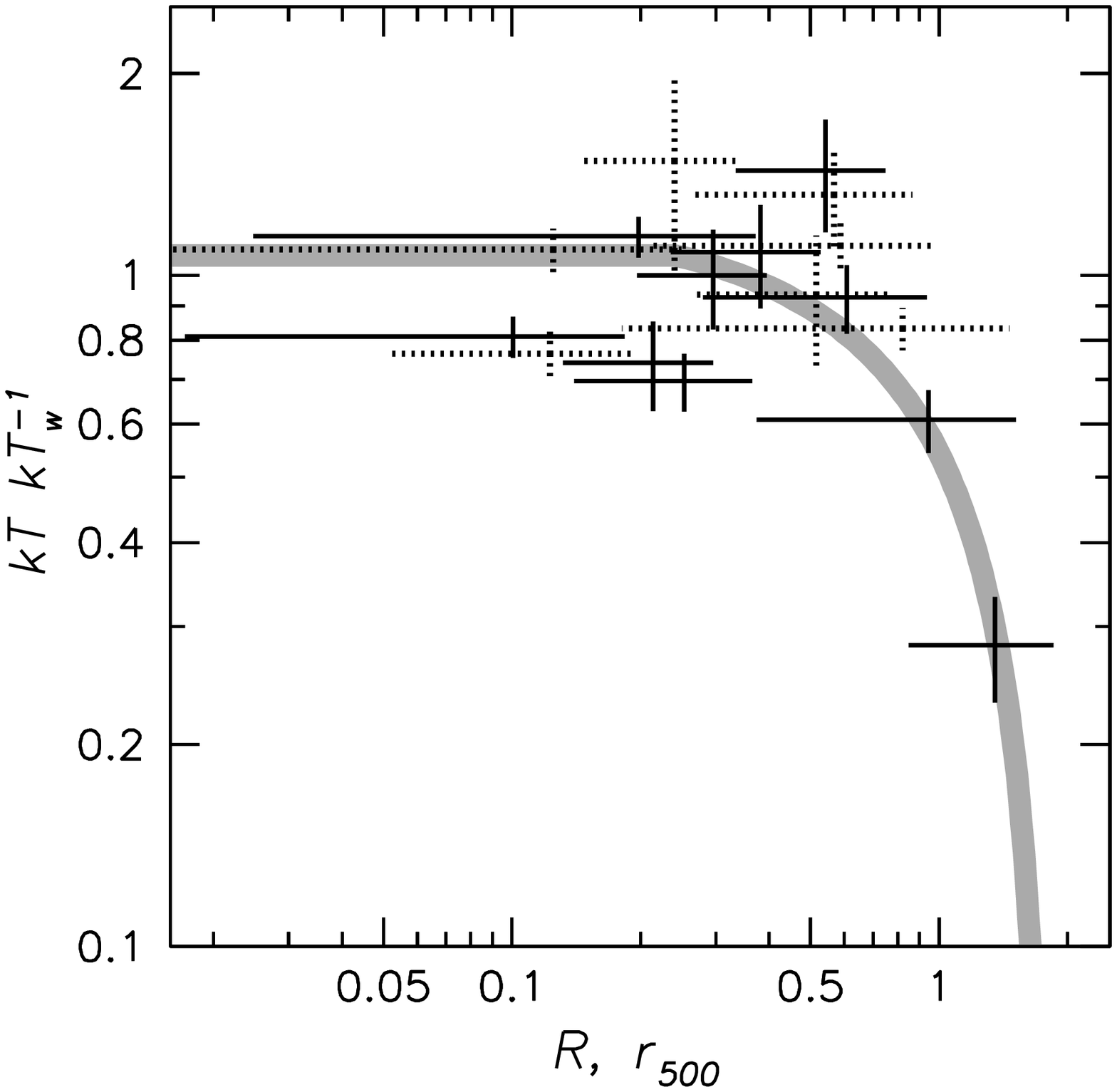}

\caption{RXCJ1131.9--1955. The figure caption is similar to that of
Fig.\ref{f:cl01}.  The surface brightness image is overlaid with contours
showing the spectral extraction regions with numbers corresponding to those
in Table \ref{t:cl10:t}.}
\label{f:cl10}%

\end{figure*}

\begin{table*}[ht]
{
\begin{center}
\footnotesize
{\renewcommand{\arraystretch}{0.9}\renewcommand{\tabcolsep}{0.09cm}
\caption{\footnotesize
Properties of main regions of RXCJ1131.9--1955.}
\label{t:cl10:t}%

\begin{tabular}{cccccccccccc}
 \hline
 \hline
N  &$kT$ &  $\rho_e$ & S & P, $10^{-12}$ & $M_{\rm gas}$ & $r_{\rm min}$ &
$r_{\rm max}$& Remarks\\
  & keV     &  $10^{-4}$ cm$^{-3}$& keV cm$^2$ & ergs cm$^{-3}$ & $10^{12} M_\odot$  & Mpc & Mpc  & \\
\hline
 1&$ 8.5\pm0.6$&$ 32.4\pm 0.8$&$ 387\pm  28$&$ 43.9\pm 3.2$&$ 9.4\pm0.2$&0.03&0.39& \\
 2&$ 7.4\pm1.3$&$ 29.1\pm 1.8$&$ 363\pm  64$&$ 34.5\pm 6.3$&$ 2.3\pm0.1$&0.21&0.41& \\
 3&$ 5.1\pm0.5$&$ 33.2\pm 2.9$&$ 231\pm  27$&$ 27.3\pm 3.6$&$ 1.6\pm0.1$&0.15&0.38& \\
 4&$ 5.5\pm0.8$&$ 45.2\pm 4.0$&$ 201\pm  33$&$ 39.7\pm 7.0$&$ 1.4\pm0.1$&0.14&0.31& \\
 5&$ 6.0\pm0.4$&$ 79.1\pm 2.5$&$ 151\pm  11$&$ 76.0\pm 5.9$&$ 2.3\pm0.1$&0.02&0.19& \\
 6&$ 2.1\pm0.4$&$  2.7\pm 0.5$&$ 503\pm 107$&$  0.9\pm 0.2$&$34.7\pm5.9$&0.89&1.95& \\
 7&$ 4.5\pm0.5$&$  4.3\pm 0.2$&$ 797\pm  89$&$  3.1\pm 0.4$&$58.5\pm2.6$&0.39&1.59& \\
 8&$ 6.9\pm0.8$&$ 10.1\pm 0.3$&$ 680\pm  81$&$ 11.2\pm 1.3$&$20.6\pm0.6$&0.29&0.98& \\
 9&$10.6\pm2.0$&$ 13.3\pm 0.6$&$ 875\pm 169$&$ 22.7\pm 4.4$&$ 4.7\pm0.2$&0.35&0.79& \\
10&$ 8.0\pm1.4$&$ 15.5\pm 1.3$&$ 599\pm 111$&$ 19.9\pm 3.9$&$ 3.5\pm0.3$&0.25&0.55& \\
\hline
\end{tabular}
}
\end{center}
}
%\vspace*{0.2cm}
\end{table*}

\subsection{RXCJ1131.9--1955 (``whirlpool'', A1300)}

The most remarkable feature of this cluster is an unusually elongated
pressure map. We have cast the name ``whirlpool'' cluster for this object
due to the observed propeller-like temperature structure. Three temperature
rims are observed to be filled with colder blobs. In the pressure this makes
some small-scale structure, while on large-scales the structure is seen
mainly in the entropy map. There is a parcel of low entropy gas at the
center of the southern part of the cluster.

On the largest scale, there is an elongation towards the north, practically
in all the maps in the direction where we also see the second largest galaxy
concentration in the optical (B\"ohringer et al. 2002). This is probably due
to the response of the gas pressure to the joint potential well of the two
clusters that make up A1300.  There is a central E-W ridge of high
temperature, that may reflect the compression of the central region due to
the approximately N--S merge of the cluster.

The spectroscopic analysis is reported in Table \ref{t:cl10:t} and
Fig.\ref{f:cl10}. The cluster has a complex temperature structure in the
core, which we assign to the distorted cool core, which partially preserves
both the low entropy and high pressure. Also the temperature decreases
strongly towards cluster outskirts. The large scatter in the entropy
profile is most probably due to the contribution of the substructure,
associated with the second optical component. The outmost bin has an entropy
strongly deviating from the scaling. The analysis confirms the statistical
significance of the low entropy blobs in the south and reveals a pressure
enhancement in the north. Combining these two features together, a plausible
interpretation is that the subcluster has lost its gas on approaching the
cluster from the south and is currently located north to the main cluster.

\begin{figure*}
\includegraphics[width=8cm]{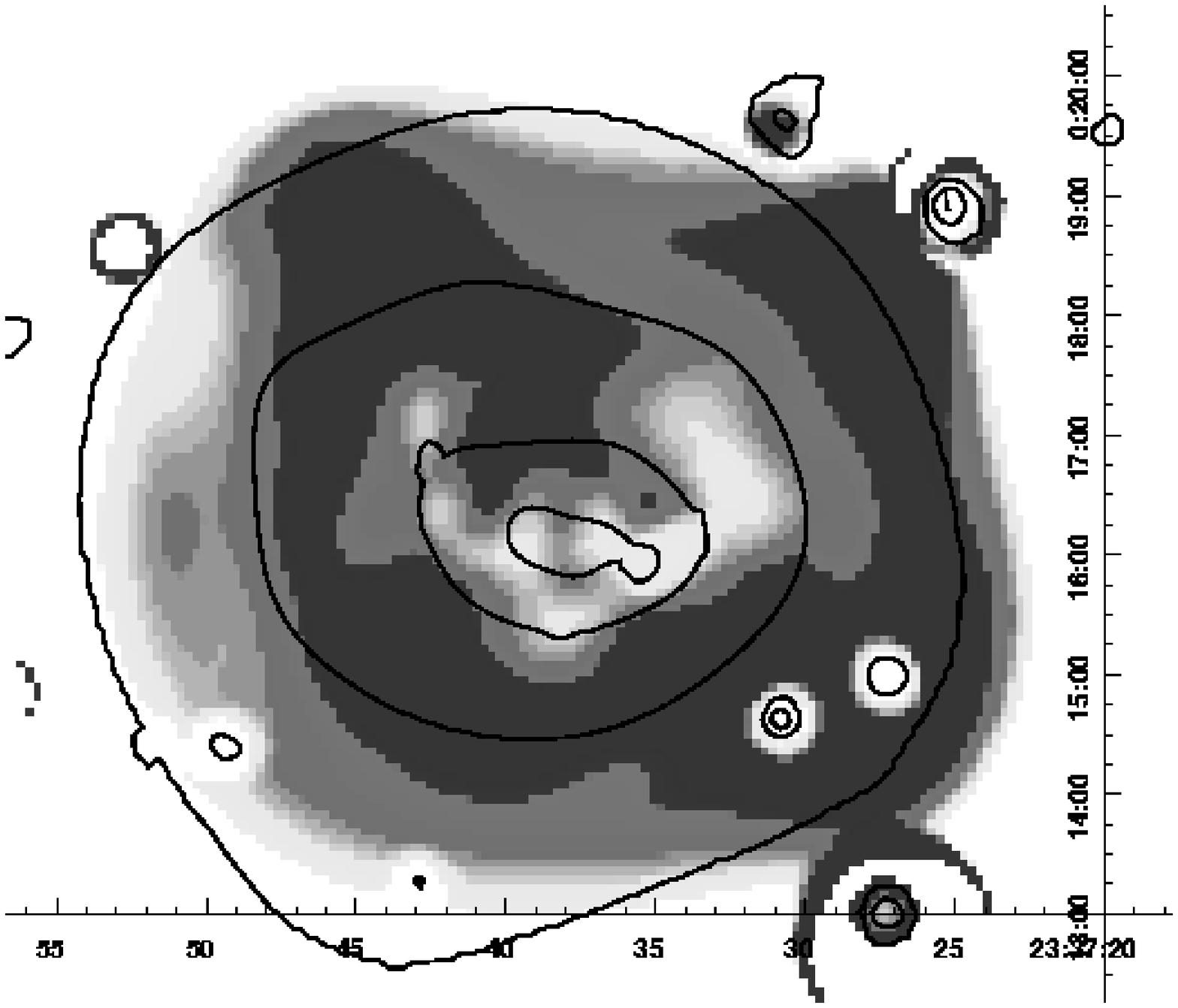}\hfill\includegraphics[width=8cm]{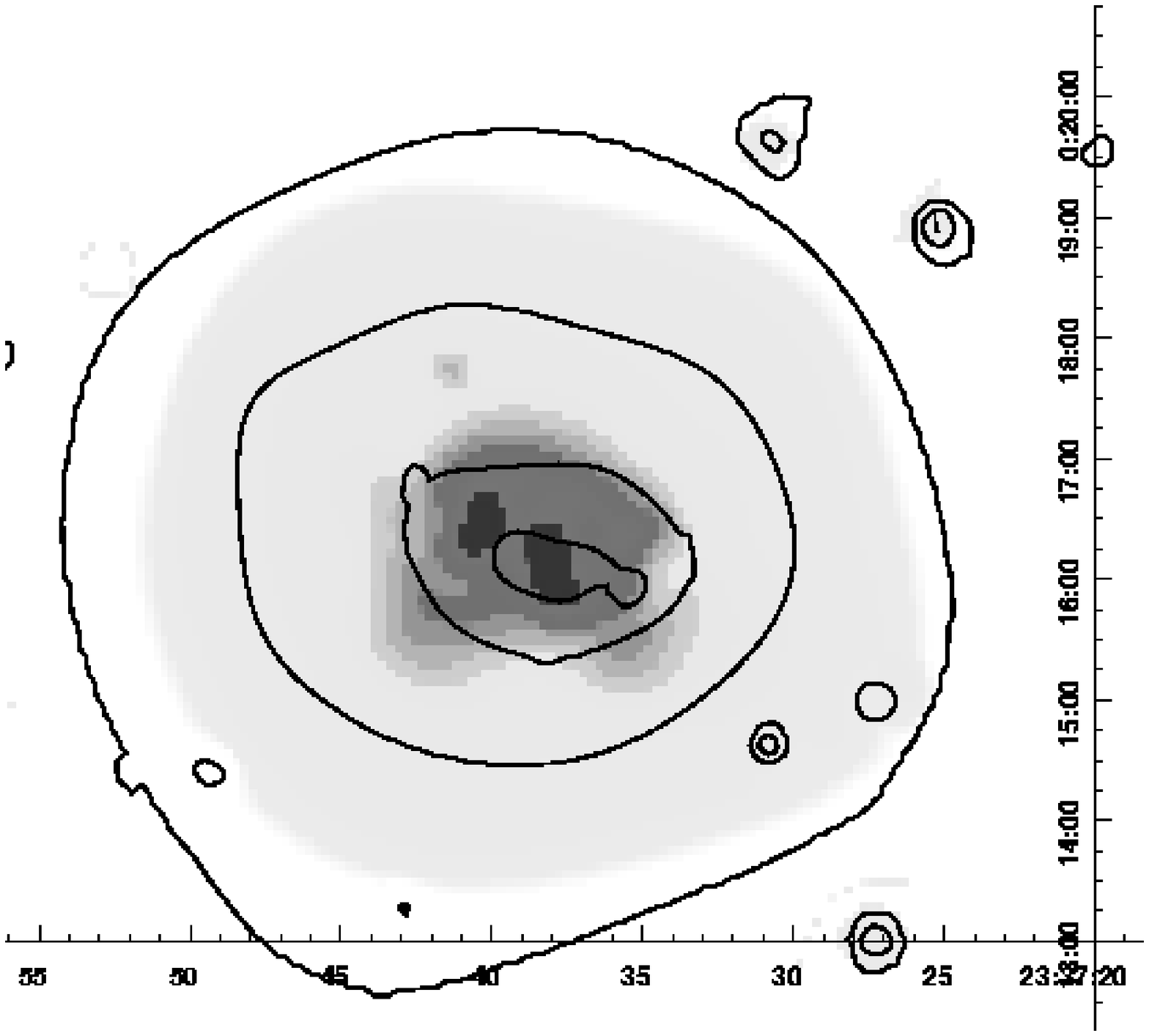}

\includegraphics[width=8cm]{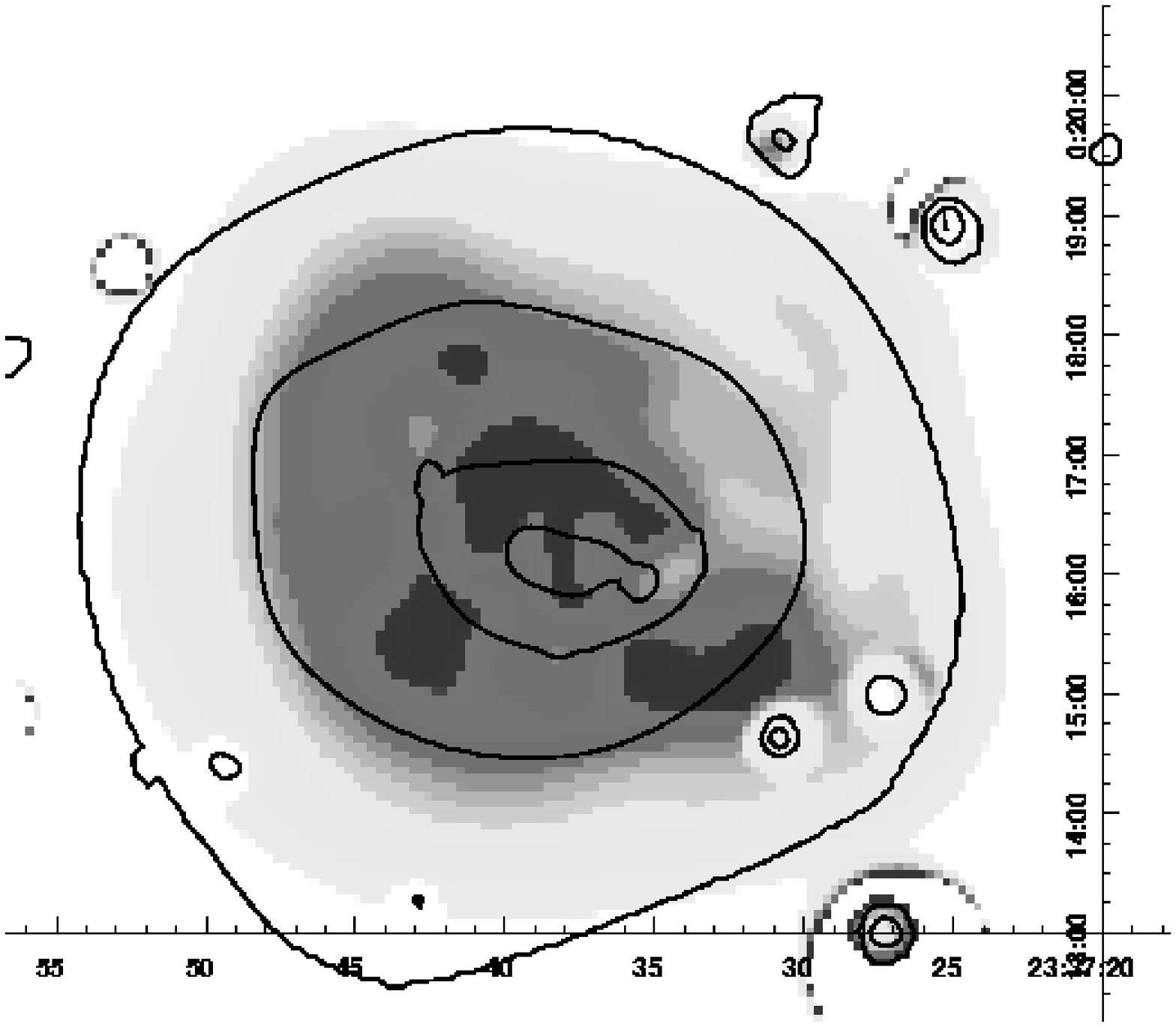}\hfill\includegraphics[width=8cm]{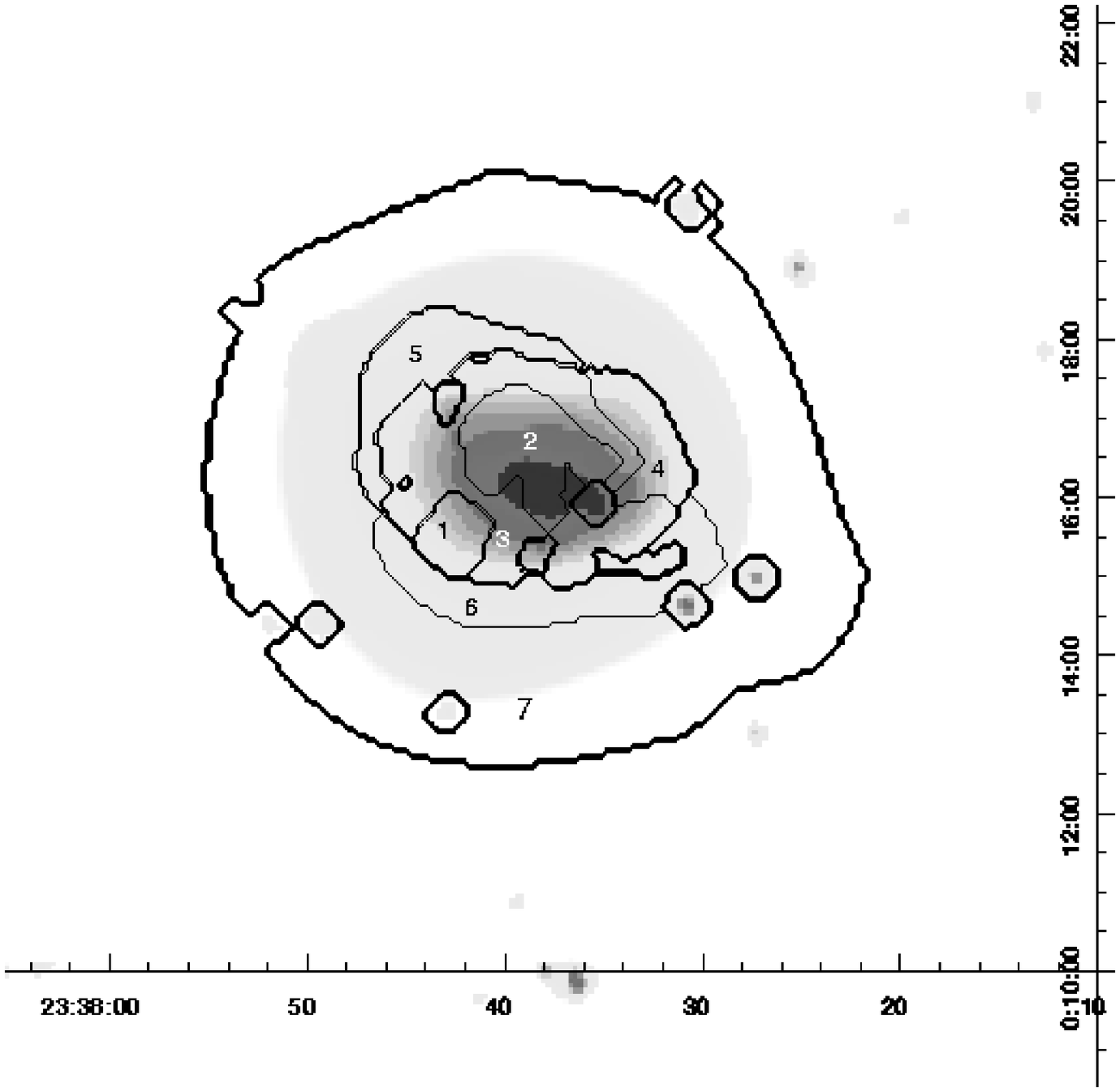}

\includegraphics[width=6cm]{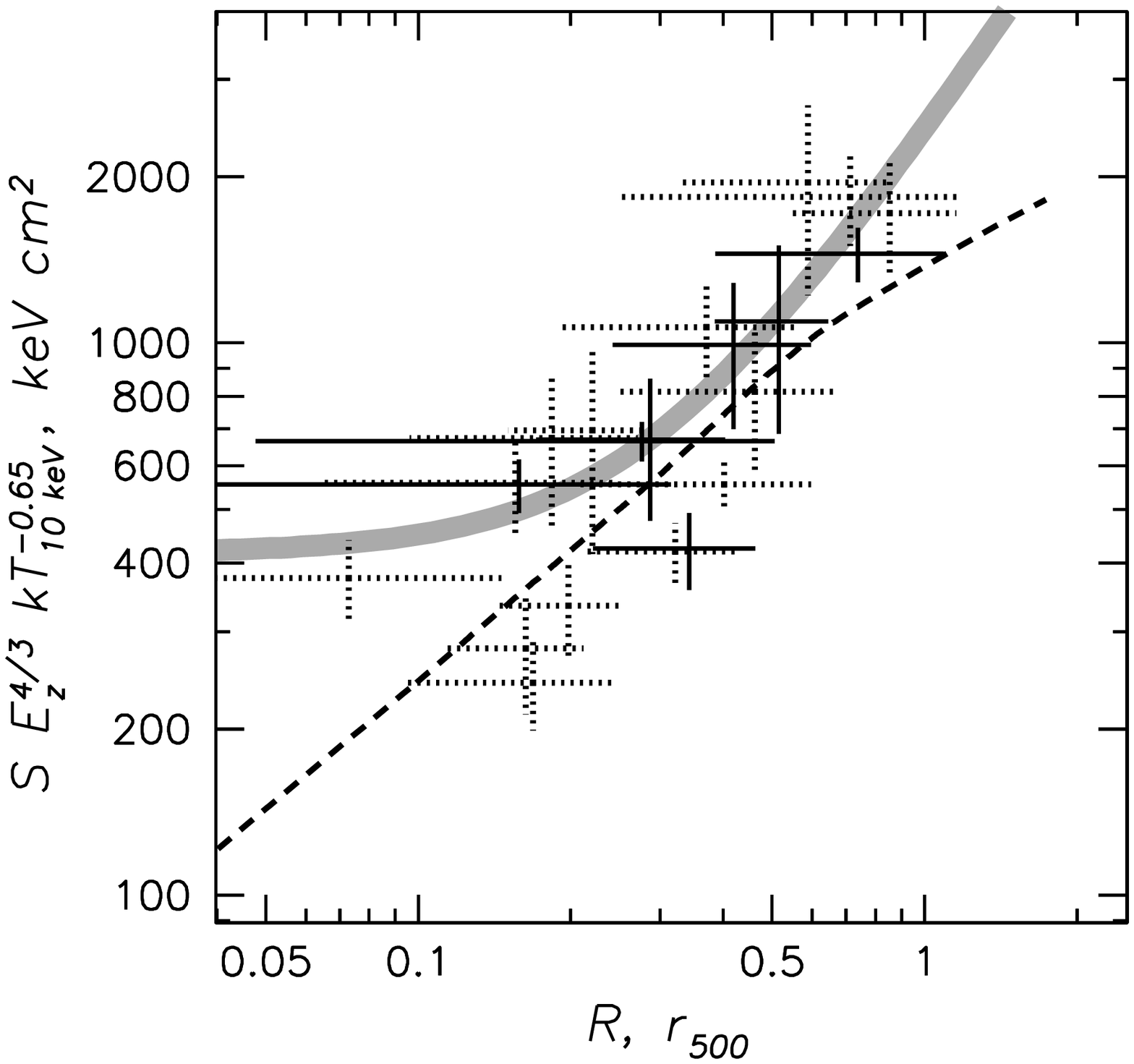}\hfill\includegraphics[width=6cm]{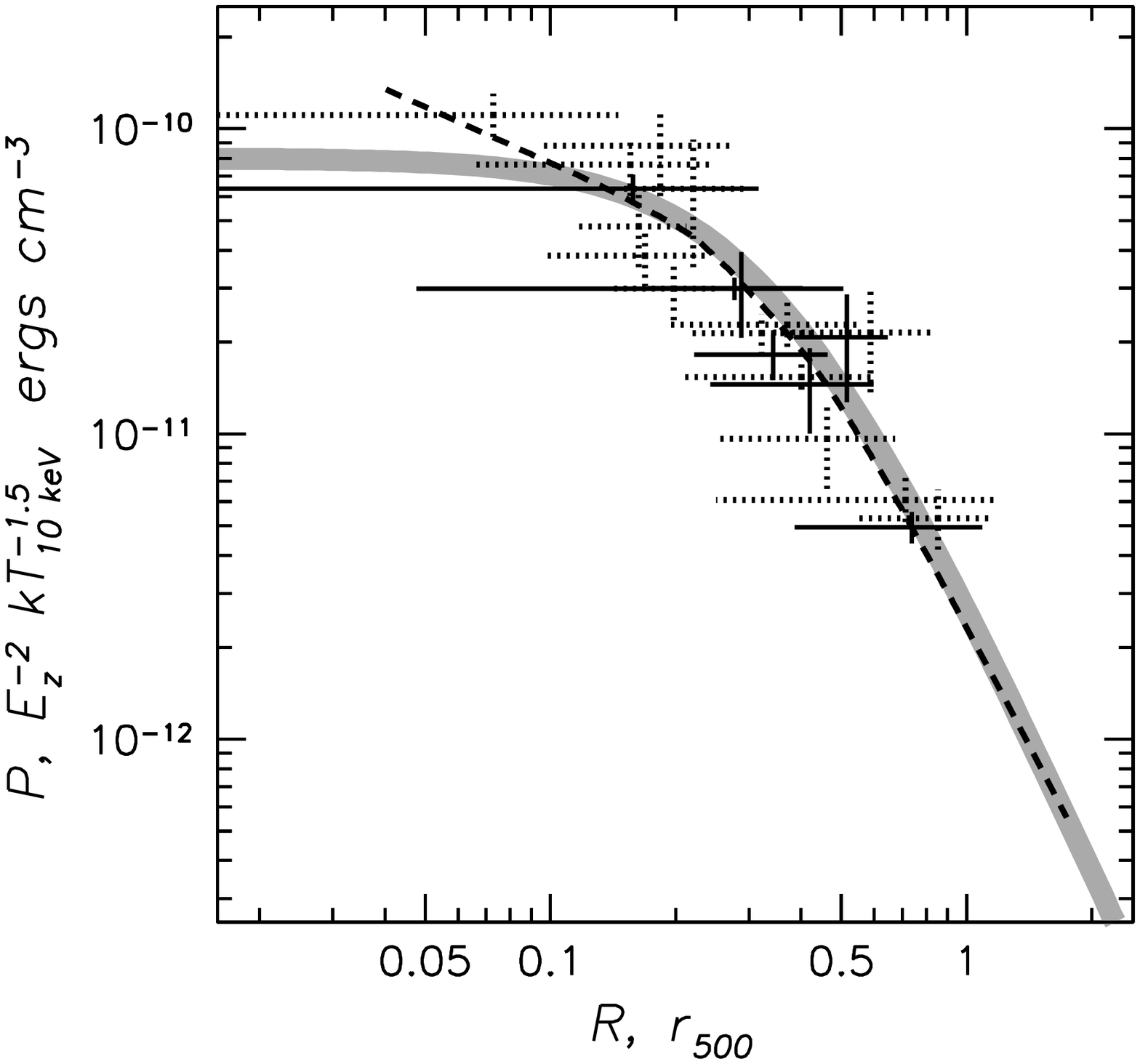}\hfill\includegraphics[width=6cm]{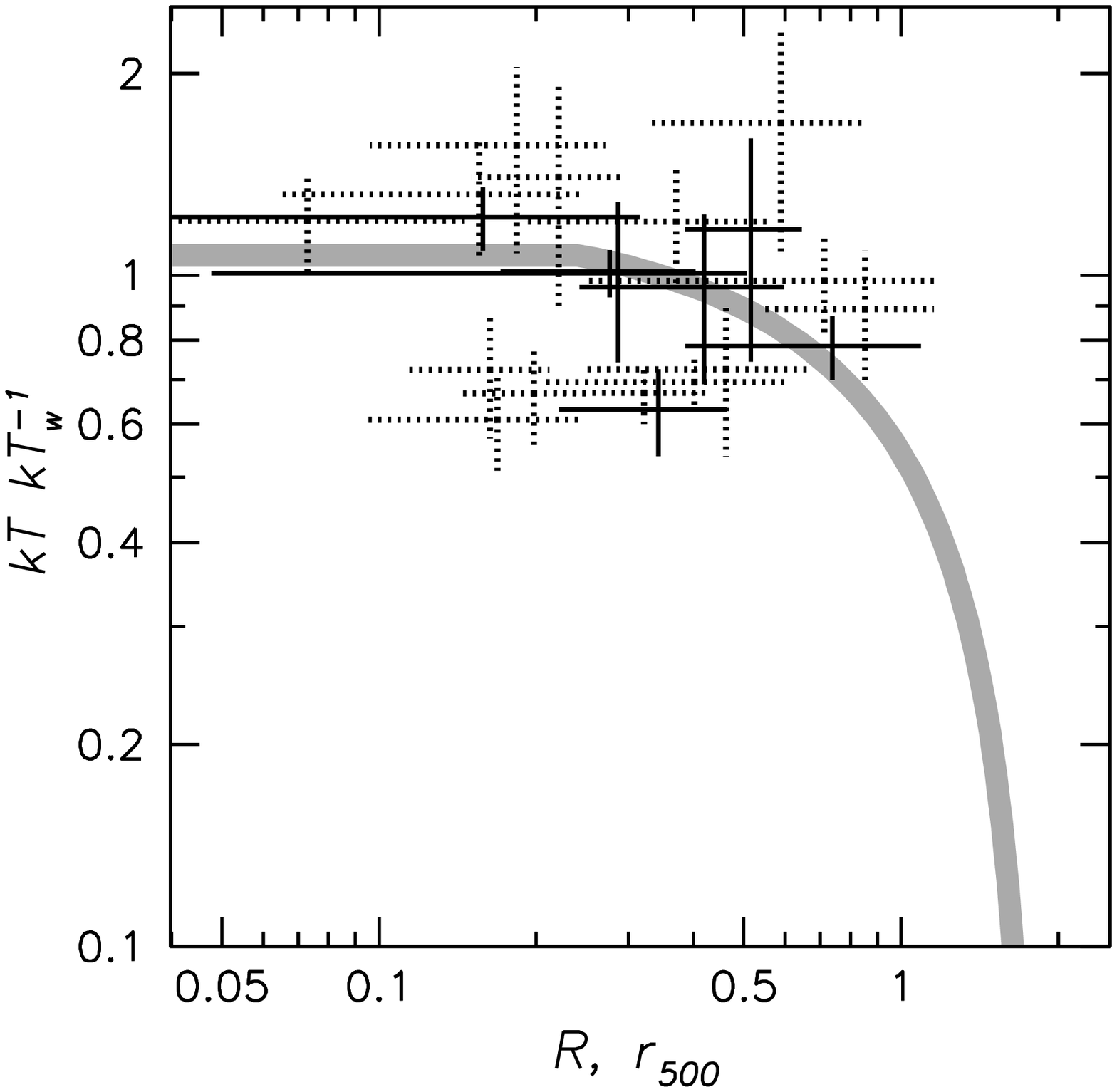}

\caption{RXCJ2337.6+0016.  The figure caption is similar to that of
Fig.\ref{f:cl01}.  The surface brightness image is overlaid with contours
showing the spectral extraction regions with numbers corresponding to those
in Table \ref{t:cl13:t}.}
\label{f:cl13}%

\end{figure*}

\begin{table*}[ht]
{
\begin{center}
\footnotesize
{\renewcommand{\arraystretch}{0.9}\renewcommand{\tabcolsep}{0.09cm}
\caption{\footnotesize
Properties of main regions of RXCJ2337.6+0016.}
\label{t:cl13:t}%

\begin{tabular}{cccccccccccc}
 \hline
 \hline
N  &$kT$ &  $\rho_e$ & S & P, $10^{-12}$ & $M_{\rm gas}$ & $r_{\rm min}$ &
$r_{\rm max}$ & Remarks\\
  & keV     &  $10^{-4}$ cm$^{-3}$& keV cm$^2$ & ergs cm$^{-3}$ & $10^{12} M_\odot$  & Mpc & Mpc   &\\
\hline
 1&$ 7.6\pm2.0$&$ 21.3\pm 3.4$&$ 460\pm 133$&$ 25.9\pm 8.1$&$ 1.5\pm0.2$&0.18&0.43& \\
 2&$ 9.2\pm1.0$&$ 37.3\pm 1.1$&$ 381\pm  42$&$ 54.7\pm 6.2$&$ 6.0\pm0.2$&0.00&0.34& \\
 3&$ 7.6\pm0.6$&$ 21.3\pm 0.5$&$ 456\pm  38$&$ 25.9\pm 2.2$&$13.0\pm0.3$&0.05&0.54& \\
 4&$ 4.7\pm0.7$&$ 20.7\pm 1.9$&$ 292\pm  47$&$ 15.7\pm 2.7$&$
 2.9\pm0.3$&0.24&0.50& disrupted core\\
 5&$ 8.8\pm3.2$&$ 12.7\pm 1.5$&$ 751\pm 281$&$ 17.8\pm 6.9$&$
 3.5\pm0.4$&0.41&0.69& secondary pressure peak?\\
 6&$ 7.2\pm2.0$&$ 10.9\pm 1.4$&$ 681\pm 202$&$ 12.5\pm 3.9$&$ 4.0\pm0.5$&0.26&0.64& \\
 7&$ 5.9\pm0.6$&$  4.5\pm 0.2$&$ 997\pm 113$&$  4.3\pm 0.5$&$41.2\pm1.8$&0.41&1.17& \\
\hline
\end{tabular}
}
\end{center}
}
%\vspace*{0.2cm}
\end{table*}

\subsection{RXCJ2337.6+0016 }

The image shows an East-West elongated core on the small scale, while on the
large scale it appears to be symmetric. The pressure map has two maxima and
two elongations towards the south-west and south-east. There are some
ring-like structures in the temperature map. The entropy state of the
cluster ICM appears to correspond to the late stage of a core disruption
with filaments of the low-entropy gas spread over a large volume. The
position of the entropy minimum is offset from the peak in the pressure.

The cluster has a clover leaf structure in the entropy map like the
0532.9--3701. Two pressure maxima could indicate the core rebounce. The
symmetry in the pressure map is regained at 1.5 arcminute radius.

\begin{figure*}
\includegraphics[width=17cm]{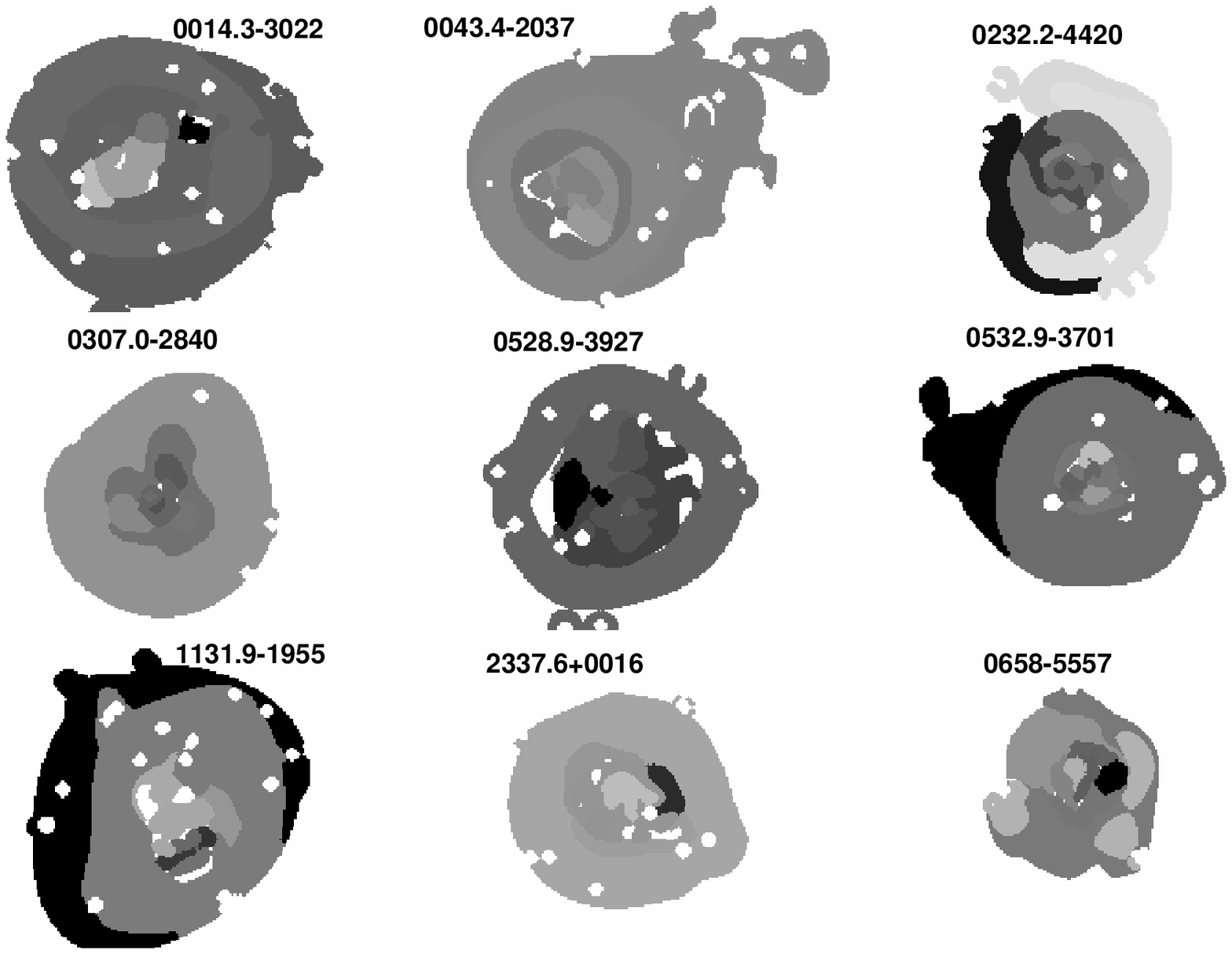} 

\caption{Maps of the ratio between the observed entropy and the average
trends measured in the DXL sample. The different shades of grey stand for a
different value of the ratio, light -- 1.9 (seen only in RXCJ 0232.2-4420
cluster), grey -- 1.1 (a dominant color of both RXCJ 0043.4-2037 and RXCJ
0307.0-2840 clusters), dark grey -- 0.9 (a dominant color of RXCJ
0014.3-3022), black -- 0.4 (e.g. tail of RXCJ 0532.9-3701).}
\label{f:maps}%
\end{figure*}

The spectroscopic analysis is reported in Table \ref{t:cl13:t} and
Fig.\ref{f:cl13}. The scenario of the disrupted core (region 4) is supported
by the spectral analysis (see also Fig.\ref{f:maps}). The pressure
enhancement (region 5) is marginal.

\begin{figure*}
\includegraphics[width=8cm]{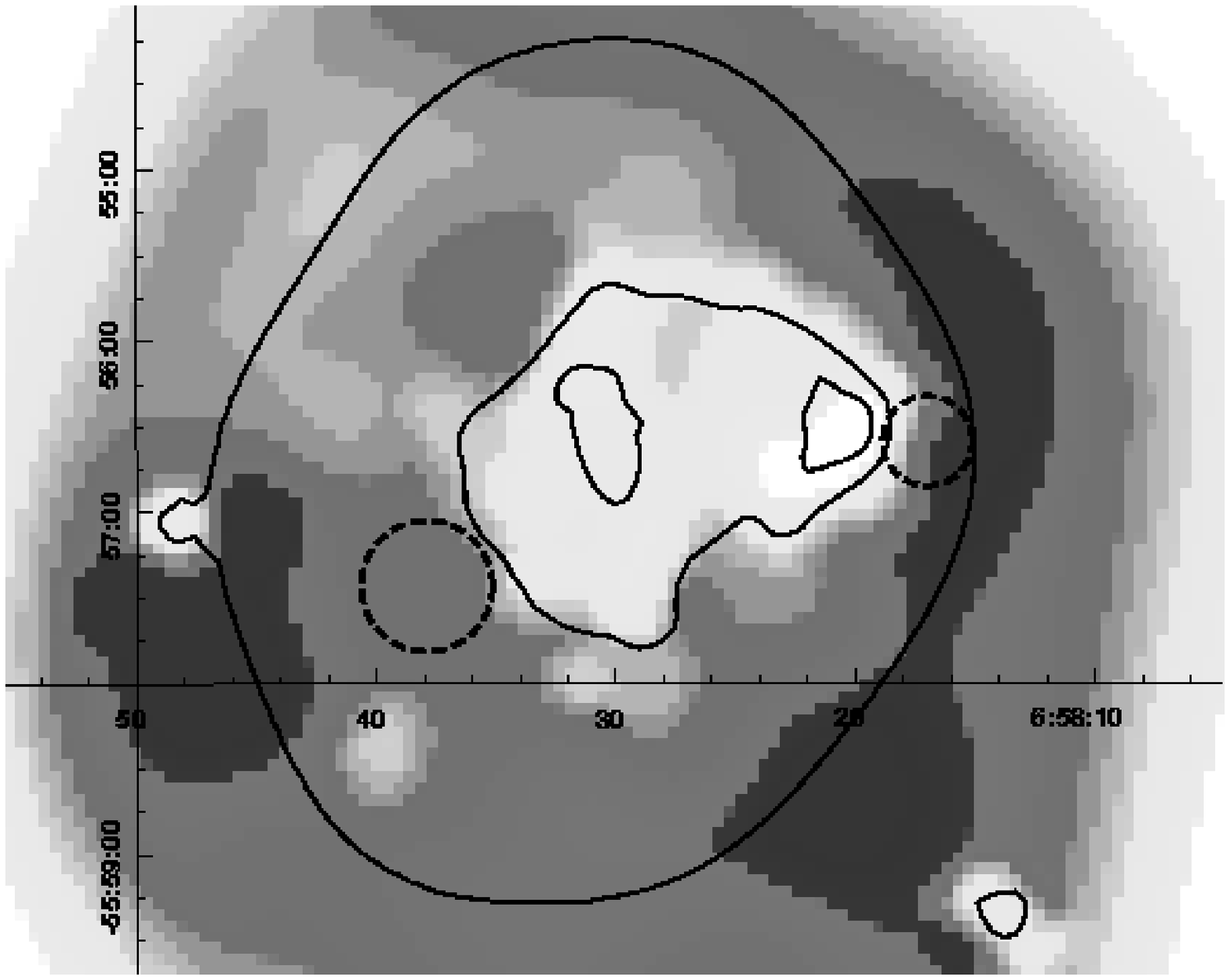}\hfill\includegraphics[width=8cm]{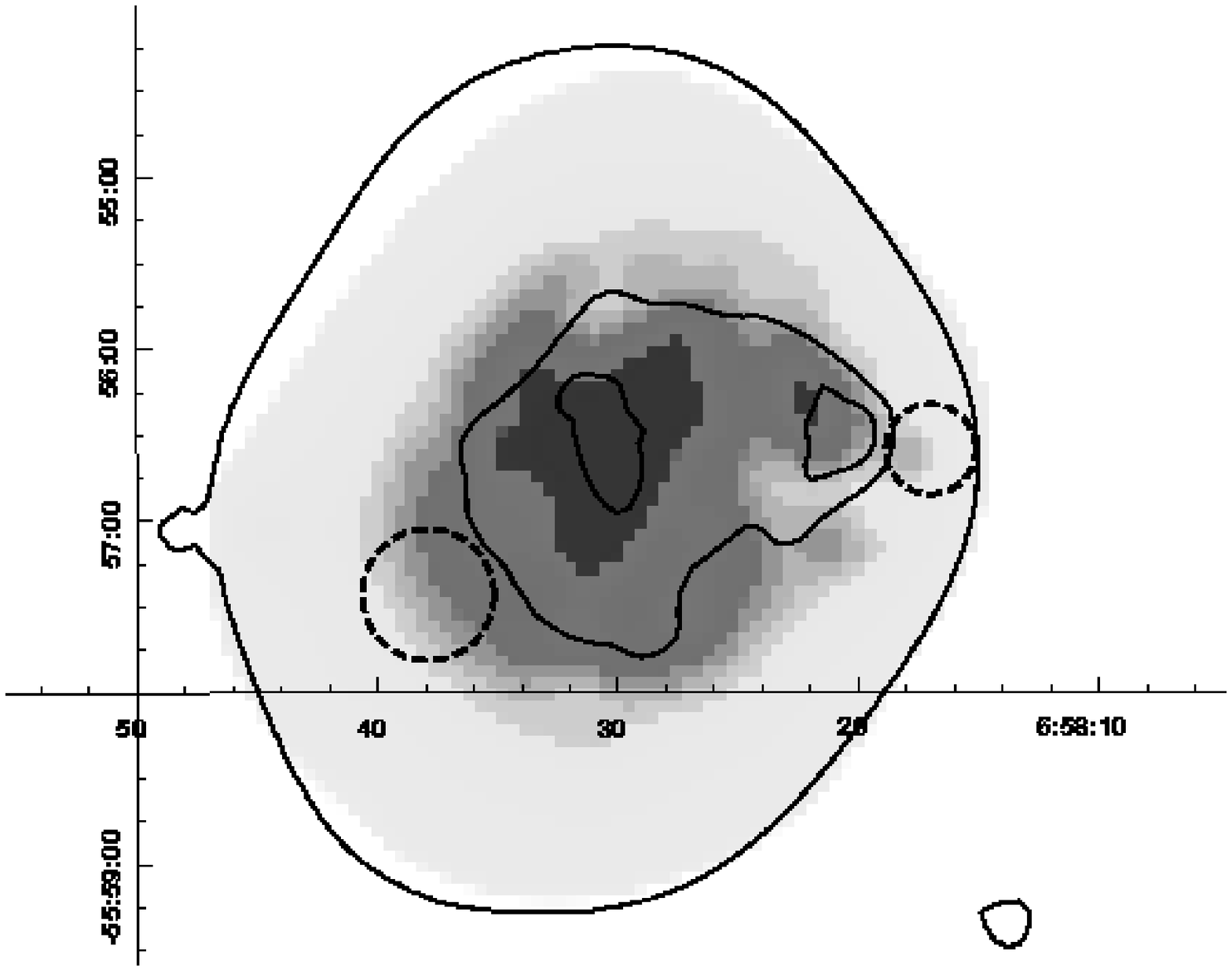}

\includegraphics[width=8cm]{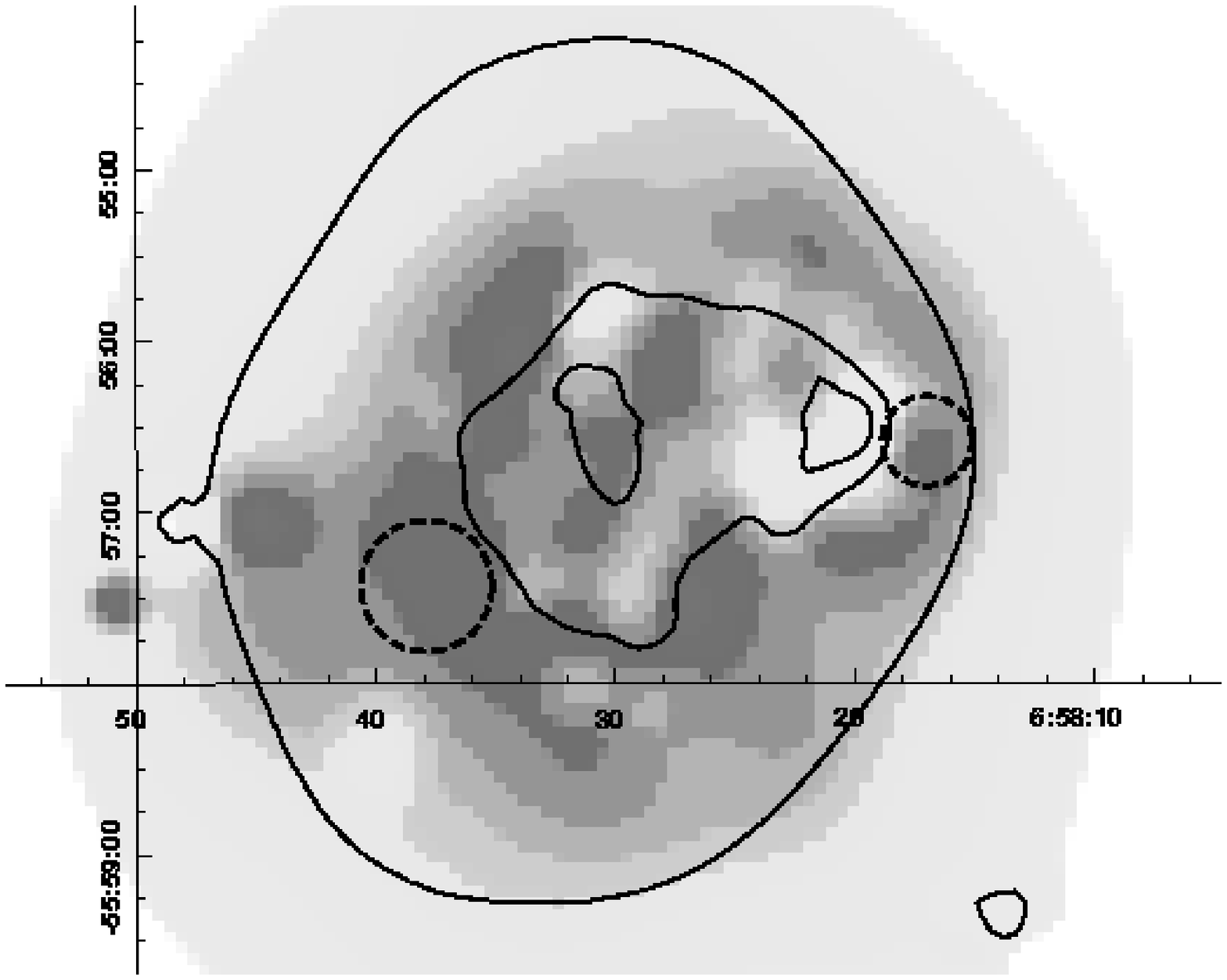}\hfill\includegraphics[width=8cm]{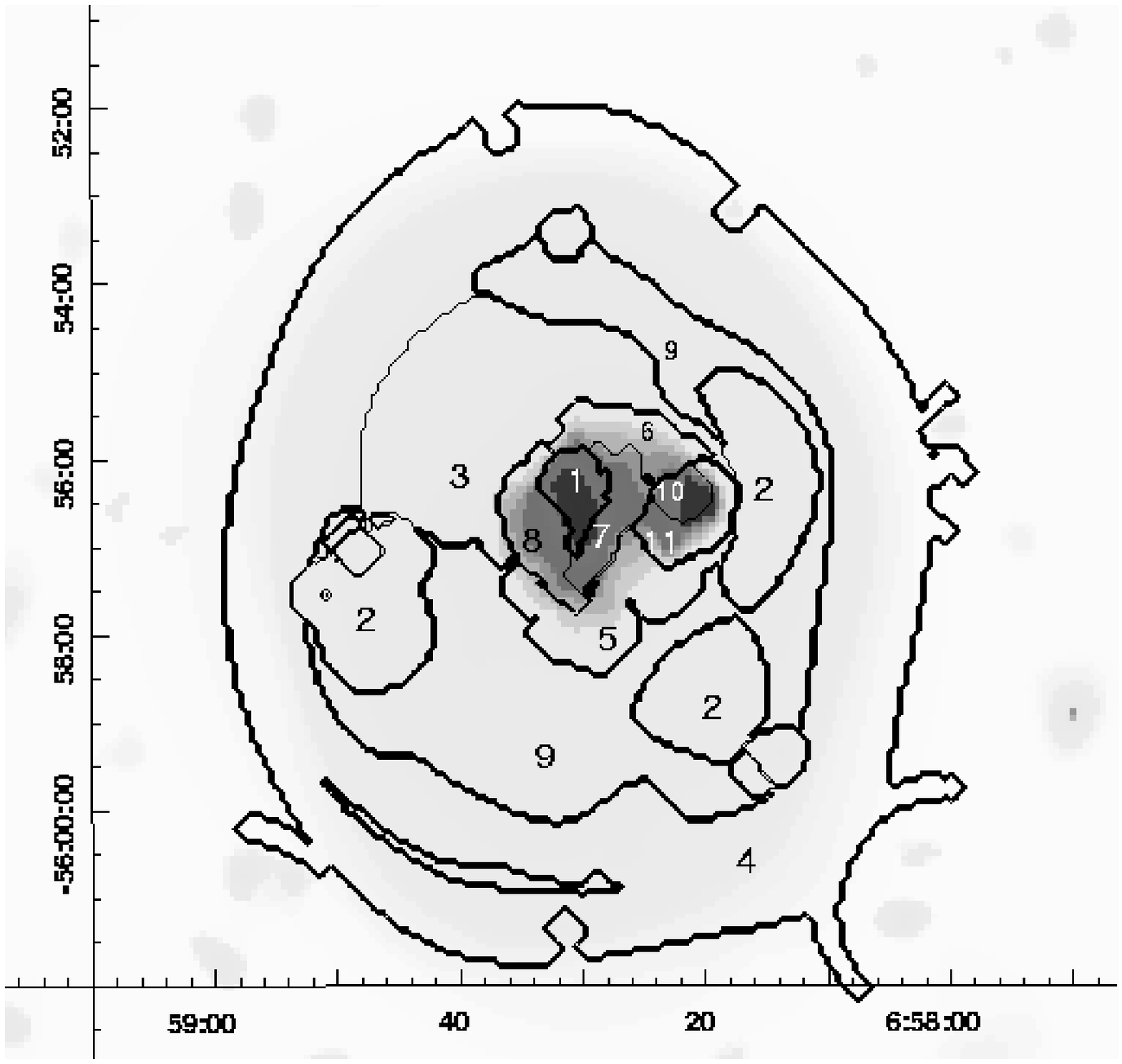}

\includegraphics[width=6cm]{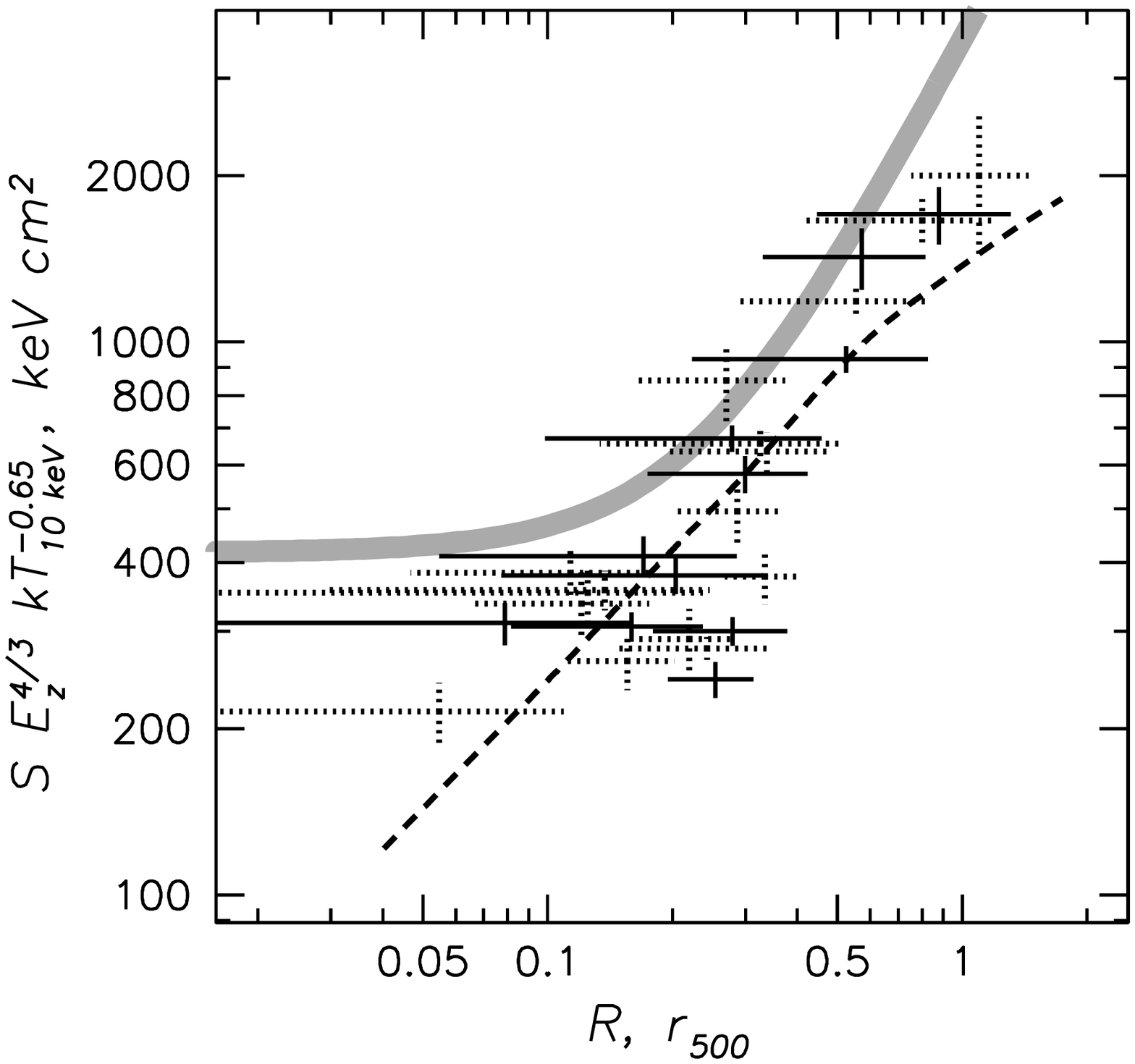}\hfill\includegraphics[width=6cm]{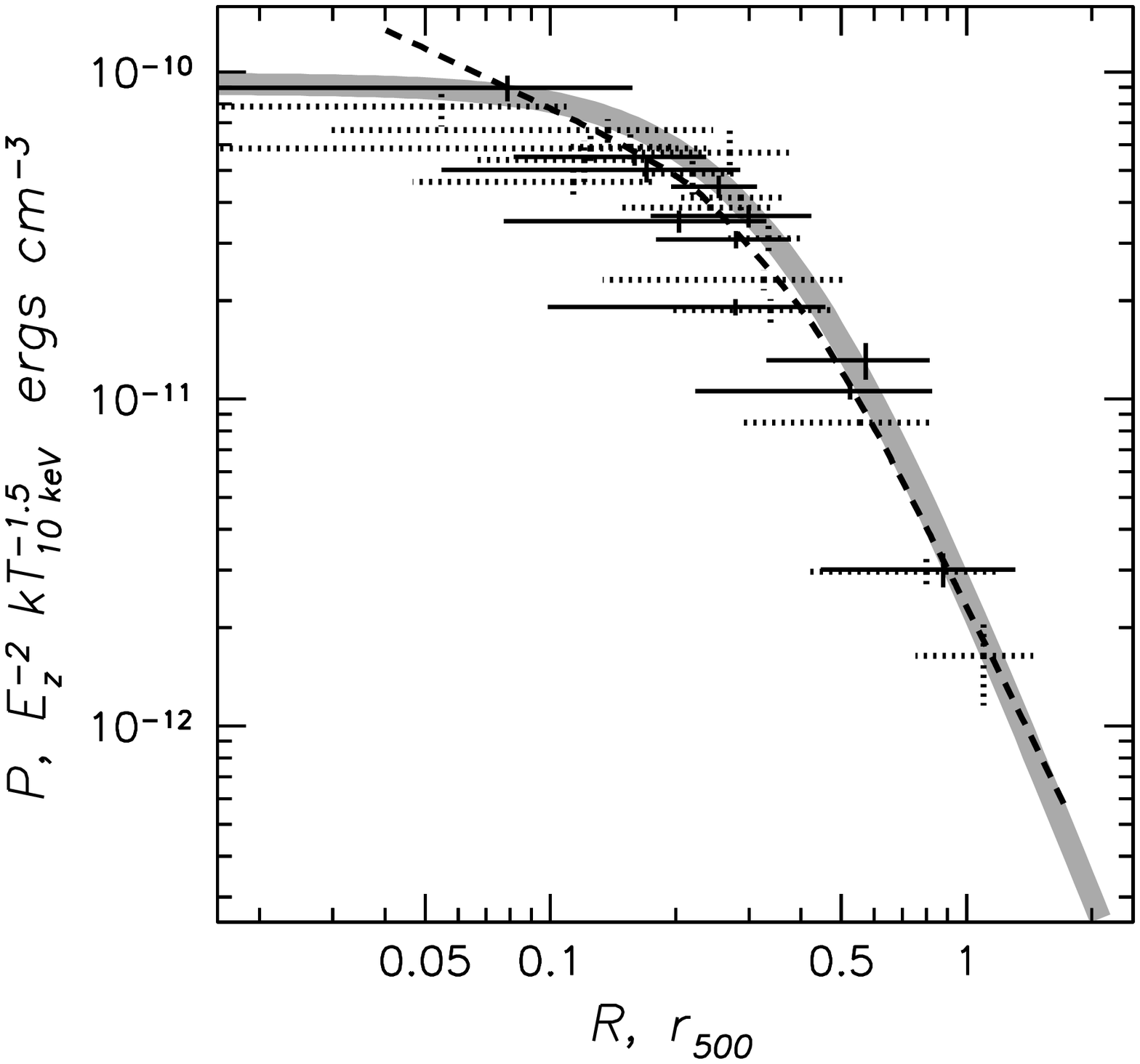}\hfill\includegraphics[width=6cm]{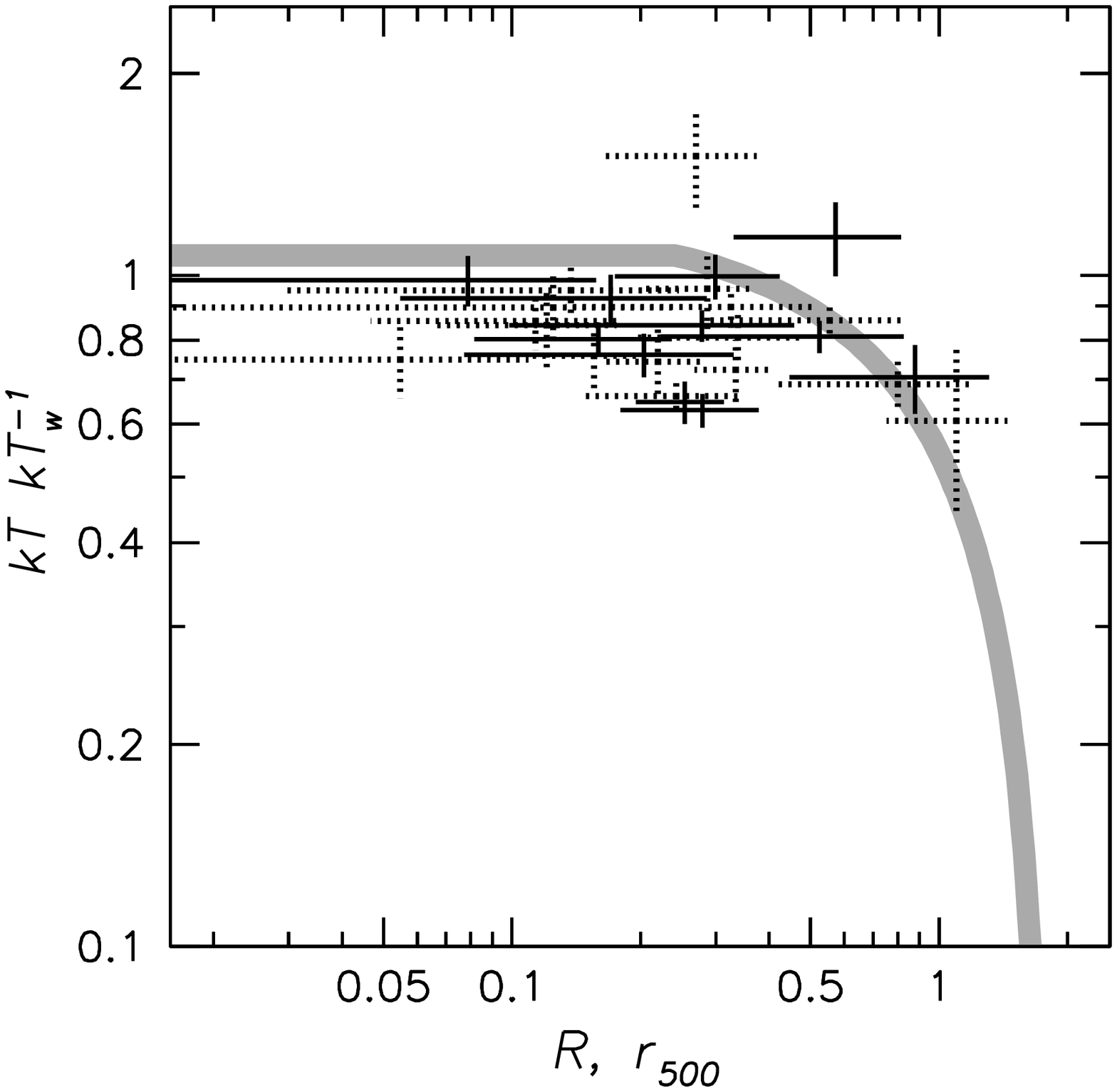}

\caption{RXCJ0658--5557.  Figure caption is similar to that of
Fig.\ref{f:cl01}.  The surface brightness image is overlaid with contours
showing the spectral extraction regions with numbers corresponding to those
in Table \ref{t:cl14:t}. Two dashed line circles indicate the positions of
the two peaks reveals in the weak lensing analysis of Clowe et al. (2004)}
\label{f:cl14}%

\end{figure*}

\subsection{RXCJ0658--5557 (the ``bullet'' cluster)}

\begin{table*}[ht]
{
\begin{center}
\footnotesize
{\renewcommand{\arraystretch}{0.9}\renewcommand{\tabcolsep}{0.09cm}
\caption{\footnotesize
Properties of main regions of RXCJ0658--5557.}
\label{t:cl14:t}%

\begin{tabular}{cccccccccccc}
 \hline
 \hline
N  &$kT$ &  $\rho_e$ & S & P, $10^{-12}$ & $M_{\rm gas}$ & $r_{\rm min}$ &
$r_{\rm max}$& Remarks\\
  & keV     &  $10^{-4}$ cm$^{-3}$& keV cm$^2$ & ergs cm$^{-3}$ & $10^{12} M_\odot$   & Mpc & Mpc  &\\
\hline
 1 &$12.1\pm1.0$&$ 85.2\pm1.9$&$ 290\pm  26$&$165.3\pm14.8$&$ 2.7\pm0.1$&0.00&0.22& P core-1    \\
 2 &$14.0\pm1.8$&$ 10.8\pm0.3$&$1332\pm 169$&$ 24.3\pm 3.1$&$18.6\pm0.5$&0.45&1.12& shock       \\
 3 &$10.4\pm0.6$&$ 21.3\pm0.3$&$ 626\pm  35$&$ 35.3\pm 2.0$&$20.6\pm0.3$&0.14&0.63& tail        \\
 4 &$ 8.7\pm1.0$&$  4.0\pm0.1$&$1593\pm 190$&$  5.6\pm 0.7$&$59.6\pm1.5$&0.61&1.79& main-2      \\
 5 &$12.2\pm0.9$&$ 34.2\pm0.6$&$ 540\pm  42$&$ 67.1\pm 5.3$&$ 8.8\pm0.2$&0.24&0.58& P core-3    \\
 6 &$ 9.4\pm0.7$&$ 43.1\pm0.9$&$ 354\pm  27$&$ 64.6\pm 5.0$&$ 4.4\pm0.1$&0.11&0.45&bullet tail N\\
 7 &$ 9.9\pm0.6$&$ 64.3\pm1.1$&$ 286\pm  17$&$101.8\pm 6.2$&$ 4.0\pm0.1$&0.11&0.32&bullet tail W\\
 8 &$11.4\pm1.0$&$ 51.0\pm1.0$&$ 383\pm  33$&$ 92.9\pm 8.0$&$ 5.1\pm0.1$&0.07&0.39& P core-2    \\
 9 &$10.0\pm0.5$&$ 12.2\pm0.2$&$ 871\pm  48$&$ 19.5\pm 1.1$&$42.0\pm0.6$&0.30&1.13& main-1      \\
10 &$ 8.0\pm0.6$&$ 64.7\pm1.9$&$ 230\pm  17$&$ 82.5\pm 6.5$&$ 1.6\pm0.0$&0.27&0.43& bullet peak \\
11 &$ 7.7\pm0.5$&$ 45.8\pm0.9$&$ 281\pm  17$&$ 56.9\pm 3.5$&$ 4.1\pm0.1$&0.25&0.52& bullet front\\
\hline
\end{tabular}
}
\end{center}
}
%\vspace*{0.2cm}
\end{table*}

Famous for its Chandra image (Markevitch et al. 2002), the bullet cluster
has some distinct features, which also allow us to understand the
observation of other clusters. With a Mach number of 3, deduced from the
shape of the bullet itself (angle of the Mach cone), the subcluster makes an
entropy enhancement in front of it. There are two other large entropy peaks
behind and to the south from the bullet. Apart from the small-scale
structure in the center, there appears to be a lack of features on the
pressure map, which we attribute to the propagation of the shock out to
large radii, thus strongly reducing the contrast. Therefore, the bullet
indicates a situation of a strong merger that is just completed in the
center and now moves to outskirts.  The entropy structure of the core of the
main cluster appears disrupted, yet the minimum is retained, while becoming
shallow. In the temperature map we see clear signatures of turbulence, as
indicated by the stochastic fluctuations, which in other clusters correspond
to a late stage of merging. This once again demonstrates that the time
scales for the relaxation are very different for the cluster center and
outskirts. On the largest scale the pressure as well as the image appears to
be quite smooth.

A weak lensing mass reconstruction of Clowe et al. (2003) shows that the
cluster exhibits three dark matter peaks, with only the weakest of them
corresponding to an X-ray peak, yet all of them are preceded by a shocked
zone, seen in the entropy map. The main pressure peak is approximately
located at the position of the center of the mass distribution from the weak
lensing reconstruction. This center is adopted for the volume calculations
and reported in Table \ref{t:ol}. The entropy dip of the bullet is offset
from the potential minimum and there is no entropy dip associated with the
potential minimum of the main cluster. There are, however, entropy
fluctuations in the pressure core, possibly associated with debris of the
entropy core of the main cluster.

The spectroscopic analysis is reported in Table \ref{t:cl14:t} and
Fig.\ref{f:cl14}. It reveals temperature fluctuations by a factor of 1.5.
The temperature of the bullet is only slightly lower than the bulk of the
cluster. However it exhibits a distinctly low entropy, which also allows us
to trace the tail of the bullet.  The zone assigned to bullet can be seen as
negative deviation in entropy profile in Fig.\ref{f:cl14}. The bullet
pressure peak is confirmed; it amounts to 20\% and is located behind the
zone of lowest entropy in the bullet.

By combining together all the high-entropy zones, associated with the shock
heating, we have achieved 99\% significance in the temperature variation,
from 10. to $14\pm2$ keV. This corresponds to a Mach number of $1.4\pm0.2$.
This estimate is lower, compared to the shock parameters deduced from the
image showing the Mach cone. A higher Mach number would be obtained from the
entropy enhancement: $2.6\pm0.2$. It is plausible that the extraction region
captures both shock and postshock gas. The later has lower pressure, but
records its state in the entropy. As was noted above, the observed shock is
located in front of the outward moving dark matter potential. Since the
potentials carry no longer any gas, they do not cause this shock, but just
travel at the same speed. This implies that we observe the initial forward
shock propagating through the cluster.

The entropy ratio also shows that the eastern part of the cluster has lower
entropy, as due to the stripping of the bulk of the bullet cluster.

\section{Discussion and Summary}

An analysis of the two-dimensional structure in the REFLEX clusters, as seen
in the images and spectral hardness ratio maps, reveals statistically
significant substructure, probably originating from different stages of
cluster merger. We are able to see the substructure even at very late merger
stages, where for example the X-ray image appears to be quite symmetric. We
identify the entropy to be most sensitive to both late stage mergers with
the associated slow buoyancy action of relaxation of the cluster and to
strong shocks, which change the entropy.  Two mergers with large Mach
numbers are found.

A statistical analysis of the substructure in the pressure and entropy maps,
reveals significant fluctuations around the mean profile. Typically,
pressure fluctuations are found on the 30\% level, while the entropy
fluctuations are at the 20\% level. Apparently, smoother appearance of the
pressure maps should be attributed to the larger dynamical range of the map,
covering typically two orders of magnitude. A comparison of our sample with
a similar analysis of hydro-dynamical simulations by Finoguenov et
al. (2005) reveals a similar distribution of clusters vs the level of the
substructure in both entropy and pressure.

A number of clusters exhibit a presence of low entropy gas in the outskirts,
deviating by at least an order of magnitude from the prescription of
gravitational heating. Surprisingly enough, these regions have gas pressures
similar to that of the cluster at a similar distance from the center. This
argues in favor of these regions being embedded in the cluster gas and
maintaining the pressure equilibrium, therefore revealing a medium survived
from the accretion shock heating. Existence of this effect has been
suggested by the simulations (e.g. Motl et al. 2004), but has so far only
been reported for A85 (Durret et al. 2005).

Incomplete (in a sense of being on-going) shock propagation in clusters soon
after the major merging event could also be a cause of the low entropies
seen at the outskirts.  In fact, the clusters in the advanced stage of
interaction have systematically higher entropy at $r_{500}$ compared to the
average trend.

\begin{acknowledgements}
The paper is based on observations obtained with XMM-Newton, an ESA science
mission, with instruments and contributions directly funded by ESA Member
States and the USA (NASA). The XMM-Newton project is supported by the
Bundesministerium f\"ur Bildung und Forschung/Deutsches Zentrum f\"ur Luft-
und Raumfahrt (BMFT/DLR), the Max-Planck Society and the
Heidenhain-Stiftung, and also by PPARC, CEA, CNES, and ASI. The authors
thank the referee, Florence Durret, for the constructive comments and her
appreciation of their efforts. AF thanks Alastair Sanderson and Joe Mohr for
useful discussions. A partial support from NASA grant NNG04GF68G to UMBC is
acknowledged. AF acknowledges support from BMBF/DLR under grant 50 OR 0207
and MPG.
\end{acknowledgements}

\end{document}